\documentclass[a4paper,11pt]{article}
\pdfoutput=1 

\usepackage{jheppub} 
                    
\usepackage{arydshln}
\usepackage[T1]{fontenc} 

\usepackage{graphicx,subfigure}
\usepackage{amsmath,amssymb,amsfonts}
\usepackage{bm}
\usepackage{color}
\usepackage{lineno}
\usepackage{comment}
\graphicspath{{./plot/}}

\newcommand{\Eq}[1]{Eq.~\eqref{#1}}

\newcommand{\bs}[1]{\boldsymbol #1}
\newcommand{\ftotal}{{\mathcal{F}}}
\newcommand{\ftotalbis}{{\widetilde{f}}}

\def \be {\begin{equation} }
\def \ee {\end{equation}}
\def \bes {\begin{subequations} }
\def \ees {\end{subequations}}

\def \a {\alpha}

\def \d {\delta}

\def \o {\omega}

\def \vp {\bm{p}}
\def \vq {\bm{q}}

\def \vk {\bm{k}}
\def \phase {(2\pi)^3}
\def \phaseAll {\vp', \vk', \vk}
\def \<{\langle}
\def \>{\rangle}
\def \+{\dagger}
\def \({\left(}
\def \){\right)}

\def \[{\left[}
\def \]{\right]}

\def \ra {\rightarrow}

\def \tnew {\tilde{t}}
\def \unew {\tilde{u}}
\def\Im{{\rm Im}}

\def \no {\nonumber}

\def \qperp
\def \CC {{\cal C}}
\def \CJ {{\cal J}}
\def \CM {{\cal M}}

\def \qperp {q_{\perp}} 
\def \CT {\Delta t}

\def \bQ {\bar{Q}}
\def \all {{\rm all}}
\def \hard {{\rm hard}}
\def \single{{\rm single}}
\def\PT {\mathcal{P}}

\newcommand{\GeV}{\text{GeV}}

\def\lra{\leftrightarrow}
\def\ca{C_{\rm A}}
\def\cf{C_{\rm F}}
\def\da{d_{\rm A}}
\def\df{d_{\rm F}}

\def \min {\rm min}

\def \kT {k_{T}}
\def \tnew {\tilde{t}}
\def \unew {\tilde{u}}
\def \snew {\tilde{s}}
\def \Mnew {{\cal M}}
\def \pinit {p_{\rm in}}

\begin{document}

\vspace{5mm}
\preprint{MIT-CTP-4997}

\title{\bf
Moli\`ere Scattering in Quark-Gluon Plasma:
Finding Point-Like Scatterers in a Liquid
}

\author[a,b]{Francesco  D'Eramo,}
\author[c]{Krishna Rajagopal,}
\author[c]{Yi Yin}
\affiliation[a]{Dipartimento di Fisica e Astronomia, Universit\`a di Padova, Via Marzolo 8, 35131 Padova, Italy}
\affiliation[b]{INFN, Sezione di Padova, Via Marzolo 8, 35131 Padova, Italy}
\affiliation[c]{Center for Theoretical Physics, Massachusetts Institute of Technology, Cambridge, MA 02139, USA}

\emailAdd{francesco.deramo@pd.infn.it}
\emailAdd{krishna@mit.edu}
\emailAdd{yiyin3@mit.edu}

\abstract{
By finding rare (but not exponentially rare) large-angle deflections of partons within a jet produced in a heavy ion collision, or of such a jet itself, experimentalists can find the weakly coupled short-distance quark and gluon particles (scatterers) within the strongly coupled liquid quark-gluon plasma (QGP) produced in heavy ion collisions. This is the closest one can come to probing QGP via a scattering experiment and hence is the best available path toward learning how a strongly coupled liquid emerges from an asymptotically free gauge theory. The short-distance, particulate, structure of liquid QGP can be revealed in events in which a jet parton resolves, and scatters off, a parton from the droplet of QGP. The probability for picking up significant transverse momentum via a single scattering was calculated previously, but only in the limit of infinite parton energy which means zero angle scattering. Here, we provide a leading order perturbative QCD calculation of the Moli\`ere scattering probability for incident partons with finite energy, scattering at a large angle.  We set up a thought experiment in which an incident parton with a finite energy scatters off a parton constituent  within a ``brick'' of QGP, which we treat as if it were weakly coupled, as appropriate for scattering with large momentum transfer, and compute the probability for a parton to show up at a nonzero angle with some energy. We include all relevant channels, including those in which the parton that shows up at a large angle was kicked out of the medium as well as the Rutherford-like channel in which what is seen is the scattered incident parton. The results that we obtain will serve as inputs to future jet Monte Carlo calculations and can provide qualitative guidance for how to use future precise, high statistics, suitably differential measurements of jet modification in heavy ion collisions to find the scatterers within the QGP liquid.
}

\date{\today}
\maketitle

\section{Introduction
\label{sec:intro}
}

When the short-distance structure of quark-gluon plasma is resolved, it 
must consist of weakly coupled
quarks and gluons because QCD is asymptotically free. 
And yet, at length scales of order its inverse temperature $1/T$ and longer,
these quarks and gluons become
so strongly correlated as to form a liquid.
Heavy ion collisions at the Relativistic Heavy Ion Collider (RHIC) and the Large Hadron Collider (LHC) produce droplets of this liquid QGP
whose expansion and cooling is well described
by relativistic viscous hydrodynamics with an unusually small viscosity 
relative to its entropy density. (For reviews, see Refs.~\cite{Heinz:2013th,Romatschke:2017ejr,Busza:2018rrf}.)
This discovery poses a question: how does this strongly coupled liquid emerge (as a function of coarsening resolution scale) from an asymptotically free gauge theory? 
In other contexts, the path to addressing a question like this about some newly discovered complex strongly correlated 
form of matter would begin
with doing scattering experiments, and in particular would begin with doing scattering experiments
in which the momentum transfer is large enough that the microscopic constituents (in our case, weakly coupled at short distance scales) are resolved.  Some analogue of such high resolution scattering experiments are a necessary first step toward understanding the microscopic structure and inner workings of QGP.
Since the droplets of QGP produced in heavy ion collisions rapidly cool and turn into an explosion of ordinary hadrons, the closest that anyone can come to doing scattering experiments off QGP is to look for the scattering of energetic ``incident'' partons that are produced  in the same collision as the droplet of QGP itself.  Since such energetic partons shower to become jets, this  provides
one of the motivations for analyzing how jets produced in heavy ion collisions are modified via their passage through QGP.  
Pursuing such measurements with the goal
of understanding the microscopic workings of QGP 
has been identified~\cite{Akiba:2015jwa,Geesaman:2015fha,LongRangeEU} as a central goal for the field once
higher statistics jet data anticipated in the 2020s, at RHIC
from the coming sPHENIX detector~\cite{Adare:2015kwa} and at the LHC from higher luminosity running, are in 
hand.

The short-distance,
particulate, structure of liquid QGP can be revealed by seeing 
events in which a jet parton resolves, and scatters off,
a parton from the droplet of QGP.
If the QGP were a liquid at all length scales, with no particulate microscopic constituents at all, as for example is the case in the infinitely strongly coupled
conformal plasma of ${\cal N}=4$ supersymmetric Yang-Mills (SYM) theory, 
then the probability for
an energetic parton plowing through it to pick up some momentum $\qperp$ transverse
to its original direction is Gaussian distributed in $\qperp$~\cite{Liu:2006ug,DEramo:2010wup,DEramo:2012uzl}, meaning that large-angle, large
momentum transfer, 
scattering is exponentially (maybe better to say ``Gaussianly'') rare.
The $\qperp$ distribution should similarly be Gaussian for the case of an energetic parton plowing through the QGP of QCD --- as long as $\qperp$ is not too large.  One way to see this is to realize that as long as $\qperp$ is small enough the energetic parton probes the QGP on long enough wavelengths  and ``sees'' it as a liquid.  Another way to reach the same conclusion is to imagine the not-too-large
$\qperp$ as being built up by multiple soft (low momentum transfer; strongly coupled) interactions with the QGP.
The key point, though, is that in QCD, unlike in ${\cal N}=4$ SYM theory, this cannot be the full
story:
real-world QGP must be particulate when its short-distance structure is resolved. This
means that large-angle, high momentum transfer, scattering may be rare but  is not Gaussianly rare, as
Rutherford would have understood.
So, if experimentalists can detect rare (but not Gaussianly rare) large-angle deflections of 
jet partons plowing through QGP, referred to as ``Moli\`ere scattering'' after the person who
first discussed the QED analogue~\cite{Moliere:1947zza,Moliere:1948zz,Moliere:1955zz}, 
they can find its weakly coupled quark and gluon 
constituents~\cite{DEramo:2012uzl,Kurkela:2014tla} and begin to study how the strongly coupled liquid emerges from its microscopic structure.

One idea for how to look for large angle scattering is to look for deflections
of an entire jet~\cite{DEramo:2012uzl} by looking for an increase in the ``acoplanarity'' of dijets or gamma-jets (meaning the angle by which the two jets or the photon and jet 
are not back-to-back) in heavy ion collisions relative
to that in proton-proton collisions.  
The acoplanarity is already quite significant in proton-proton collisions because many dijets (or gamma-jets) are not back-to-back because they are two jets (or a photon and a jet) in an event with more jets.  This makes 
it challenging to detect a rare increase in acoplanarity due to rare large-angle scattering, but
these measurements have been pursued by CMS~\cite{Chatrchyan:2012gt,Sirunyan:2017qhf}, ATLAS~\cite{ATLAS:2016} 
and ALICE~\cite{Adam:2015doa} at the LHC and
by STAR~\cite{Adamczyk:2017yhe} at RHIC, and it will be very interesting to see their precision increase
in future higher statistics measurements.
The same study can be done using events with one (or more, unfortunately) jets produced (only approximately) back-to-back with a Z-boson, albeit with lower statistics~\cite{Sirunyan:2017jic}.
It was realized in Ref.~\cite{Kurkela:2014tla} that Moli\`ere scattering can also be found by looking for rare large-angle scattering of partons within a jet shower, rather than of the entire jet.
We shall see that this is advantageous in that it allows one to consider energetic partons within a jet with only,
say 20 or 40 GeV in energy, whose kinematics allow for larger angle scattering than is possible
if one considers the deflection of (higher energy) entire jets.
However, the jet substructure observables
needed to detect rare large angle scattering of partons within a jet (via measuring 
their modification in jets produced in heavy ion collisions) are of necessity more complicated
than acoplanarity.  It is very important that such observables are now
being measured~\cite{Acharya:2017goa,Sirunyan:2017bsd,Sirunyan:2018gct,Aiola:2017uym,Acharya:2018uvf} and analyzed in heavy ion collisions, as it remains to
be determined which substructure observables, defined with which grooming prescription,
will turn out to be most effective.
Quantitative predictions for experimental observables, whether acoplanarities or substructure observables, require analysis of jet production and showering at the level of a jet Monte Carlo, 
first in proton-proton collisions and then embedded within a realistic hydrodynamic model for the expanding cooling droplet of 
matter produced in a heavy ion collision.
We shall not do such a study here; our goal is to provide a key theoretical input for future phenomenological
analyses, not to do phenomenology here.
Nevertheless, we expect that at a qualitative level our results
can provide some guidance for planning experimental measurements to come.

%
%
%
\begin{figure} 
\center
\includegraphics[width=0.55\textwidth]{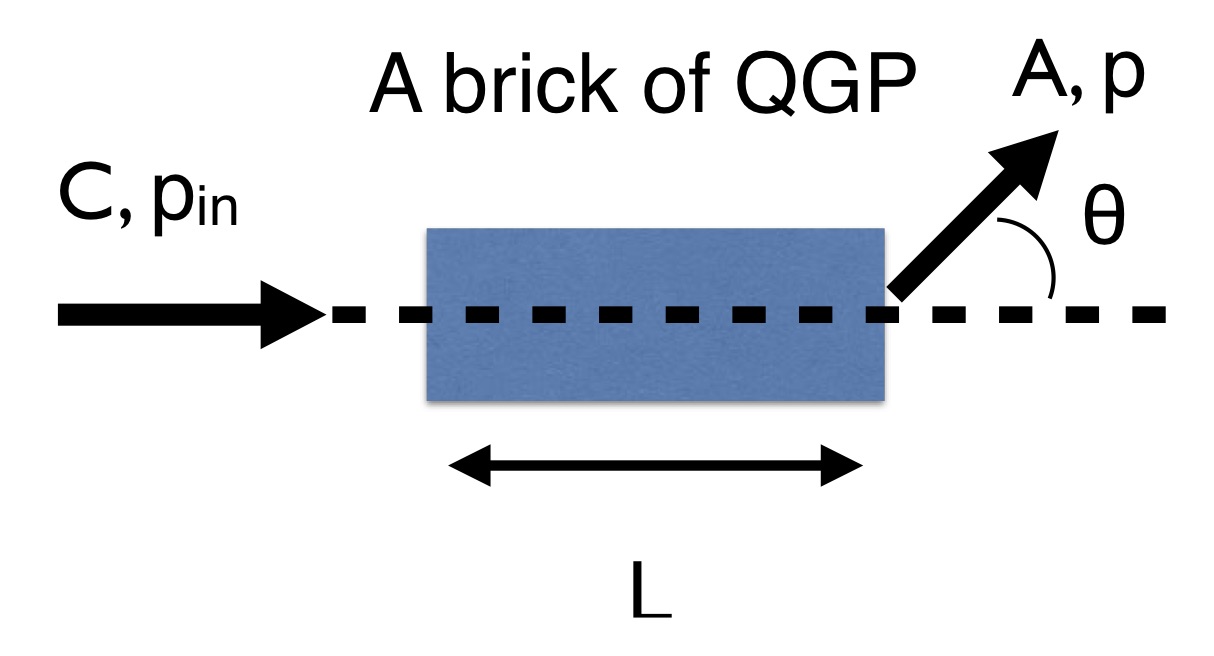}
\caption{
\label{fig:kinematic}
Kinematics of the thought experiment that we analyze. An incident parton of ``type'' $C$ (type meaning gluon or quark or antiquark) with energy $p_{\rm in}$ impinges on a ``brick'' of QGP with thickness $L$.   An outgoing parton of type $A$ with energy $p$ is detected at an angle $\theta$ relative to the direction of the incident parton.  We shall calculate the probability distribution of $p$ and $\theta$ for a given $p_{\rm in}$ and for all possible choices $A$ and $C$.   
  }
 \end{figure}
 %
 %
%

In this paper, we set up a thought experiment in which we ``shoot'' a single energetic
parton (quark or antiquark or gluon) with initial energy
$\pinit$ through a static ``brick'' of QGP of thickness $L$ in thermal equilibrium at a constant temperature $T$, c.f.~Fig.~\ref{fig:kinematic}.
For simplicity, we shall model the medium within our brick as a cloud of massless quarks and gluons, with Fermi-Dirac and Bose-Einstein momentum distributions, respectively.  This
is surely only of value as a benchmark.  Although treating the partons as massless is appropriate if the momentum transfer is high enough, as we shall quantify in Section~\ref{sec:validity}, adding thermal masses would surely be a worthwhile extension of our study.  Also, our calculations could be repeated in future using any
proposed model for the momentum distributions of the quarks and gluons as seen by a
high-momentum probe. 
Indeed, it is hard to imagine a better possible future than the prospect of making experimental measurements
that reveal the presence of rare large-angle Moli\`ere scattering, seeing quantitative
disagreements with predictions obtained via incorporating our calculation within a jet Monte Carlo analysis, and reaching the conclusion that the momentum distributions of the quarks and gluons
seen by a high-momentum probe differ from the benchmark distributions that we have chosen.

We shall then compute
$F(p,\theta)$,
the probability  distribution for finding an outgoing hard parton with energy $p$ and angle $\theta$ relative to the direction of the incident hard parton. 
We choose to normalize the distribution $F\(p,\theta\)$ as
\begin{eqnarray}
\label{N-min}
\int^{\pi}_{\theta_{\min}} d\theta\, \int^{\infty}_{p_{\min}} dp\, F\(p,\theta \)
= N_{\rm hard}\(\theta_{\min}\)\, , 
\end{eqnarray}
where $N_{\rm hard}\(\theta_{\min}\)$ denotes the number of outgoing hard partons in a specific region of the phase space $\theta\geq\theta_{\min}$, $p\geq p_{\min}$ per single incident parton.
We have introduced a somewhat arbitrary hard energy scale $p_{\min}$ so that we can
refer to a parton with $p>p_{\min}$ as a hard parton. 
We will specify $p_{\min}$ as needed in Sec.~\ref{sec:results}, and will always choose $p_{\min}$ to
be significantly greater than $T$.
$F\(p,\theta\)$ will depend on the temperature of the plasma, $T$, on the energy of the incident parton, $\pinit$, on the time that the parton
spends traversing the brick of QGP, $\CT\equiv L/c$, 
as well as on whether the incident parton and the
outgoing parton are each a quark, antiquark or gluon, but we shall
keep all these dependences implicit in our notation in this Introduction.

It should be evident that our thought experiment is only that.  The droplet of QGP produced in a heavy ion collision expands and cools rapidly; its dynamics is certainly not that of a constant temperature static brick.  And, a jet shower is made up from many partons and has a complex showering dynamics of its own.
In order to do phenomenology, our results for $F(p,\theta)$ must be incorporated within
a Monte Carlo calculation of jet production and showering, with the jets embedded within
a realistic hydrodynamic description of a droplet of QGP.  Such a future calculation, in which
the dynamics of a jet  (including the splitting and propagation) and of the droplet of plasma is described $\Delta t$ by $\Delta t$ by $\Delta t$, for some small value of $\Delta t$, after each $\Delta t$ our result for $F(p,\theta)$ could be applied to each parton in the shower.
In this way, our results can be used to add large-angle Moli\`ere scattering to a jet Monte Carlo 
calculation which does not currently include it, like for example the Monte Carlo calculations done within the hybrid model in Refs.~\cite{Casalderrey-Solana:2014bpa,Casalderrey-Solana:2015vaa,Casalderrey-Solana:2016jvj,Hulcher:2017cpt}.  In the case of a Monte Carlo calculation in which
hard two-to-two scattering is already included, for example those done within 
JEWEL~\cite{Zapp:2008af,Zapp:2008gi,Zapp:2012ak,Zapp:2013vla,Zapp:2013zya,KunnawalkamElayavalli:2017hxo},
MARTINI~\cite{Schenke:2009gb} or LBT~\cite{Wang:2013cia,He:2015pra,Cao:2017zih},
our results can be used in a different way, namely as a benchmark against which to compare 
for the purpose of identifying observable consequences of large-angle scattering.
The other way in which the results of our calculation will be of value is as a qualitative
guide to experimentalists with which to assess how large the effects of interest 
may turn out to be, namely as a qualitative guide to what the probability is that a parton
with a given energy in a jet could scatter by an angle $\theta$.
In Section~\ref{sec:NhardEstimates} we shall illustrate our results by plotting what we obtain for 
partons with $\pinit=25\, T = 10$~GeV and $\pinit=100\, T=40$~GeV and $\pinit=250\, T=100$~GeV incident on
a brick
with $T=0.4$~GeV and $\Delta t=3$~fm.

Although we believe that our results will be of value as a qualitative guide for planning and assessing future experiments,
giving a sense of just how rare it should be for a parton in a jet to scatter at a large enough angle that the jet grows a new prong that can be discerned via high-statistics measurements of suitably defined jet substructure observables, there should be no illusion that this will be a straightforward program.  We do not anticipate any smoking guns to be found. As an object lesson, it is worth considering the question of 
how to detect evidence, in experimental data, for the Gaussian distribution of transverse kicks
$\qperp$ that all the partons in a jet must pick up as they traverse the plasma.  As we noted
above, the probability distribution for small $\qperp$ is Gaussian, with a width often denoted by 
$\hat q L$, after passage through plasma over a distance $L$ and this can be understood either via holographic calculations at strong coupling or as a consequence of multiple scattering in a weakly coupled picture.  
Constraints on the measured value of $\hat q$ all come from comparing calculations of
energy loss (not transverse kicks themselves) to experimental data on observables that are
sensitive to energy loss within a weakly coupled formalism in which $\hat q$ also controls
parton energy loss~\cite{Burke:2013yra}.  There is at present no clear experimental detection of the Gaussian distribution of
transverse kicks themselves.
The natural way to look for them is to look for increases in the angular width of jets, jet broadening, due
to propagation through plasma, as all the partons in a jet accumulate Gaussian-distributed transverse kicks.  In fact, it is with this in mind that these kicks are typically referred to
as transverse momentum broadening.
There are many extant measurements of the modification of jet shape observables in heavy ion collisions~\cite{Adamczyk:2013jei,Chatrchyan:2013kwa,Khachatryan:2015lha,Adam:2015doa,Khachatryan:2016erx,Acharya:2017goa,Adamczyk:2017yhe,Sirunyan:2018jqr,Sirunyan:2018gct,Acharya:2018uvf}, and many theorists have made efforts to turn these measurements
into constraints on transverse momentum broadening, for example see Refs.~\cite{Wang:2013cia,Chen:2016vem,Mueller:2016gko,Mueller:2016xoc,Casalderrey-Solana:2016jvj,KunnawalkamElayavalli:2017hxo,Milhano:2017nzm,Luo:2018pto}, but
there are two significant confounding effects that obscure transverse momentum broadening~\cite{Casalderrey-Solana:2016jvj}.
The first effect is that the energy and momentum ``lost'' by the jet becomes
a wake in the plasma which then in turn becomes soft particles spread over a large range of angles around the jet direction, carrying momentum in the jet direction. Some of this momentum gets reconstructed as a part of the jet, meaning that this contributes to jet broadening unless soft particles are groomed away~\cite{Tachibana:2014lja,Floerchinger:2014yqa,He:2015pra,Tachibana:2015qxa,Cao:2016gvr,Casalderrey-Solana:2016jvj,Tachibana:2017syd,KunnawalkamElayavalli:2017hxo,Milhano:2017nzm}.  
The second effect arises from the interplay between the fact that higher energy jets are less
numerous than lower energy jets and the tendency for narrow jets to lose less energy
than wide jets. (This tendency is seen at weak coupling~\cite{CasalderreySolana:2010eh,Milhano:2015mng}, in holographic models for jets at strong coupling~\cite{Chesler:2015nqz}, and in the hybrid model~\cite{Casalderrey-Solana:2016jvj}.)
As a consequence, the jets that remain in any given energy bin after an ensemble of jets
passes through a droplet of QGP tend to be narrower than the jets in that energy bin would
have been absent the QGP: wider jets are pushed into lower energy bins, where they
are much less numerous than the narrower jets found there~\cite{Milhano:2015mng,Rajagopal:2016uip,Casalderrey-Solana:2016jvj,Brewer:2017fqy}.
So, even though individual jets may broaden, at the ensemble level there is a strong
tendency for the jets with a given energy to be narrower after passage through the plasma
than jets with that energy would have been.
Before an experimental measurement of transverse momentum broadening can be made,
careful work must be done to find ways to evade, or precisely measure, both of these confounding effects.  Relative to our goals in this paper, this is a cautionary tale. Although what
we are looking for (jets sprouting an extra prong due to a parton within the jet
scattering at a large angle) sounds more distinctive, because such events will be rare
the effort will require high statistics, judicious choice of observables, and a very considerable phenomenological modeling effort.
Our results provide an initial input for such an effort.

The probability for picking up a given transverse momentum $\qperp$ via a single hard scattering 
off a parton in the plasma was calculated previously~\cite{Arnold:2008zu,DEramo:2012uzl}, but only in the limit of infinite parton energy which means zero angle scattering.  That is, these authors calculated the probability that an infinite energy parton picks up some
significant transverse momentum $\qperp$ in a Moli\`ere scattering, without changing its
direction.  
Since what is most relevant to any experimental observable is the scattering angle, it is hard
to use these results per se to gain guidance for what to expect in future experimental 
measurements.
Here, we remedy this by providing a
leading order perturbative QCD calculation of the Moli\`ere scattering probability 
for incident partons with finite energy, computing the probability distribution
for both the scattering angle and the energy of the outgoing parton.

The computation of $F(p,\theta)$ in weakly coupled QGP, even a static brick of weakly coupled QGP,
 is a multiscale problem and, in addition, there are different phase space regions where $F(p,\theta)$ is governed by different processes, as discussed schematically in Ref.~\cite{Kurkela:2014tla}. 
We specifically focus here on the kinematic regime in which 
the angle $\theta$ is sufficiently 
large that the dominant process is a single binary 
collision between the incident hard parton and a medium parton (a scatterer in the medium). 
For sufficiently large $\theta$,
the contribution from multiple scattering is not relevant since one single collision is more likely to give a large angle than multiple softer collisions in sum.  At smaller values of $\theta$, multiple softer collisions do add up and dominate, yielding a Gaussian distribution in the momentum transfer as 
discussed above.  We shall focus on the large $\theta$ regime which is more likely to be populated via a single Moli\`ere scattering 
\begin{eqnarray}
\label{2to2}
\textrm{incident parton}+\textrm{target medium parton}\to \textrm{outgoing parton}+X\, . 
\end{eqnarray}
The  second important way in which our calculation extends what has been done
before is that we include all relevant channels.  
The parton that
is scattered by a large angle need not be the incident parton, 
as in Rutherford scattering or deep inelastic scattering;
it could be a parton from
the medium that received a kick from the incident parton. We include this channel as well, and 
we shall see that in some kinematic regimes it is dominant.
That is, in Eq.~\eqref{2to2}
the outgoing hard parton (the one that we imagine detecting via its contribution to some jet substructure observable or, if the incident parton represents an entire jet, via 
its contribution to an acoplanarity),
as well as the $X$ which goes undetected in our thought experiment, can each 
be either the deflected incident parton or the recoiling parton from the medium that received a kick. 
$F(p,\theta)$ describes the energy and momentum transfer of the incident parton to the medium and contains information about the nature of the scatterers in QGP.

In this work, we shall evaluate $F(p,\theta)$ for sufficiently large $\theta$ by following the standard methods of perturbative QCD. 
We then determine the probability distribution $P(\theta)$ for the angle of an outgoing hard parton by integration over $p$:
\begin{equation}
\label{P-Theta}
P(\theta)=\int^{\infty}_{p_{\min}}\, dp\, F(p,\theta)\, . 
\end{equation}
Finally, we integrate $P(\theta)$ over $\theta$ to obtain $N_{{\rm hard}}\(\theta_{\min}\)$, see Eq.~\eqref{N-min}.
Our calculation allows us to estimate how rare large angle scatterings with some specified $\theta$ are and in this way can be used to provide qualitative guidance for the ongoing experimental search for evidence of point-like scatterers in QGP. 

This paper is organized as follows. 
In Section~\ref{sec:formalism}, 
we derive the expressions which relate $F(p,\theta)$ to a summation over all possible $2\lra 2$ scattering process and obtain a compact expression involving the phase-space integration over the scattering amplitudes weighted by the appropriate thermal distribution function. 
We then  describe how to sum over the individual processes 
as well as how to simplify the phase-space integration. 
The reader only interested in results, not in their derivation, can jump to Sec.~\ref{sec:results},
where we present our results and
compare them to previous studies, including the computations
done in the $\pinit\to \infty$ limit in Refs.~\cite{DEramo:2010wup,DEramo:2012uzl}. 
By considering incident partons with finite energy and including all relevant channels, our goal is to provide a quantitative tool for incorporation in future jet Monte Carlo calculations as well as
qualitative guidance for how to
use future precise, high statistics, suitably differential measurements of jet modification in heavy ion collisions to find the scatterers within the QGP liquid.

\section{Kinetic Theory Set-up and Calculation Details 
\label{sec:formalism}
}

In this Section, we explain how we derive the probability distribution 
$F(p,\theta)$ for finding an outgoing parton with energy $p$ at an angle $\theta$ relative
to the direction of the incident parton.
Our key ingredient is the phase-space distribution $f_{a}({\bs p}, t)$ 
\be
\begin{split}
f_{a}({\bs p}, t)  \; \equiv \;  & \text{Probability of finding an energetic parton of species $a$} \\ &  
\text{in a phase-space cell with momentum ${\bs p}$ at the time $t$,} \\ & \text{averaged over helicity and color states,} 
\end{split}
\label{eq:fdef}
\ee
where $a$ can be $u$, $\bar{u}$, $d$, $\bar{d}$, $s$, $\bar{s}$ or $g$. 
As emphasized in the definition, we neglect the dependence on helicity and color configurations. Although the phase-space distribution in principle can depend also on these variables, we assume that the medium is unpolarized and has no net color charge. Furthermore, if we average over the possible helicity and color configurations for the incoming hard probe, we are allowed to use the averaged distribution introduced in \Eq{eq:fdef}.   We shall set our calculation up as a calculation of the time evolution of $f_a({\bs p},t)$ in kinetic theory in which this distribution initially has delta-function support, describing the incident hard parton, and later describes the probability of finding an energetic parton
of species $a$ that has ended up with momentum ${\bs p}$ after a binary collision.

\subsection{Initial conditions}

We imagine a static brick of quark-gluon plasma, and we then imagine shooting 
an energetic parton with energy $p_{\rm in}$ and momentum ${\bs p}_{\rm in}$ at it. The on-shell condition reads $p_{\rm in}^2 = {\bs p}_{\rm in}^2$, therefore $p_{\rm in}$ denotes both the energy and the magnitude 
of the momentum for the incoming parton. 
(We shall assume that this parton does not radiate, split or shower during the time $\Delta t$ that it is traversing our brick of plasma, since our goal is to focus on large-angle scattering caused by a single binary collision. In future phenomenological studies in which our results are used within a jet Monte Carlo, results from our calculation would be used $\Delta t$ by $\Delta t$ by $\Delta t$, with the value of $\Delta t$ chosen small enough that radiation or splitting is negligible during a single $\Delta t$.)
If the energetic parton of species $a$ enters the medium at the initial time $t_I$, the initial condition for the phase space distribution function reads
\be
f_a({\bs p}, t_I) \equiv \frac{1}{\nu_a} f_I({\bs p}) 
\equiv \frac{1}{\nu_a} \frac{1}{V}\frac{4 \pi^2}{\pinit^2} \, \delta(p - p_{\rm  in}) \, \delta(\cos\theta - \cos\theta_{\rm in})  \ ,
\label{eq:BC}
\ee
where $V$ is a unit volume that will not appear in any results.
Here, we have fixed the initial energy and direction. Without any loss of generality we can take the $z$-axis to lie along the direction of the incident parton, which fixes $\cos\theta_{\rm in} = 1$. We normalize the expression in \Eq{eq:BC} in such a way that the incoming flux is one incoming parton per unit volume. The degeneracy factor $\nu_a$ is defined as 
\be
\label{nu-def}
\nu_a = \left\{ \begin{array}{llll}
2 \times (N_c^2 - 1) & & & a = \text{gluon} \\
2 \times  N_c & & & a = \text{quark or antiquark} \ ,
\end{array} \right. \ ,
\ee 
accounting for helicity and color configurations, with $N_c$  the number of colors.  And, for later convenience we have introduced the definition of a function $f_I(\bs p)$, where $I$ 
refers to initial and is not an index, that describes the species-independent momentum-distribution in the initial condition.

\subsection{Evolution of the phase-space distribution}
\label{sec:evolution}

We wish to answer the following question: if an incoming parton
enters the medium at the time $t_I$, what is the probability of finding an energetic parton of species $a$ (not necessarily the same as that of the incident parton) exiting on the other side with a given energy and at a given scattering angle? In order to give a quantitative answer, we need to track the evolution of the function $f_{a}({\bs p}, t)$. 
At time $t=t_I$, $f_a$ is zero for all $\bs p$ other than ${\bs p}={\bs p}_{\rm in}$; at later times, because the incident parton can scatter off partons in the medium $f_a$ can be nonzero at 
other values of $\bs p$, and in particular at nonzero angles $\theta$.
Henceforth, we shall evaluate  $f_{a}({\bs p}, t)$ at some nonzero angle $\theta$, meaning that $a$ labels the species of the energetic parton detected there.

The calculation of the time evolution of $f_{a}({\bs p}, t)$ is performed in Appendix~\ref{app:Boltz}, we report only the final result here. We assume that the probe scatters off a constituent of the medium at most once during its propagation through the medium over a time $\Delta t$. We will later come back to this approximation and check when it is legitimate, namely when $\Delta t$ is sufficiently small and/or when $\theta$ is sufficiently large so that no summation over multiple scattering is needed. Within this approximation, the phase space distribution at the time $t_I + \Delta t$ when the parton exits the medium takes the form
\be
\begin{split}
f_{a}({\bs p} , t_I + \Delta t) = \frac{\Delta t}{\nu_a} \sum_{\rm processes} & \,    \frac{1}{1 + \delta_{cd}} \int_{\phaseAll} \left| \mathcal{M}_{a b \leftrightarrow c d} \right|^2 \, \times \\ &
\left[n_{c}(p^\prime) f_{d}({\bs k}^\prime, t_I) + f_{c}({\bs p}^\prime, t_I) n_{d}(k^\prime)\right] \left[ 1 \pm n_{b}(k) \right] \ . 
\label{eq:finaldistr}
\end{split}
\ee
The form of this expression can be readily understood for all scattering processes 
except $q \bar q \leftrightarrow gg$ or $q \bar{q} \lra q^\prime \bar{q}^\prime$, where $q$ and $q^\prime$ are different flavors, as follows (although it applies to those processes too). 
Our convention is that the parton $a$ detected in the final state comes from parton $c$
in the initial state, and the undetected parton $b$ comes from parton $d$.
So, the $n_c f_d$ term in the result (\ref{eq:finaldistr}) 
corresponds to the case where the outgoing hard parton $a$ that
is detected came from the medium, having been kicked out of the medium
by the incident parton $d$, whereas
the $f_c n_d$ term corresponds to the case where the detected parton $a$ 
came from the incident parton $c$, which scattered off parton $d$ from the medium. 
The $\left[ 1 \pm n_B \right]$ factor (where the sign is $+$ if $b$ is a boson and $-$ if $b$ is a fermion) describes Bose enhancement or Pauli blocking 
and depends on
the occupation of the mode in which the undetected particle of species $b$
in the final state is produced.
The sum runs over all possible binary processes $a b \leftrightarrow c d$, with $\vp', \vk'$ ($\vp, \vk$) the momenta of $c,d$ ($a,b$). The phase space integral is written in a compact form
\begin{eqnarray}
\label{intphasespace}
\int_{\phaseAll} &\equiv& \frac{1}{2 p}\, \int \frac{d^{3}\vk}{2k \, \phase}\,\int \frac{d^{3}\vp'}{2p' \, \phase}\,\int \frac{d^{3}\vk'}{2k' \, \phase}\, 
\no\\ &\times& (2\pi)^{4}\, \d^{(3)}\(\vp+\vk-\vp'-\vk'\)\, \d\(p+k-p'-k'\) \ .
\end{eqnarray}
The squared matrix elements are summed over initial {\it and} final helicity and color configurations, without any average. The term with the Kronecker delta function accounts for the cases when $c$ and $d$ are identical particles. Finally, we must specify the ``soft'' medium distribution functions $n_a(p)$.  As we discussed in Section~\ref{sec:intro},
we shall choose to use distributions as if the quarks and gluons seen  in the QGP by a high-momentum
probe were massless, noninteracting, and in thermal equilibrium, meaning that $n_a(p)$ depends only on the
statistics and energy of the particle in the medium that is struck and is given by
\be
n_a(p) = \left\{ \begin{array}{llll}
\left[\exp(p/T) - 1\right]^{-1} & & & a = \text{gluon} \\
\left[\exp(p/T) + 1\right]^{-1} & & & a = \text{quark or antiquark} 
\end{array} \right. \ ,
\label{eq:BEandFD}
\ee
Note that we are considering a medium in which the chemical potential for baryon number vanishes, meaning that the equilibrium distributions for quarks and antiquarks are identical.
For this locally isotropic medium, the equilibrium distributions depend on the parton energy $p$ but not on the direction of its momentum. They are also time-independent, since we are considering a static brick of plasma with a constant $T$.
By taking a noninteracting gas of massless quarks, antiquarks and gluons, in thermal equilibrium,
as our medium we are defining a benchmark, not an expectation.  As we noted in Section~\ref{sec:intro}, we look forward to the day when comparisons between experimental data
and predictions made using our results incorporated within a jet Monte Carlo are being used
to determine how $n_a(p)$ for QGP differs from the benchmark that we have employed here.
A future program along these lines could be thought of as the analogue, for a thermal medium, of
determining the parton distribution functions for a proton.

Initially, at time $t_I$, $f_a$ takes on the form (\ref{eq:BC}) and is zero for all $\bs p$ except for
$\bs p=\bs p_{\rm in}$.  The expression (\ref{eq:finaldistr}) encodes the fact that after the incident parton has propagated through the medium
for a time $\Delta t$, because there is some nonzero probability that a $2\to 2$ scattering event occurred there is now some nonzero probability of finding a parton with any $\bs p$.

\subsection{QCD matrix elements}
\label{sec:matrixelements}

\begin{table} 
\begin{center}
\renewcommand{\arraystretch}{1.3}
\begin{tabular}{|c | c | c | c | c | c|}
\hline 
$n$ & Process & $\left|{\cal M}^{(n)}\right|^2 / g_s^4$ & $w^{(n)}_Q$ & $w^{(n)}_{\bar Q}$ & $w^{(n)}_G$ \\ \hline  \hline
$1$ & $q q \lra q q$ & $8\,  \frac{\df^2 \, \cf^2} {\da} \left( \frac{s^2+u^2}{t^2} + \frac{s^2+t^2}{u^2} \right) + 16\, \df \, \cf \left( \cf {-} \frac{\ca}{2} \right) \frac{s^2}{tu}$ &  
$1$ & $0$ & $0$ \\
$2$ & $\bar{q} \bar{q} \lra \bar{q} \bar{q}$ & $\left|{\cal M}^{(1)}\right|^2 / g_s^4$ &  $0$ & $1$ & $0$\\ \hdashline
$3$ & $q \bar{q} \lra q \bar{q}$ & $8\,  \frac{\df^2 \, \cf^2}{\da}   \left( \frac{s^2+u^2}{t^2} + \frac{t^2+u^2}{s^2} \right)
	    + 16\, \df \, \cf \left( \cf {-} \frac{\ca}{2} \right) \frac{u^2}{st}$ &  $1$ & $1$& $0$\\ \hdashline
$4$ & $q q^\prime \lra q q^\prime$ & $8\,  \frac{\df^2 \, \cf^2}{\da} \left( \frac{s^2+u^2}{t^2} \right)$ & $N_f - 1$ & $0$ & $0$ \\ 
$5$ & $\bar{q} \bar{q}^\prime \lra \bar{q} \bar{q}^\prime$ & $\left|{\cal M}^{(4)}\right|^2 / g_s^4$ &  $0$ & $N_f - 1$ & $0$ \\
$6$ & $q \bar{q}^\prime \lra q \bar{q}^\prime$ & $\left|{\cal M}^{(4)}\right|^2 / g_s^4$ &  $N_f - 1$ & $N_f - 1$ & $0$ \\ \hdashline
$7$ & $q \bar{q} \lra q^\prime \bar{q}^\prime$ & $8\, \frac{\df^2 \, \cf^2}{\da} \left( \frac{t^2 + u^2}{s^2} \right)$ &  $N_f - 1$ & $N_f - 1 $& $0$ \\ \hdashline
$8$ & $q \bar{q} \lra g g$ & $8\, \df \, \cf^2 \left( \frac{t^2 + u^2}{t u}  \right) - 8\, \df \, \cf \, \ca \left( \frac{t^2+u^2}{s^2} \right)$ &  $1$ & $1$ & $N_f$ \\ \hdashline
$9$ & $q g \lra q g$ & $-8\, \df \, \cf^2 \left( \frac{u}{s}  +  \frac{s}{u} \right) + 8\, \df \, \cf \, \ca \left( \frac{s^2 + u^2}{t^2} \right)$ &  $1$ & $0$ & $N_f$\\
$10$ & $\bar{q} g \lra \bar{q} g$ & $\left|{\cal M}^{(9)}\right|^2 / g_s^4$ &  $0$ & $1$ & $N_f$\\  \hdashline
$11$ & $g g \lra g g$ & $ 16\, \da \, \ca^2 \left( 3 - \frac{su}{t^2} - \frac{st}{u^2} - \frac{tu}{s^2} \right)
$ &  $0$ & $0$& $1$ \\
\hline
\end{tabular}
\end{center}
\caption{List of the binary collision processes that can produce a hard parton in the final state with large transverse momentum with respect to the incoming probe. 
Here, $q$ and $q'$ are quarks of distinct flavors, $\bar q$ and $\bar{q}'$ the associated antiquarks, and $g$ is a gauge boson (gluon). The third column lists explicit leading order expressions for the corresponding QCD squared matrix elements, in vacuum, summed over initial {\it and final} polarizations and colors, as a function of the standard Mandelstam variables $t=-2\(p' p-\vp'\cdot\vp\)$, $u=-2\(p' k -\vp'\cdot\vk\)$ and $s = -t -u$. (See Ref.~\cite{Arnold:2002zm}.) In a SU($N_{c}$) theory with fermions in the fundamental representation, we have for the dimensions of the representations and the Casimir factors $\df= \ca = N_{c}$, $\cf=\(N^2_{c}-1\)/(2N_{c})$, and $\da = 2\, \df \cf=N^2_{c}-1$. For SU(3) (i.e. QCD), $\df = \ca = 3$, $\cf = 4/3$, and $\da = 8$. Finally, we give the degeneracy factors $w^{(n)}_C$ appearing in \Eq{eq:probfinal}. Here, $N_f$ is the number of light flavors; we take $N_f=3$ throughout.
} 
\label{tab:QCDprocesses}
\end{table}

The formalism set up so far is valid for a generic theory with arbitrary degrees of freedom and arbitrary interactions giving rise to binary scattering processes, and relies principally just on the kinematics of the binary collisions. 
The specific dynamics becomes relevant only when we have to specify the matrix elements in \Eq{eq:finaldistr}. We do so here, in so doing specializing to QCD. We collect the results for the matrix elements for all processes relevant to our study  in Table~\ref{tab:QCDprocesses}. 
We label each process with an integer index ($n = 1, 2, \ldots, 11$), and we write the associated matrix element summing   over initial {\it and} final colors and polarizations. We also assign to each process a degeneracy factor $w^{(n)}$, different for each degree of freedom involved in the collision, which will be useful shortly. With these matrix elements in hand, we can evolve the initial phase-space distribution given in \Eq{eq:BC} by plugging it into \Eq{eq:finaldistr}. In this way, we obtain the phase-space probability after the incident parton has spent a time $\Delta t$ in the medium. 

In addition to  neglecting 
all medium-effects in the distribution functions (\ref{eq:BEandFD}) as we
discussed in Section~\ref{sec:evolution}, we shall do the same
in the QCD matrix
elements for $2\to 2$  collisions.  This means that we are assuming weak coupling throughout
and furthermore means that we can only trust our results in the kinematic regime
in which the energy and momentum transferred between the incident parton and
the parton from the medium off which it scatters is much larger than the  Debye mass.
We shall check this criterion quantitatively in Section~\ref{sec:validity}.

\subsection{Probability distribution after passage through the medium
\label{sec:probability}
}

Having derived the evolution of the phase-space distribution in \Eq{eq:finaldistr}, we can now define and compute
the probability distribution, which is the main result of this paper. Thus far, we have denoted different parton species with lower case letters (i.e. $a = u, \bar{u}, d, \bar{d}, s, \bar{s}, g$). It is convenient to introduce uppercase indices denoting different types of partons: gluons, quark and antiquarks (i.e. $A=G, Q, \bar{Q}$). We use this notation to define the probability distribution that we introduced in Fig.~\ref{fig:kinematic}:
\be
\begin{split}
F^{C \rightarrow A}(p, \theta; p_{\rm in})  \; \equiv \;  & \text{Probability of finding a parton of type $A$ with energy $p$} \\ &
\text{at an angle $\theta$ with respect to the direction of} \\ & \text{an incoming parton of type $C$ with energy $p_{\rm in}$.} 
\end{split}
\label{eq:Fdef}
\ee
This quantity is given by the sum over all possible processes with $C$ and $A$ in the initial and final state, respectively. Its explicit expression reads
\be
F^{C \rightarrow A}(p, \theta; p_{\rm in}) = 
V\,\frac{p^{2}\,\sin\theta}{\(2\pi\)^{2}} \sum_{a \in A}  \nu_{a} 
f_{a}(p,\theta; t_I + \Delta t) \ .
\label{eq:ProbMain}
\ee
The prefactor in front of the sum is the Jacobian of the phase-space integration
\be
V\, \frac{d^{3}\vp}{\(2\pi\)^{3}} = \frac{p^2 dp \, d\cos\theta \,d\phi}{\(2\pi\)^{3}} \qquad \qquad \Rightarrow \qquad \qquad 
V\,\frac{p^2 \sin\theta}{\(2\pi\)^{2}} \, dp \, d\theta \ ,
\ee
The sum runs over all the lowercase indices corresponding to parton species of the type $A$. For example, if $A$ stands for a quark, the sum runs over the values $a = u, d, s$. The degeneracy factor $\nu_a$ appears because our distribution functions are averaged over colors and polarizations; the detector cannot resolve these quantum numbers, we account for all of them by this multiplicative factor. Finally, we note that the distribution function $f_{a}(p,\theta; t_I + \Delta t)$ appearing in Eq.~(\ref{eq:ProbMain}) is the time-evolved quantity given in \Eq{eq:finaldistr}, evolved from
an initial condition at time $t_I$ given by
\be
\label{finit-case}
f_{a}(p_{\rm in}, t_I) = \left\{ \begin{array}{lccccl}
f_I({\bs p}_{\rm in}) / \nu_C & & & & & \textrm{for one value of } a \in C \\
0 & & & & & \textrm{for all other values of }  a
\end{array}
\right.
\ee
where the function $f_I({\bs p}_{\rm in})$ was defined in \Eq{eq:BC}.   (For example, if $C=Q$ meaning that the incident parton is a quark then $f_a$ is nonzero for either $a=u$ or $a=d$ or $a=s$, and the flavor of the incident quark makes no difference to our calculation.)

We have defined the probability (\ref{eq:Fdef}) such that it does not distinguish between quarks of different flavors, but it does distinguish between quarks, antiquarks and gluons. 
So, if our goal is to find the total probability of finding {\it any} energetic parton in the final state
with energy $p$ and angle $\theta$, we have to sum over the different types of partons. 
As an example, if we consider an incoming quark, the probability of getting {\it any} energetic parton in the final state reads
\be
\label{F-sum-Q}
F^{Q \rightarrow {\rm all}}(p, \theta; p_{\rm in}) = F^{Q \rightarrow Q}(p, \theta; p_{\rm in}) + F^{Q \rightarrow \bar{Q}}(p, \theta; p_{\rm in}) + F^{Q \rightarrow G}(p, \theta; p_{\rm in})  \ .
\ee

In the last step in our derivation, we directly plug the expression for the time-evolved 
phase-space distribution given in \Eq{eq:finaldistr} into our  expression for the probability distribution~\eqref{eq:ProbMain}. Before doing that, it is useful to introduce some notation to make our final expression more compact. We define the generalized Kronecker delta functions 
$\tilde{\d}_{a, G} \equiv \delta_{a, g}$, $\tilde{\d}_{a, Q}$ which equals $1$ if $a=u$ or $d$ or $s$
and which vanishes for other values of $a$, and $\tilde{\d}_{a, \bar Q}$ which equals
$1$ if and only if $a =\bar u$ or $\bar d$ or $\bar s$.
Moreover, we define the generalized medium ``soft'' distribution function
\be
n_{a}(p) = 
\tilde{\delta}_{a,G} \,n_{\rm B.E.}(p)+\left(\tilde{\delta}_{a,Q} +\tilde{\delta}_{a,\bar Q} \right)\,n_{\rm F.D.}(p) 
\ee
where $n_{\rm B.E.}(p)$ and $n_{\rm F.D.}(p)$ are the Bose-Einstein and Fermi-Dirac distributions from \Eq{eq:BEandFD}, respectively. With this notation in hand, we can now write the complete leading order expression for the probability function defined in Eq.~(\ref{eq:Fdef}):
\be
\begin{split}
F^{C \rightarrow A}(p, \theta; p_{\rm in}) = 
V\, \frac{\kappa}{T} \, & \, \frac{p^{2}\,\sin\theta}{\(2\pi\)^{2}} \sum_n  \, w_C^{(n)}  \, \frac{\tilde{\d}_{a, A}}{1 + \delta_{cd}} \int_{\phaseAll} 
\frac{\left| \mathcal{M}^{(n)}_{a b \leftrightarrow c d} \right|^2}{g_s^4} \, \times \\ &
\frac{1}{\nu_C} \left[ \tilde{\d}_{d, C} \;  f_I({\bs k}^\prime) \; n_{c}(p^\prime) +  \tilde{\d}_{c, C} \;  f_I({\bs p}^\prime) \; n_{d}(k^\prime)\right] \left[ 1 \pm n_{b}(k) \right] \ .
\end{split}
\label{eq:probfinal}
\ee
Here, we have defined a dimensionless parameter $\kappa$ multiplying the overall expression via
\begin{eqnarray}
\label{kappa-def}
\kappa \equiv g_s^4 \, T \, \Delta t \ .
\end{eqnarray}
$\kappa$ becomes large either for a thick brick (large $T \Delta t$) or for 
a large value of the QCD coupling constant $g_s$ that controls the magnitude of
all the matrix elements for binary collision processes.
Note that the $V$ in the prefactor of Eq.~\eqref{eq:probfinal} cancels the $1/V$ from Eq.~\eqref{eq:BC}, meaning that no $V$ will appear in any of our results.  Henceforth we shall not write the factors of $V$.
Note also that neglecting multiple scattering as we do is only valid when
$N_{\rm hard}$, the integral over $F^{C \rightarrow A}(p, \theta; p_{\rm in})$ defined
in Eq.~(\ref{N-min}), is small.  
For any given choice of $p$ and $\theta$, if $\kappa$ is too large multiple scattering cannot be neglected and our formalism breaks down.  Equivalently, for any given $\kappa$ our formalism will be valid in the regime of $p$ and $\theta$, in particular for large enough $\theta$, where $F^{C \rightarrow A}(p, \theta; p_{\rm in})$ is small and multiple scattering can be neglected.

The sum over $n$ in \Eq{eq:probfinal} runs over all the $11$ processes in Table~\ref{tab:QCDprocesses}. The delta $\tilde{\d}_{a, A}$ ensures that only processes with a parton of type $A$ present in the final state are accounted for. Crucially, each process is multiplied 
by the $C$-dependent weight factor $w_C^{(n)}$, given explicitly in the last three columns of Table~\ref{tab:QCDprocesses}. As an example, if we are considering the production of $A = Q$ from an incident gluon, $C=G$, via $gg \rightarrow \bar{q} q $, the weight factor $w_G^{(8)}$ is $N_f$ since we can produce this final state by pair-production of any flavor of light quark.
Thus, this multiplicative factor accounts for the multiple ways a given process can produce the energetic parton $A$ in the final state. When such an outgoing parton originates from an incident parton $c$, the matrix element has to be multiplied by the thermal weight $\tilde{\d}_{c, C} \;  f_I({\bs p}^\prime) \; n_{d}(k^\prime)$, whereas when the incoming parton is $d$ this factor is $\tilde{\d}_{d, C} \;  f_I({\bs k}^\prime) \; n_{c}(p^\prime)$.

The expression \Eq{eq:probfinal} is the central result of this paper, albeit 
written in a compact and hence relatively formal fashion.
We note again that 
this relation is valid only as long as $\Delta t$ is much shorter than the characteristic time between those binary collisions between the incident parton and constituents of the medium that produce scattered partons with a given $p$ and $\theta$. 
We will see in Section~\ref{sec:results} that this is true as long as the scattering angle is larger than some $\theta_{\rm min}$, where $\theta_{\rm min}$ will depend on $p$, $p_{\rm in}$ and $
\kappa$. 
Before turning to results in Section~\ref{sec:results}, 
in Section~\ref{sec:sum}  we shall 
write the expression (\ref{eq:probfinal}) more explicitly in specific cases and
in 
Section~\ref{sec:integration} we shall
describe some of details behind the computations via which we obtain our results.

\subsection{How to sum over different processes} 
\label{sec:sum}

In order to write the expression (\ref{eq:probfinal}) more explicitly and in particular in order
to sum the various different phase space integrals over various different matrix elements
that contribute to a given physical process of interest,
it is convenient to define the following set of phase space integrals:
\bes
\label{ave-def}
\begin{eqnarray}
\label{ave-def-a}
\<\,\(n\)\,\>_{D,B} &\equiv &
\frac{1}{T} \frac{p^{2}\,\sin\theta}{\(2\pi\)^{2}}\, \int_{\phaseAll}\, \frac{\left| \mathcal{M}^{(n)} \right|^2}{g_s^4} \, f_I(\vp')\, n_{D}(k')\, 
\[1\pm n_{B}\(k\)\]\, ,
\\
\label{ave-def-b}
\<\,\(\widetilde{n}\)\,\>_{D,B} &\equiv &
\frac{1}{T} \frac{p^{2}\,\sin\theta}{\(2\pi\)^{2}}\, \int_{\phaseAll}\, \frac{\left| \mathcal{M}^{(n)} \right|^2}{g_s^4} \, f_I(\vk')\, n_{D}(p')\, 
\[1\pm n_{B}\(k\)\]\, ,
\end{eqnarray}
\ees
where the index $n$ spans  the 11 different binary collision processes listed in Table~\ref{tab:QCDprocesses}. The $\pm$ sign in both equations correspond to the cases where $B$ is a boson or a fermion, respectively. For processes with identical incoming partons
(and also for process 8 in Table~\ref{tab:QCDprocesses}),
 we have $\<\(n\)\>_{D,B} =\<\(\widetilde{n}\)\>_{D,B}$. More explicitly, we have
\begin{eqnarray}
\label{M-symmetry}
\<\(n\)\>_{D,B} =\<\(\widetilde{n}\)\>_{D,B} \, , \qquad n = 1, 2, 7, 8, 11 \ .
\end{eqnarray}
If we look back at \Eq{eq:probfinal}, we notice that we can always express $F^{C \rightarrow A}(p, \theta; p_{\rm in})$ as a weighted sum over $\<\,\(n\)\,\>_{D,B}$ and $\<\,\(\widetilde{n}\)\,\>_{D,B}$. Obtaining such expressions is the goal of this Section.  There are $3 \times 3 = 9$ different cases, corresponding to three options for both the incoming and outgoing parton: quark, antiquark or gluon. 
We shall first list 4 cases, corresponding to choosing either quark or gluon. Replacing quarks by antiquarks gives 3 more cases, with identical results. We shall end with the 2 cases where the incoming and outgoing partons are quark and antiquark or vice versa. The brick of quark-gluon plasma is assumed to not carry a net baryon number, therefore the results for these last 2 cases are also identical. 
In the remainder of this subsection, we give explicit expressions for these 5 independent results. For each case, we define the partial contributions as follows
\be
F^{C \rightarrow A}(p, \theta; p_{\rm in}) \equiv \sum_n F_{(n)}^{C \rightarrow A}(p, \theta; p_{\rm in}) \ .
\ee
That is, we decompose the total probability that we are interested in into a sum of up to 11 different terms, one for each of the processes listed in Table~\ref{tab:QCDprocesses}. As we will see shortly, only a subset of them will actually contribute in each case. For example, in order to understand which ones are relevant to $F^{Q\rightarrow Q}(p,\theta;\pinit)$ we need to look at Table~\ref{tab:QCDprocesses} and identify those processes with at least one quark in the initial and in the final states. The final result for each case can then be expressed in terms of the functions defined in Eqs.~\eqref{ave-def-a} and \eqref{ave-def-b}.  Individual processes in Table~\ref{tab:QCDprocesses} can contribute in more than one case; for example, process 9, quark-gluon scattering,  contributes to the probabilities for four cases: $F^{Q\rightarrow Q}(p,\theta;\pinit)$, $F^{G\rightarrow Q}(p,\theta;\pinit)$, $F^{Q\rightarrow G}(p,\theta;\pinit)$ and $F^{G\rightarrow G}(p,\theta;\pinit)$.

\begin{description}
\item[$F^{Q \rightarrow Q}(p, \theta; p_{\rm in})$ (``incident quark, outgoing quark''):] 
We start from the case where both the incoming and the outgoing parton are quarks. The relevant processes are the ones labeled by $n = 1, 3, 4, 6, 7, 9$ in Table~\ref{tab:QCDprocesses} with individual expressions given as follows.  First,
\bes
\label{FQQ-n}
\begin{eqnarray}
F_{(1)}^{Q \rightarrow Q}(p, \theta; p_{\rm in}) &= &\frac{\kappa}{\nu_q} \frac{w^{(1)}_{Q}}{2} \left[ \<\,\(1\)\,\>_{Q,Q}+\<\,\(\widetilde{1}\)\,\>_{Q,Q} \right]  
\no\\
&=&\frac{\kappa}{2\nu_q} \left[ \<\,\(1\)\,\>_{Q,Q}+\<\,\(\widetilde{1}\)\,\>_{Q,Q} \right] 
= \frac{\kappa}{\nu_q}\,\<\,\(1\)\,\>_{Q,Q}\, ,   
\end{eqnarray}
where the factor $1/2$ is a symmetry factor (see Eq.~\eqref{eq:probfinal}), and $w^{(1)}_{Q}$ is read from Table.~\ref{tab:QCDprocesses}. 
In the last step, we have used the fact that $ \<\,\(1\)\,\>_{Q,Q}=\<\,\(\widetilde{1}\)\,\>_{Q,Q}$ according to the relation~\eqref{M-symmetry}.
Likewise,
\begin{eqnarray}
F_{(3)}^{Q \rightarrow Q}(p, \theta; p_{\rm in}) &= & 
\frac{\kappa}{\nu_q} \, \<\,\(3\)\,\>_{Q,Q}   \ , \\  
F_{(4)}^{Q \rightarrow Q}(p, \theta; p_{\rm in}) 
&=&\frac{\kappa}{\nu_q} (N_f - 1) \left[ \<\,\(4\)\,\>_{Q,Q}+\<\,\(\widetilde{4}\)\,\>_{Q,Q} \right]  \ , \\ 
F_{(6)}^{Q \rightarrow Q}(p, \theta; p_{\rm in}) &=& \frac{\kappa}{\nu_q} (N_f - 1) \, \<\,\(6\)\,\>_{Q,Q} 
\no \\
&=& \frac{\kappa}{\nu_q} (N_f - 1) \, \<\,\(4\)\,\>_{Q,Q} \ , 
\end{eqnarray}
since the squared matrix elements for the processes $4$ and $6$ are identical.
And, 
\begin{eqnarray}
F_{(7)}^{Q \rightarrow Q}(p, \theta; p_{\rm in}) &= & \, \frac{\kappa}{\nu_q} (N_f - 1) \, \<\,\(7\)\,\>_{Q,Q} \ , \\ 
F_{(9)}^{Q \rightarrow Q}(p, \theta; p_{\rm in})
&=& \, \frac{\kappa}{\nu_q} \,  \<\,\(9\)\,\>_{G,G}   \ .
\end{eqnarray}
\ees
Upon summing the above, we find the final result
\be
\label{FQQ}
\begin{split}
F^{Q \rightarrow Q}(p, \theta; p_{\rm in}) =  \frac{\kappa}{\nu_q} \left\{ \right. & \left.  \<\,\(1\)\,\>_{Q,Q}+  \<\,\(3\)\,\>_{Q,Q}  + \<\,\(9\)\,\>_{G,G} \, + \right. \\ &
\left. (N_f - 1) \left[ 2 \<\,\(4\)\,\>_{Q,Q} +  \<\,\(\widetilde{4} \)\,\>_{Q,Q}  + \<\,\(7\)\,\>_{Q,Q} \right]  \right\}  \ .
\end{split}
\ee

\item[$F^{Q \rightarrow G}(p, \theta; p_{\rm in})$ (``incident quark, outgoing gluon''):] This case gets contributions from the processes labeled by $n = 8, 9$. We identify again the individual contributions to the total probability
\bes
\begin{eqnarray}
\label{F8-QG}
F_{(8)}^{Q \rightarrow G}(p, \theta; p_{\rm in}) &= & \, \frac{\kappa}{\nu_q} \,\,\[ \<\,\(8\)\,\>_{Q,G} + \<\,\(\widetilde{8}\)\,\>_{Q,G} \]
\no \\
&=&
 \frac{2\kappa}{\nu_q}\, \<\,\(8\)\,\>_{Q,G} \, ,
 \end{eqnarray}
 where we have used the relation~\eqref{M-symmetry}. 
 And, 
\begin{eqnarray}
F_{(9)}^{Q \rightarrow G}(p, \theta; p_{\rm in}) &= & \, \frac{\kappa}{\nu_q} \,\<\,\(\widetilde{9}\)\,\>_{G,Q}   \, , 
\end{eqnarray}
\ees
which add up to give the final result for this case
\be
\label{FQG}
F^{Q \rightarrow G}(p, \theta; p_{\rm in}) = \frac{\kappa}{\nu_q} \,  \[2 \<\,\(8\)\,\>_{Q,G}  + \<\,\(\widetilde{9}\)\,\>_{G,Q} \] \ . 
\ee

\item[$F^{G \rightarrow Q}(p, \theta; p_{\rm in})$ (``incident gluon, outgoing quark''):] The calculation for this case is analogous to the previous one. The partial contributions read
\bes
\begin{eqnarray}
F_{(8)}^{G \rightarrow Q}(p, \theta; p_{\rm in}) &= & \,\frac{\kappa}{\nu_g} \,N_{f}\, \<\,\(8\)\,\>_{G,Q}  \, . 
\\ 
F_{(9)}^{G \rightarrow Q}(p, \theta; p_{\rm in}) &= & \, \frac{\kappa}{\nu_g} \, N_{f}\,\<\,\(\widetilde{9}\)\,\>_{Q,G}   \ , 
\end{eqnarray}
\ees
which, after summing, result in
\be
\label{FGQ}
F^{G \rightarrow Q}(p, \theta; p_{\rm in}) = \frac{\kappa}{\nu_g} \, N_f\,\left[ \<\,\(8\)\,\>_{G,Q} + \<\,\(\widetilde{9}\)\,\>_{Q,G} \right] \ . 
\ee

\item[$F^{G \rightarrow G}(p, \theta; p_{\rm in})$ (``incident gluon, outgoing gluon''):]  When both the incoming and outgoing energetic partons are gluons, the processes contributing to the probability distribution are the ones labeled by $n = 9, 10, 11$. The individual terms are
\bes
\begin{eqnarray}
F_{(9)}^{G \rightarrow G}(p, \theta; p_{\rm in}) &= & \, \frac{\kappa}{\nu_g} \, N_f \,\<\,\(9\)\,\>_{Q,Q}  \ , \\ 
F_{(10)}^{G \rightarrow G}(p, \theta; p_{\rm in}) &= & \, \frac{\kappa}{\nu_g} \, N_f  \,\<\,\(10\)\,\>_{Q,Q} 
=\frac{\kappa}{\nu_g} \, N_f \,\<\,\(9\)\,\>_{Q,Q} \, , 
\end{eqnarray}
where we have take into account the fact that processes $9$ and $10$ have identical squared matrix elements. 
And, 
\begin{eqnarray}
F_{(11)}^{G \rightarrow G}(p, \theta; p_{\rm in}) &= & \, \frac{\kappa}{\nu_g} \, \frac{1}{2} \left[ \<\,\(11\)\,\>_{G,G} + \<\,\(\widetilde{11}\)\,\>_{G,G} \right] 
 =\frac{\kappa}{\nu_g} \, \<\,\(11\)\,\>_{G,G} \,  \,
\end{eqnarray}
where once again we have used the relation~\eqref{M-symmetry}. 
\ees
Consequently, we find
\be
\label{FGG}
F^{G \rightarrow G}(p, \theta; p_{\rm in}) = \frac{\kappa}{\nu_g} \, \left\{  2 N_f \<\,\(9\)\,\>_{Q,Q} +  \<\,\(11\)\,\>_{G,G}  \right\} \ . 
\ee

\item[$F^{Q \rightarrow \bar{Q}}(p, \theta; p_{\rm in})$ (``incident quark, outgoing antiquark''):] The last case we consider is when a quark enters the medium and an energetic antiquark exits on the opposite side. The processes that contribute to this case are
\bes
\begin{eqnarray}
F_{(3)}^{Q \rightarrow \bar{Q}}(p, \theta; p_{\rm in}) &= & \, \frac{\kappa}{\nu_q} \, \<\,\(\widetilde{3}\)\,\>_{Q,Q}  \ , \\ 
F_{(6)}^{Q \rightarrow \bar{Q}}(p, \theta; p_{\rm in}) &= & \, \frac{\kappa}{\nu_q} \, (N_f - 1)  \,\<\,\(\widetilde{6}\)\,\>_{Q,Q}  
\no \\
&=&
 \frac{\kappa}{\nu_q} \, (N_f - 1)  \,\<\,\(\widetilde{4}\)\,\>_{Q,Q}\, ,  
\end{eqnarray} 
where we use the fact that processes $6$ and $4$ have identical squared matrix elements. 
In addition, 
\begin{eqnarray}
F_{(7)}^{Q \rightarrow \bar{Q}}(p, \theta; p_{\rm in}) &= & \, \frac{\kappa}{\nu_q} \, (N_f - 1) \, \<\,\(\widetilde{7}\)\,\>_{Q,Q} 
\no \\
&=& 
 \frac{\kappa}{\nu_q} \, (N_f - 1) \, \<\,\(7\)\,\>_{Q,Q} \, , 
\end{eqnarray}
where we have use the relation~\eqref{M-symmetry}. 
\ees
The total probability for this case is
\be
\label{FQQbar}
F^{Q \rightarrow \bar{Q}}(p, \theta; p_{\rm in}) = \frac{\kappa}{\nu_q} \left\{ \<\,\(\widetilde{3}\)\,\>_{Q,Q} + 
 (N_f - 1)  \left[ \,\<\,\(\widetilde{4}\)\,\>_{Q,Q} + \<\,(7)\>_{Q,Q}  \right] \right\} \ .
\ee

\end{description}

\subsection{Phase space integration
\label{sec:integration}
}

After performing the summation over different processes, our final task is to evaluate the phase space integrals in Eqs.~\eqref{ave-def-a} and \eqref{ave-def-b}. 
The expression in Eq.~\eqref{ave-def-a} involves a 9-fold integration in the phase space $\(\vp', \vk',\vp\)$. We first integrate over a 4-dimensional delta function in Eq.~\eqref{intphasespace}. 
The integration over the azimuthal angle is straightforward. Finally, we perform two more integrations by taking the advantage of the delta function in $f_{I}$. 
(See Appendix.~\ref{sec:int-derivation} for details.) Upon following techniques widely used in the literature (see e.g.~Refs~\cite{Moore:2001fga,Arnold:2003zc,Moore:2004tg}), we find
\begin{eqnarray}
\label{ave-alpha1}
\< \(n\)\>_{D, B}
&=&
\frac{1}{16\(2\pi\)^{3}}\(\frac{p\,\sin\theta}{p_{\rm in}\,q\, T}\)
\int^{\infty}_{\rm k_{\rm min}}\, d\kT\, n_{D}\(\kT\)\, \[1\pm n_{B}(k_{X}) \]\,
 \int^{2\pi}_{0} \frac{d\phi}{2\pi}\, 
\frac{\left| \mathcal{M}^{(n)} \right|^2}{g_s^4} \ .
\end{eqnarray}
Here, $\kT$ denotes the energy of the thermal parton from the medium 
whose momentum we shall denote by $\vk_T$
and $k_{X}=k+\o$ denotes the energy of the undetected 
final state parton. The integration range starts from the value
\begin{eqnarray}
\label{k-min}
k_{\rm min} = \frac{q-\o}{2}\, , 
\end{eqnarray}
corresponding to the minimum energy allowed by kinematics 
for the thermal parton from the medium. 
Moreover, $\phi$ is the angle between the two planes identified by the pair of vectors $(\vp,\vq)$ 
and $(\vq, \vk_{T})$, and we use $\o\equiv \pinit-p$ and $\vq=\vp-\pinit$ to denote energy and momentum difference between the incident parton and the outgoing parton that is detected. 
The matrix elements $\mathcal{M}^{(n)}$ that
appear in Eq.~(\ref{ave-alpha1}) are to be taken from Table~\ref{tab:QCDprocesses},
with the Mandelstam variables $t$ and $u$ occurring within them 
specified in terms of quantities $\tnew$ and $\unew$ 
that can be expressed as functions of $q$, $\o$, $\kT$ and $\phi$ as follows
\begin{eqnarray}
\label{t1}
\tnew&=&\o^{2}-q^{2}\, ,\qquad
\unew= -\snew-\tnew\, , 
\\
\label{s1}
\snew
&=&\(-\frac{\tnew}{2 q^{2}}\) \{\,\[ \(\pinit+p\)\,\(\kT+k_{X}\)+q^{2}\]-
\sqrt{\(4\pinit\,p+\tnew\)\,\(4 \kT k_{X}+\tnew\)}\cos\phi\, \}\, , 
\end{eqnarray}
where in the matrix elements in Eq.~(\ref{ave-alpha1}) 
we have simply 
 $t=\tnew$ and $u=\unew$ but where we will need to set
 $t=\unew$ and $u=\tnew$ below in our result for  $\< \( \widetilde{n} \) \>_{D,B}$.
Here, $q$, and $\tnew$ can be expressed as functions of $p$, $\pinit$ and $\cos\theta$ thus:
\begin{eqnarray}
\label{qt-in-theta}
q=\sqrt{\pinit^{2}+p^{2}-2\pinit\, p\cos\theta}\, ,
\qquad
\tnew= - 2p\, \pinit\(1-\cos\theta\)\, . 
\end{eqnarray}

Following a calculation that proceeds along similar lines, 
the quantity in Eq.~\eqref{ave-def-b} can be expressed as
\be
\begin{split}
\label{ave-alpha2}
 & \, \< \(\widetilde{n}\)\>_{D, B} = \left. \< \(n\)\>_{D, B} \right|_{\tilde{t} \leftrightarrow \tilde{u}} = \\ &
\frac{1}{16\(2\pi\)^{3}}\(\frac{p\,\sin\theta}{p_{\rm in}\,q\, T}\)
\int^{\infty}_{\rm k_{\rm min}}\, d\kT\, n_{D}\(\kT\)\, \[1\pm n_{B}(k_{X}) \]\,
 \int^{2\pi}_{0} \frac{d\phi}{2\pi}\, 
\frac{\left| \mathcal{M}^{(n)} \right|^2_{\tilde{t} \leftrightarrow \tilde{u}}}{g_s^4} \ ,
\end{split}
\ee
where the role of $\tnew, \unew$ are interchanged in the squared matrix element with respect to Eq.~\eqref{ave-alpha1}. There are two integrations left in Eqs.~\eqref{ave-alpha1} and \eqref{ave-alpha2}, over $\phi$ and $\kT$. 
Remarkably, the integration over $\phi$ can be performed analytically, as explained in Appendix~\ref{sec:phi-int}. The remaining integration over $\kT$ has to be performed numerically.

\section{Results and discussion
\label{sec:results}
}

The purpose of this work is to evaluate $F^{C \rightarrow A}(p,\theta)$, 
the probability distribution
for finding an outgoing hard parton of type A with energy $p$ and angle $\theta$ relative to the direction of an incident hard parton of type C with energy $p_{\rm in}$.   (For simplicity, here as in the Introduction we shall write $F^{C \rightarrow A}(p,\theta;\pinit)$ as just $F^{C \rightarrow A}(p,\theta)$.)
Recall that by ``type''  we mean gluon or quark or antiquark. 
We consider a static brick of a weakly interacting QGP, and have included the contributions from a single binary collision between the incident hard parton and a medium parton. 
In Section~\ref{sec:formalism},
we have presented a careful derivation of the expression for $F^{C \rightarrow A}(p,\theta)$ in Eq.~\eqref{eq:probfinal}, 
and have provided further technical details on the summation over different processes in Section~\ref{sec:sum}, 
as well as the simplification of the phase space integration in Section~\ref{sec:integration}. 
By summing over different types, we obtain the probability distribution for finding final parton of any type, 
\begin{eqnarray}
F^{C\rightarrow \all}(p,\theta)&=&F^{C \rightarrow G}(p,\theta)+F^{C \rightarrow Q}(p,\theta)+F^{C \rightarrow {\bar Q}}(p,\theta)\, . 
\end{eqnarray}
Integration of $F^{C\rightarrow \all}(p,\theta)$ over $p$ using Eq.~\eqref{P-Theta} then yields $P(\theta)$, namely the probability distribution for the angle $\theta$.

\subsection{Comparison with previous work
\label{sec:qualitative}
}

Before we present our results, we shall briefly sketch how they agree with
results obtained previously where they should. The details of this comparison
are found in Appendix~\ref{sec:compare}.
The probability distribution for an energetic parton that travels for a distance $L$ through a weakly coupled QGP  to pick up transverse momentum $q_{\perp}$, which we shall denote $\PT(q_\perp)$, was analyzed in Ref.~\cite{DEramo:2012uzl}.
These authors confirmed that for sufficiently small $L$ or for sufficiently large $q_{\perp}$, 
$\PT(q_{\perp})$ will approach $\PT_{{\rm single}}(q_{\perp})$  (denoted by $P_{{\rm thin}}(q_{\perp})$ in Ref.~\cite{DEramo:2012uzl}),
the probability distribution obtained upon including at most a single scattering between the incident parton and a scatterer from the thermal medium. 
This is expected on physical grounds since the most probable way of picking up a large $q_{\perp}$ is via a single scattering. 
Expressions for $\PT_{{\rm single}}(q_{\perp})$ 
were calculated previously
under the condition $q_{\perp}\ll T$ in Ref.~\cite{Aurenche:2002pd} 
and under the condition
$q_{\perp}\gg T$ in Ref.~\cite{Arnold:2008zu}. 
The calculations of Ref.~\cite{DEramo:2012uzl}  do not assume any ordering
between $q_\perp$ and $T$, and their results 
agree with the older results in the appropriate limits.
In all of these previous studies, however, 
the calculations are performed by first taking a limit in which $\pinit/T\to \infty$ while $q_{\perp}/T$ remains finite, meaning in a limit in which  $\theta \to 0$.
In this limit, 
Rutherford-like scattering in which an incident parton scatters off a parton from the thermal medium is dominant over all other $2\lra 2$ processes, including those in which a parton from the medium is kicked to a large angle as well as processes such as $q\bar{q}\lra gg$.  
We shall {\it not} take the $\pinit/T\to \infty$ limit, meaning that we must include all $2\lra 2$ processes and that we can describe scattering processes that produce a parton at some nonzero angle $\theta$ and hence can compute $P(\theta)$, the probability distribution for the scattering angle $\theta$.

To compare to the previous results referred to above, we take the limit $\theta\ll 1$ in our result for $P(\theta)$ and compare what we find there with $\PT_{{\rm single}}(q_{\perp})$ from Refs.~\cite{Aurenche:2002pd,Arnold:2008zu,DEramo:2012uzl}.
When we take $\theta\ll 1$, we find that $F^{C\to \all}(p,\theta)$ is peaked at $p\approx \pinit$, i.e. $\o/\pinit\ll 1$ where we have defined 
\begin{equation}
\omega\equiv \pinit - p\ .
\end{equation}
Consequently, to compare to previous results we evaluate $F^{C\to \all}(p,\theta)$ in the regime
\begin{eqnarray}
\label{limit}
\theta \ll 1\, , 
\qquad
|\o|/\pinit\ll 1\, ,
\end{eqnarray}
and then perform the necessary integrations to obtain $P(\theta)$ in this regime.
In Eq.~\eqref{PTheta-relation} in Appendix.~\ref{sec:relation}, 
we show that our results agree with those from the literature if 
$P(\theta)$ is given by $(\pinit^2 \theta / 2\pi) \PT_{\rm single} (q_\perp)$ in the regime (\ref{limit}).
In subsequent parts of 
Appendix~\ref{sec:compare}, we confirm in detail that 
our results do indeed match those found in Refs.~\cite{Aurenche:2002pd,Arnold:2008zu,DEramo:2012uzl} in the kinematic regime where they should.

In this work, we have extended the previous studies by considering finite (but large) $\pinit/T$ meaning that $\o/\pinit$ and $\theta$ need not vanish.
Consequently, 
there are new features in our computations. 
In particular, we have included all $2\lra 2$ scattering processes, as given in Table~\ref{tab:QCDprocesses}, in our evaluation of $F^{C\rightarrow A}(p,\theta)$.
Furthermore, 
when $\o/\pinit$ is finite, 
either the deflected incident parton or the recoiling thermal parton or both 
can show up with energy $p$ and angle $\theta$.
Indeed,
we shall see in the subsequent sections that $P(\theta)$ at nonzero $\theta$ differs
qualitatively from that obtained by extrapolating its behavior in the small $\theta$ limit. 
In particular, the large-angle tail of $P(\theta)$ is in reality fatter than one would guess from
such an extrapolation.  This makes the inclusion of all $2\lra 2$ processes as we do
important and interesting, not just necessary.

Next, we note that by working at finite $\pinit/T$ we introduce a kinematic cutoff on 
the momentum transfer, meaning that
when we increase $\theta$ the probability distribution
$P(\theta)$ must eventually be suppressed since (because of energy/momentum conservation) the minimum energy of the thermal parton needed to yield a specified $\theta$ will become much larger than $T$. We shall illustrate this quantitatively later, 
see the blue curves in Fig.~\ref{fig:Thetamin}.  
The analogous kinematic cutoff on $q_\perp$ in  $\PT_{\rm single}(q_\perp)$ computed
in the limit in which $\pinit/T\to \infty$ and $\theta\to 0$ is less constraining~\cite{DEramo:2012uzl}.

Finally, we note that in Ref.~\cite{He:2015pra} 
quantities analogous to $F(p,\theta)$ or integrals of
$F(p,\theta)$ have been computed in the Linear Boltzmann Transport (LBT) model for energetic partons shooting through a brick of weakly coupled QGP as in our calculation, 
albeit largely with a focus on a kinematic regime in which $p$, and hence the momentum transfer, are only a few GeV.   These authors also compute a quantity directly related to the transverse momentum distribution $\PT\(q_{\perp}\)$ using the LBT model for $q_{\perp}$ out to around 10 GeV, and provide a very interesting study of how the distribution becomes more and more Gaussian as the thickness of the brick is increased.  However, even for the thinnest brick that they consider the values of $q_{\perp}$ that they investigate are not large enough for single scattering to be dominant.
It would be interesting to extend these LBT calculations to larger $q_{\perp}$ where the probability of multiple scattering is negligible and compare them to our results, upon taking into account the appropriate Jacobian.

\subsection{Results for the  probability distributions $F(p,\theta)$ and  $P(\theta)$
\label{sec:FandP}
}
%

%
%
\begin{figure} 
\centering
%
%
\includegraphics[height=0.24\textwidth]{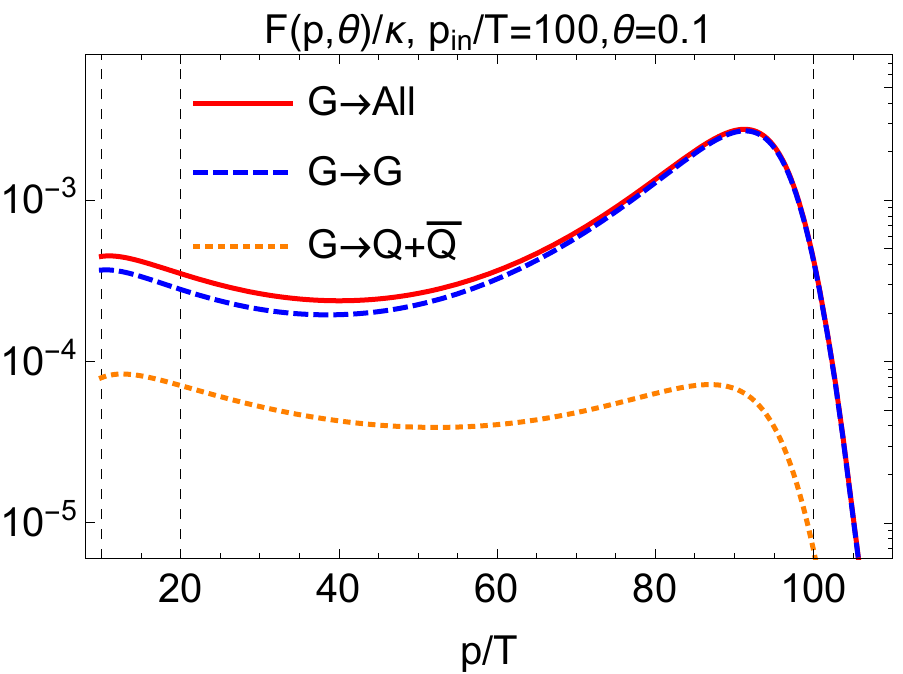}
\includegraphics[height=0.24\textwidth]{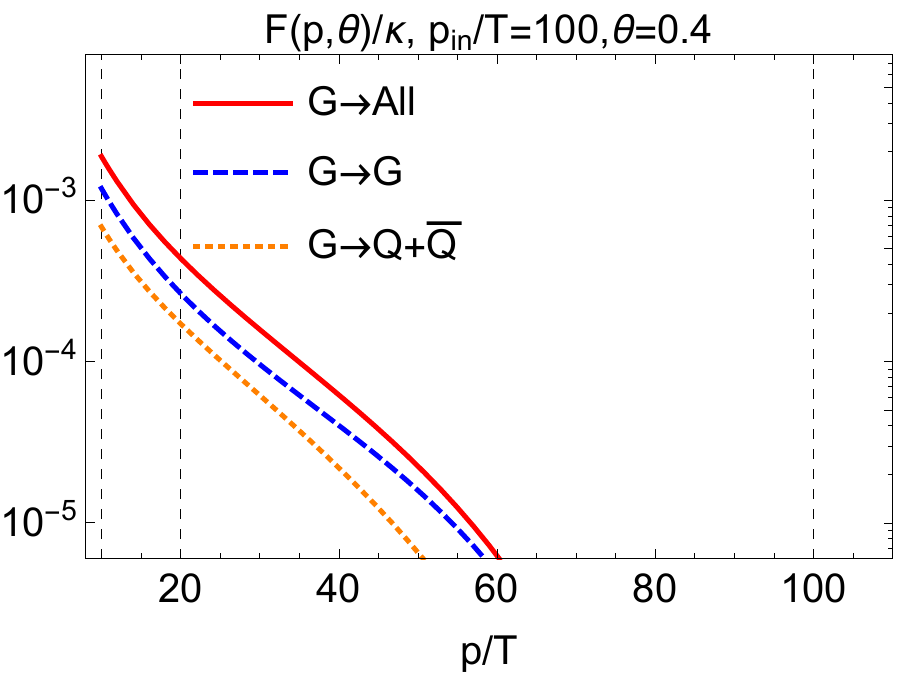}
\includegraphics[height=0.24\textwidth]{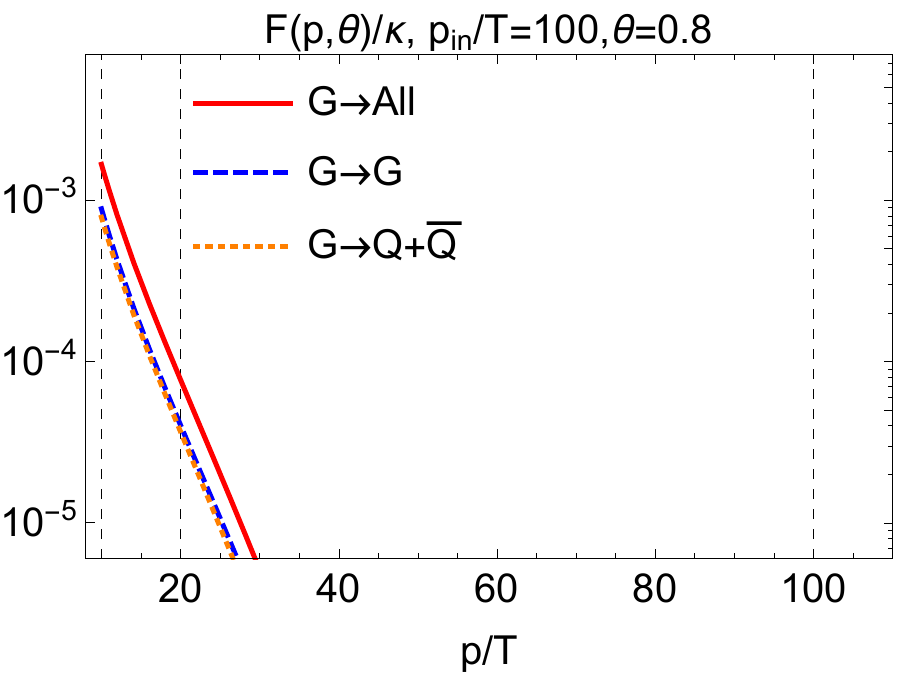}
%
%
%
%
\includegraphics[height=0.24\textwidth]{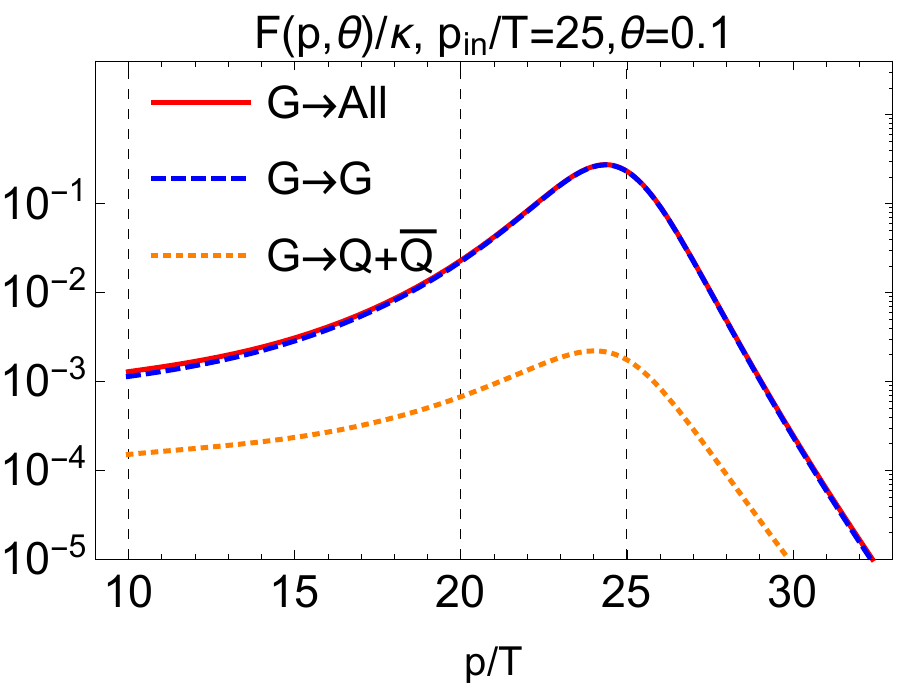}
\includegraphics[height=0.24\textwidth]{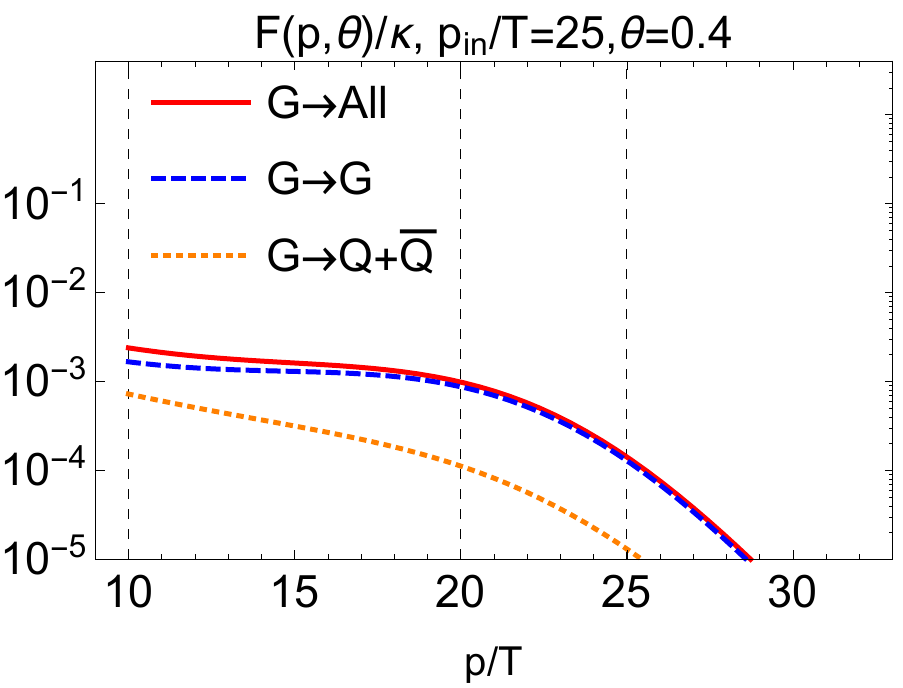}
\includegraphics[height=0.24 \textwidth]{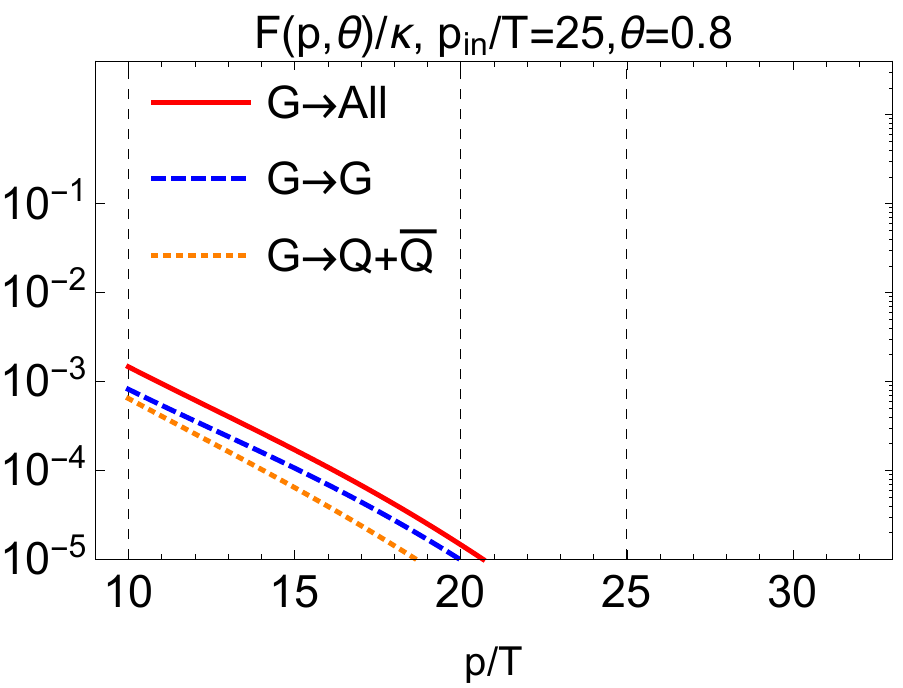}
%
%
%
\includegraphics[height=0.24\textwidth]{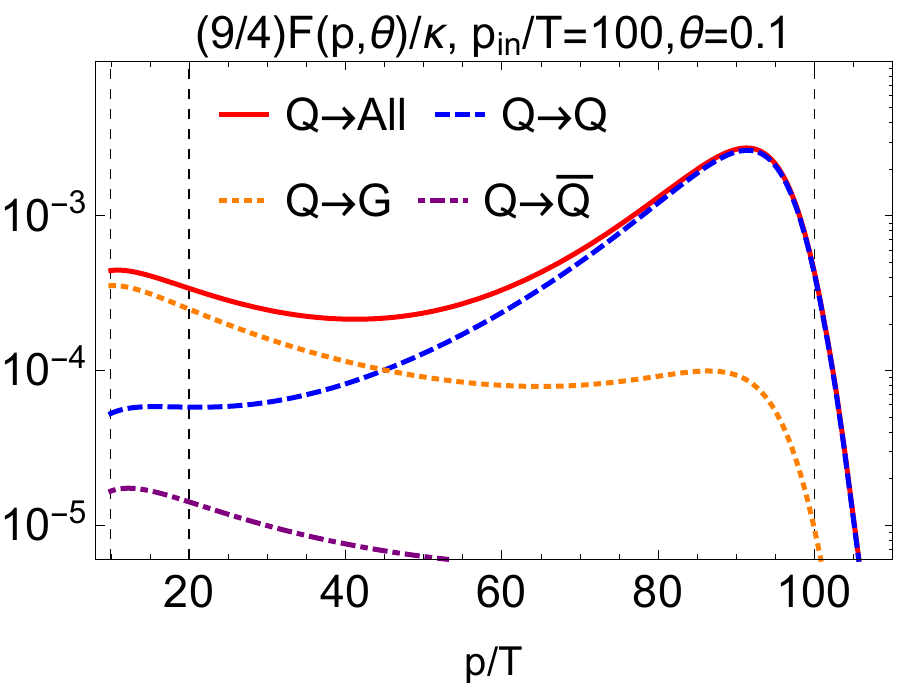}
\includegraphics[height=0.24\textwidth]{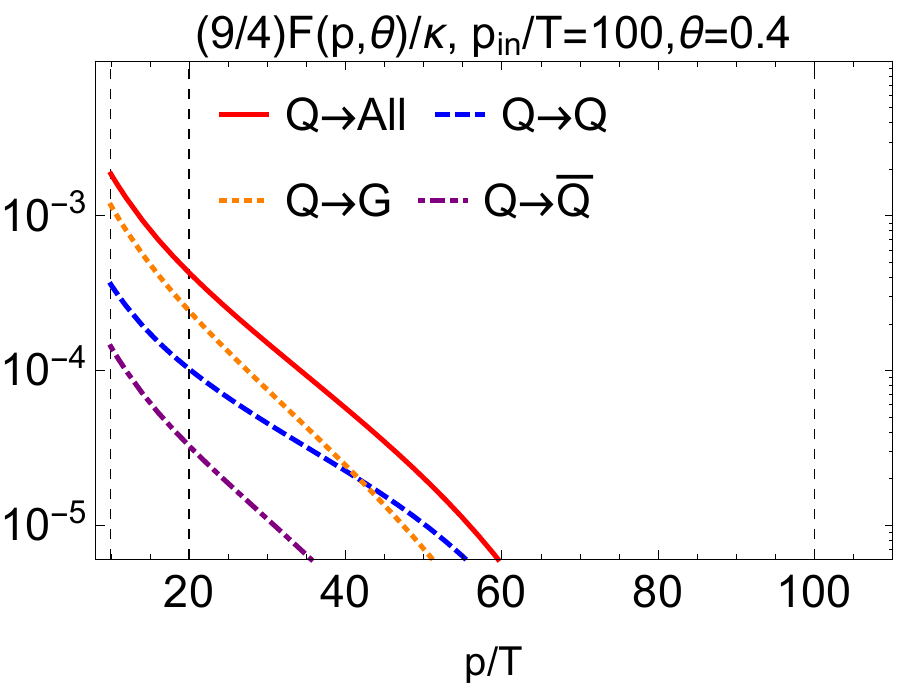}
\includegraphics[height=0.24\textwidth]{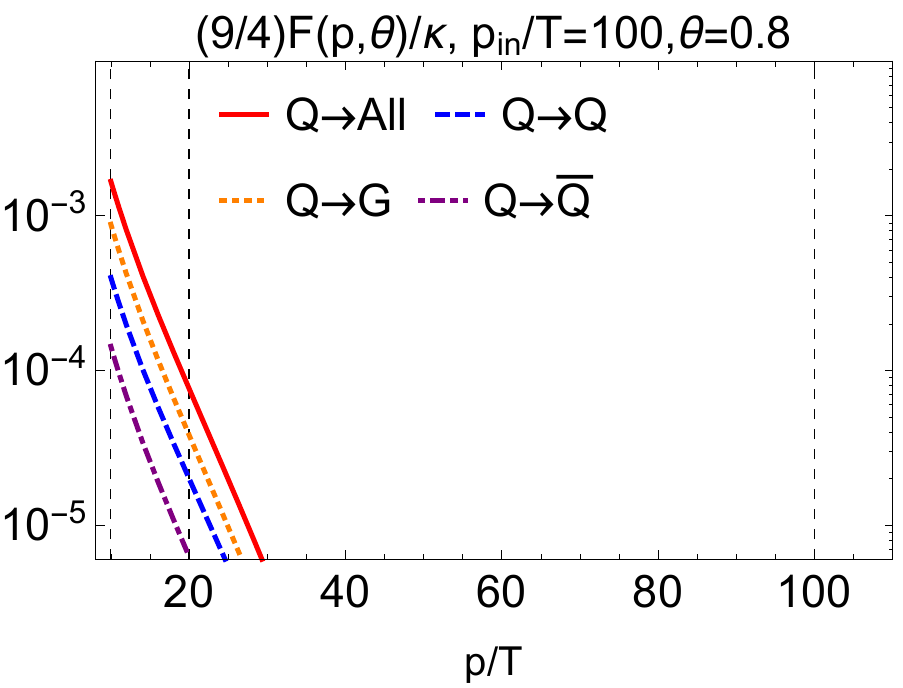}
%
%
%
%
\includegraphics[height=0.24\textwidth]{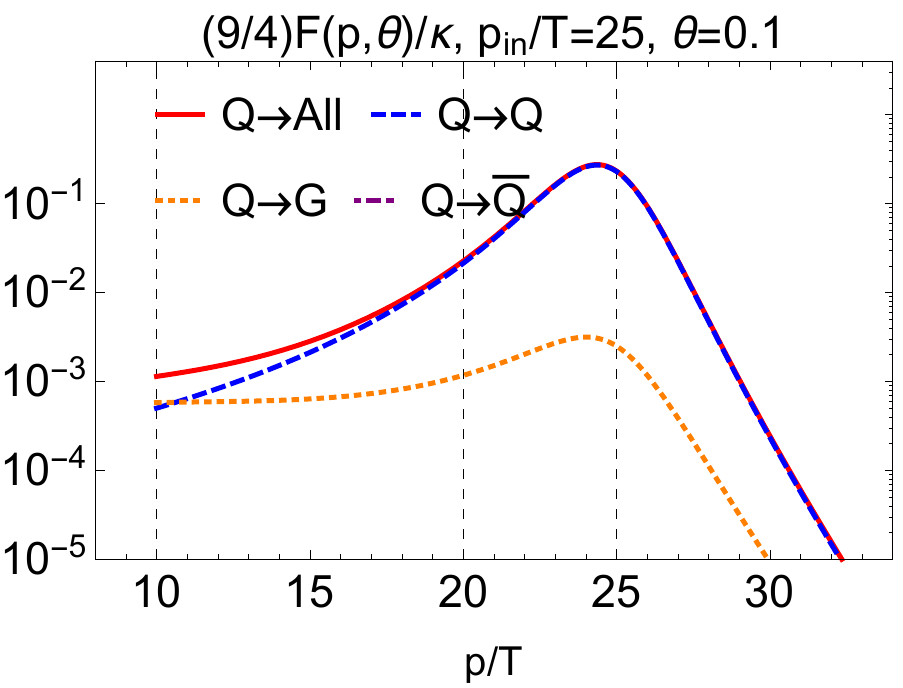}
\includegraphics[height=0.24\textwidth]{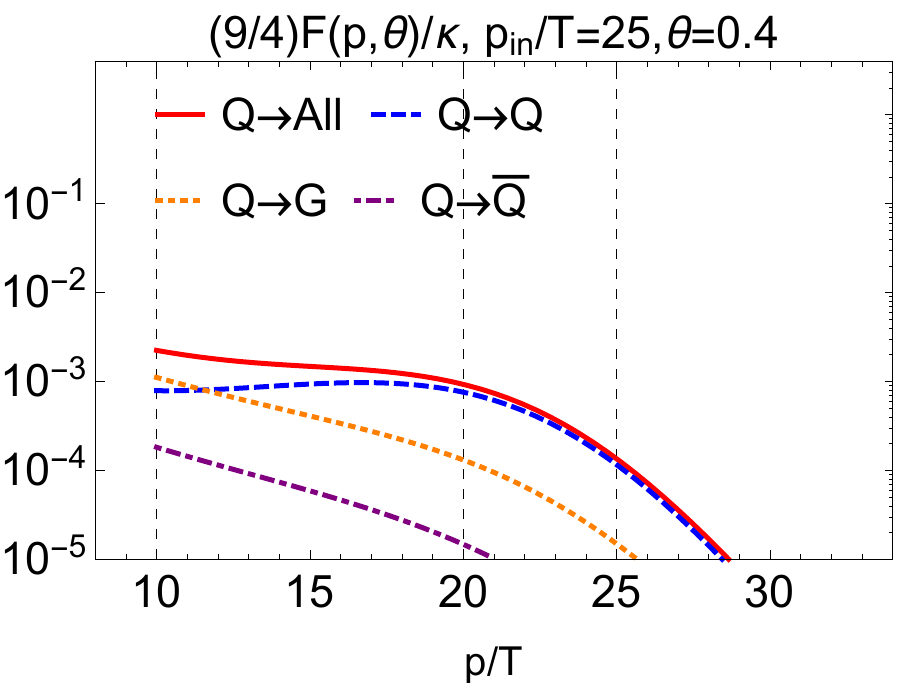}
\includegraphics[height=0.24\textwidth]{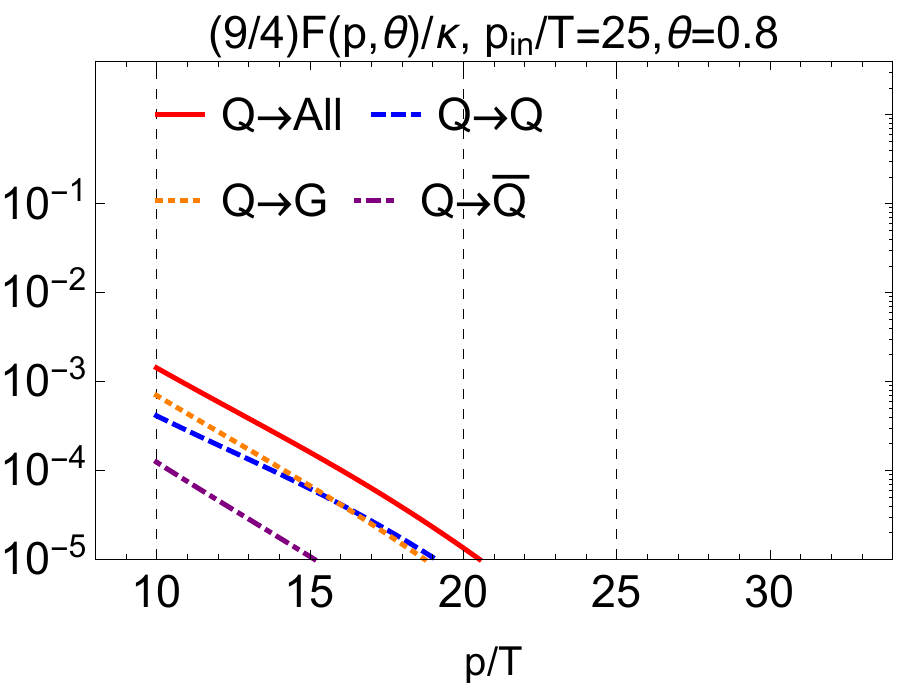}
\caption{
\label{fig:rate-fixtheta}
The probability distribution $F^{C\ra\all}\(p,\theta\)$ divided by $\kappa=g^{4}T\CT$ plotted
as functions of $p/T$ for
an ``incident gluon'' with $\pinit/T=100$ (first row of panels) or $\pinit/T=25$ (second row) 
and that for an ``incident quark'' with $\pinit/T=100$ (third row) or $\pinit/T=25$ (fourth row).
From left to right, the columns correspond to choosing $\theta=0.1$, 0.4 and 0.8.
Since we are considering a brick of QGP with net baryon number zero, $F^{G\to Q}(p,\theta)=F^{G\to {\bar Q}}(p,\theta)$, $F^{Q\to G}(p,\theta)=F^{{\bar Q}\to G }(p,\theta)$, $F^{Q\to \bar{Q}}(p,\theta)=F^{{\bar Q}\to Q }(p,\theta)$, and $F^{Q\to Q}(p,\theta)=F^{{\bar Q}\to {\bar Q}}(p,\theta)$. 
In the Figure, the curves labelled $G\to Q+\bar Q$ are the sum $F^{G\to Q}(p,\theta)+F^{G\to {\bar Q}}(p,\theta)=2\, F^{G\to {Q}}(p,\theta)$.
The vertical dashed black lines correspond, from right to left, to $\pinit/T$, and to the two different choices of $p_{\min}/T$ which we will use below in the evaluation of $P(\theta)$ as shown in Fig.~\ref{fig:dPdThetaPiT}, namely $p_{\min}/T=20$ and 10.
  }
\end{figure}
%
%

We shall now present results from our numerical calculation of $F^{C\ra\all}\(p,\theta\)/\kappa$ as well as for $P(\theta)/\kappa$,  
both of which are independent of $\kappa$.  Recall that $\kappa\equiv g_s^4 T\Delta t$.
The probability for a single $2\to 2$ scattering with any specified kinematics
is proportional to $g_s^4$ at tree-level,
 and is proportional to $\Delta t\equiv L/c$, the time that the incident parton would spend traversing the brick if it did not scatter.
Hence, increasing $\kappa$ (either via increasing the coupling or via increasing $T\Delta t$) must increase $F^{C\ra\all}\(p,\theta\)$ and $P(\theta)$.  Upon increasing $\kappa$, though, at some
point the assumption that single scattering dominates must break down, and along with it our
calculation.   The criterion here is that $N_{\rm hard}(\theta_{\rm min})$, defined in 
Eq.~(\ref{N-min}), must remain small and this defines an upper limit on the value of $\kappa$ at which our calculation can be used for angles $\theta$ greater than any specified $\theta_{\rm min}$, or a lower limit on the angle $\theta$ at which our calculation can be used for any given value of $\kappa$.
We shall illustrate this quantitatively in Section~\ref{sec:NhardEstimates}.
Note that in this Section we shall work in the weak coupling limit $g_s\to 0$ in which $\kappa\to 0$
and our expression for $F^{C\ra A}(p,\theta)$ in Eq.~\eqref{eq:probfinal} is valid for any nonzero 
$\theta$ and any finite $\CT$.

We shall consider $N_{f}=3$ throughout and we shall only consider QGP with no net baryon number, 
meaning zero baryon number chemical potential and
meaning that the distribution of quarks in our thermal medium is the same as that of antiquarks.

We begin our discussion by considering an incident gluon with $\pinit/T=100$. 
In the top row of Fig.~\ref{fig:rate-fixtheta}, 
we plot $F^{G\ra\all}\(p,\theta\)/\kappa$ vs $p/T$.
From  left to right, we have selected three different representative values of $\theta$, namely $\theta=0.1$, 0.4, and 0.8. 
For $\theta=0.1$, 
we observe that the probability distribution is peaked at $p\approx \pinit$, 
meaning that outgoing partons with a very small angle are likely to have a small value of $\o/\pinit$, where $\o=\pinit-p$. 
This implies that computing $F^{G\ra\all}(p,\theta)$ in the limit \eqref{limit} is sufficient to obtain $P(\theta)$ for $\theta\ll 1$, 
as we mentioned earlier. 
However, 
the dependence of $F^{G\ra\all}(p, \theta)$ on $p$ changes qualitatively as we increase $\theta$. 
$F(p,\theta)$ at $\theta=0.4$ and $\theta=0.8$ are both largest at small values of $p$ and decrease monotonically with increasing $p$. 
To understand this, 
let us recall that the difference between $p$ and $\pinit$, i.e.~$\o$, measures the energy transfer during a binary collision, 
with a smaller $p$ corresponding to a larger energy transfer $\o$. 
Likewise, a larger $\theta$ means a larger transverse momentum transfer. 
Since the typical energy of a thermal parton is quite soft, of order $T$, a large momentum transfer in a single collision between an incident parton and the thermal scatterer is more likely to be accompanied by a large energy transfer. 
That is why we see $F^{G\ra\all}(p,\theta)$ telling us that when we ask about scattering at 
large $\theta$ we find that it most often corresponds to scattering with a large $\o$ and hence a small $p$. 
Equivalently, although in different words, we note that in this regime the detected parton is 
most likely to be a parton from the medium that was kicked to a large angle $\theta$
by the incident parton, with the incident parton having lost only a small fraction of its energy
to the parton that is detected.  
The energy transfer defined as $\o$ is large because the detected parton is the parton from the medium, not the incident parton.

In Fig.~\ref{fig:rate-fixtheta}, in addition
to plotting $F^{G\ra\all}\(p,\theta\)$ we have also shown its separate components
corresponding to detecting an outgoing gluon or an outgoing quark or antiquark, namely
$F^{G\rightarrow G}\(p,\theta\)$ 
and $F^{G\rightarrow Q}\(p,\theta\) + F^{G\rightarrow \bQ}\(p,\theta\) $. (Note that $F^{G\rightarrow \bQ}\(p,\theta\)=F^{G\rightarrow Q}\(p,\theta\)$.)
While $F^{G\rightarrow Q}(p,\theta)\ll F^{{G\ra G}}(p,\theta)$ at small $\theta$, 
meaning that at small $\theta$ the 
outgoing parton is most likely to be a gluon when the incident parton is a gluon,
we see that $F^{G\ra Q}(p,\theta)+F^{G\ra \bar{Q}}(p,\theta)$ 
eventually becomes comparable to $F^{{G\ra G}}(p,\theta)$ at larger values of $\theta$.
This confirms that what is being seen at large values of $\theta$ and small values of $p$ 
is to a significant extent partons from the medium that have been struck by the incident parton.
The quarks and antiquarks seen in this regime also include those 
coming from the process $gg\to q\bar{q}$.  
And, this observation convincingly demonstrates that Rutherford-like scattering is not
dominant over other processes at larger values of $\theta$

We now consider an incident gluon with a lower initial energy, i.e.~$\pinit/T=25$, 
and plot $F^{G\ra\all}(p,\theta)/\kappa$ for this case in the second row 
of Fig.~\ref{fig:rate-fixtheta}. 
As before, 
we have selected three representative values for $\theta$, from left to right choosing
$\theta=0.1$, 0.4 and 0.8. 
The behavior of $F^{G\ra\all}(p,\theta)$ as a function of $p$ is qualitatively similar 
to that with $\pinit/T=100$: $F^{G\ra\all}(p,\theta)$ features a peak at $p\approx \pinit$ at small $\theta$, but it then becomes a decreasing function of $p/T$ at the larger values of $\theta$. 
At a quantitative level, 
we observe that for $\theta=0.1$, the peak value of $F^{G\ra\all}(p,\theta)$ with $\pinit/T=25$ is much larger than that with $\pinit/T=100$. 
This is 
due to the dominance of Rutherford-like scattering at small $\theta$, since the probability of Rutherford scattering decreases with increasing $q_\perp \approx \pinit \theta$ and we are comparing two values of $\pinit$ at the same small $\theta$.
As with $\pinit/T=100$, we see that when we choose $\theta=0.8$ we find
a probability that is peaked at small $p$ and we see that the contribution of
quarks and antiquarks is not much smaller than that of gluons.  Hence, at this large
value of $\theta$ we are seeing partons kicked out of the medium.  
We see that with $\pinit/T=25$ the choice of $\theta=0.4$ represents an intermediate case.

For completeness, in the third and fourth rows of Fig.~\ref{fig:rate-fixtheta}
we plot $F^{Q\ra\all}(p,\theta)$ for an incident quark with $\pinit/T=100$ (third row) and 25 (fourth row) at three values of $\theta$. 
We have multiplied our results for an incident quark by the ratio of Casimirs $C_{A}/C_{F}$, which is $9/4$ for $N_{c}=3$, to simplify the comparison to our results for an incident gluon.
After taking this Casimir scaling factor into account, 
the resulting $F^{Q\ra {\rm all}}(p,\theta)$ are very similar to those for incident gluons
with the same choice of $\pinit/T$.
Similar to what we found for gluons, if we look at small $\theta$ and $p$ close to $\pinit$, we see that the Rutherford-like $Q\rightarrow Q$ process makes the dominant contribution
whereas if we look at larger $\theta$ and small $p$ we see that $Q\rightarrow G$ is 
comparable to, and in fact slightly larger than, $Q\rightarrow Q$. This demonstrates
that Rutherford-like scattering is not dominant here and suggests that the detected parton is most often a parton that was kicked out of the medium.

 %
\begin{figure} 
\centering
\includegraphics[height=0.24\textwidth]{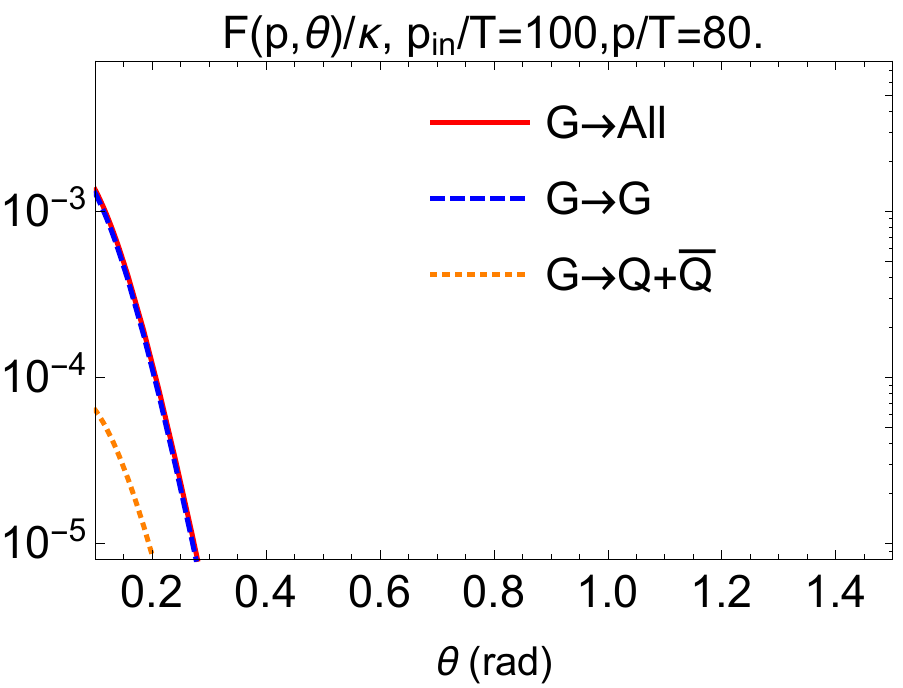}
\includegraphics[height=0.24\textwidth]{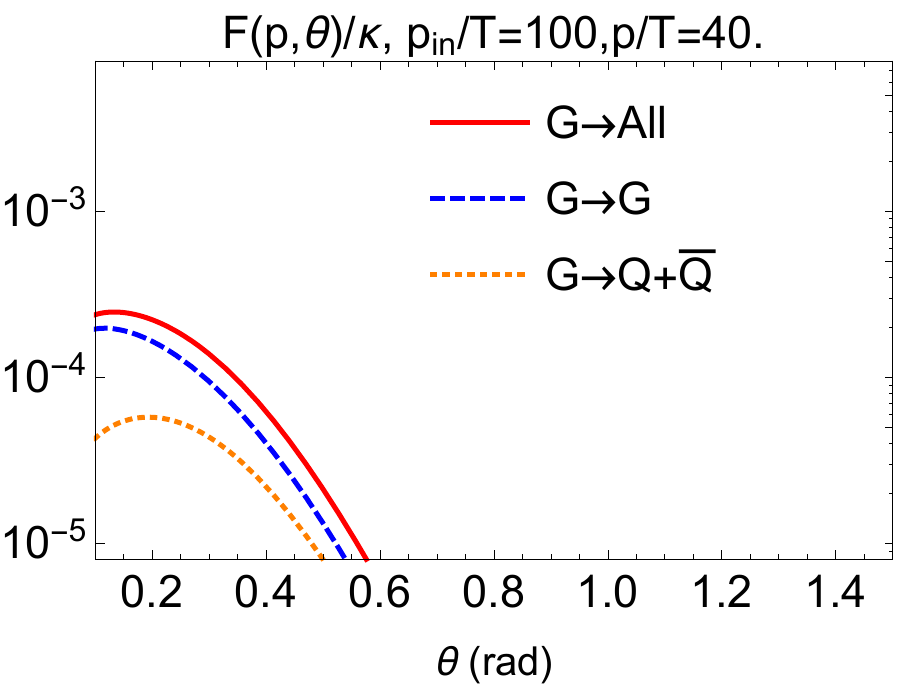}
\includegraphics[height=0.24\textwidth]{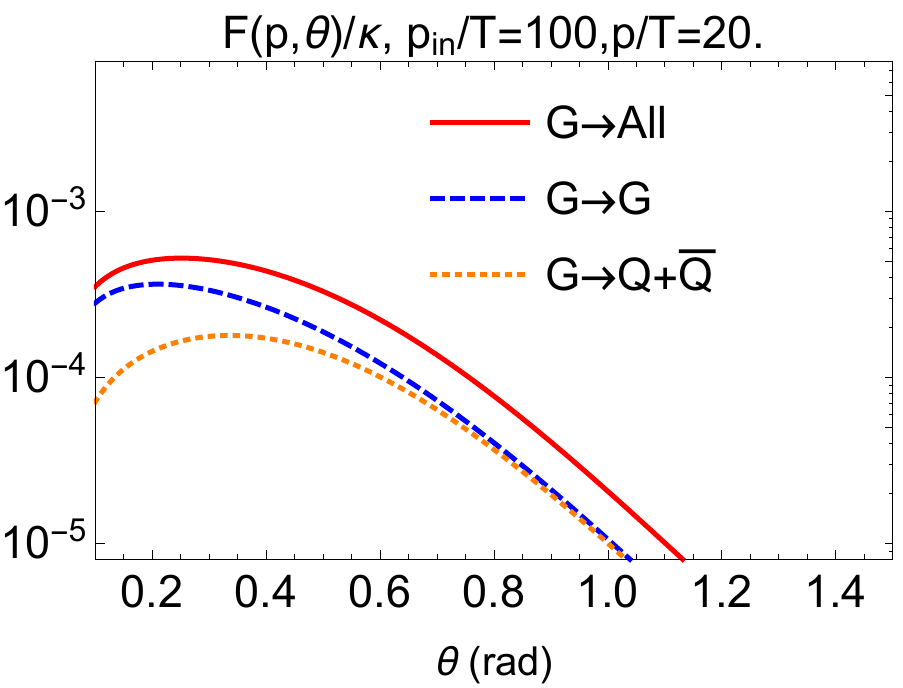}
\includegraphics[height=0.24\textwidth]{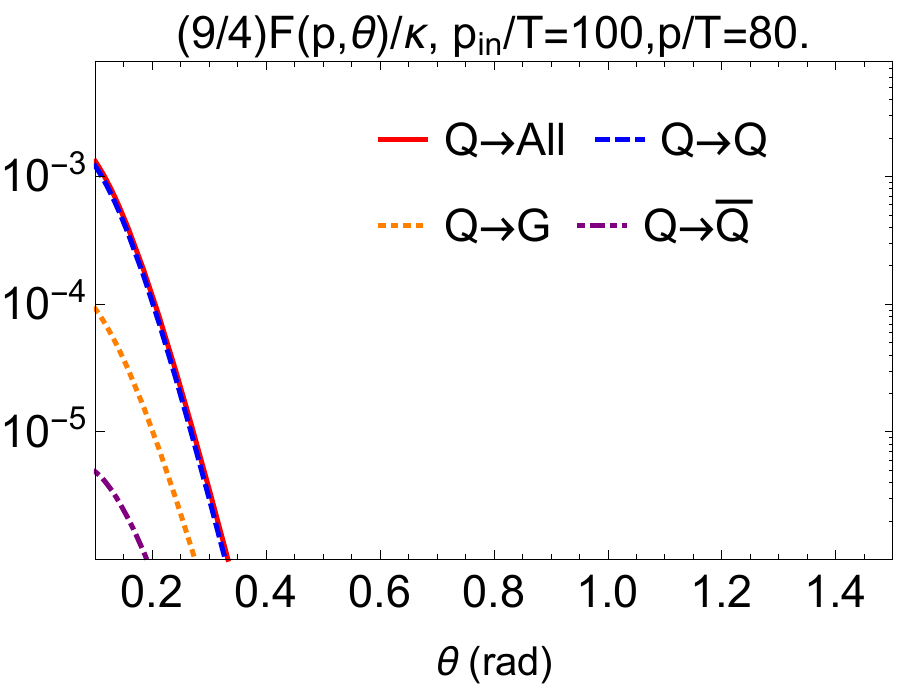}
\includegraphics[height=0.24\textwidth]{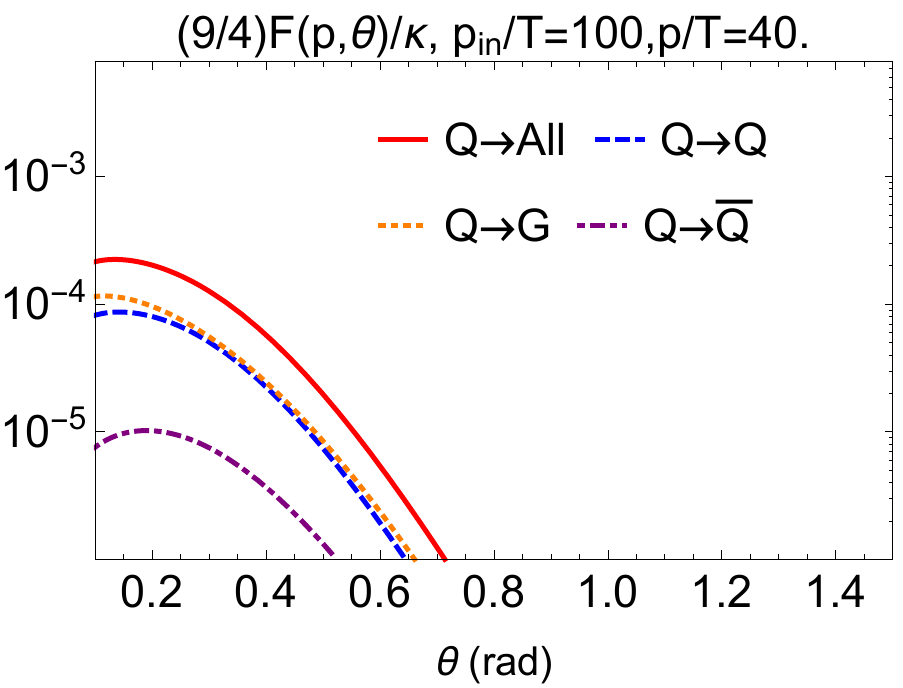}
\includegraphics[height=0.24\textwidth]{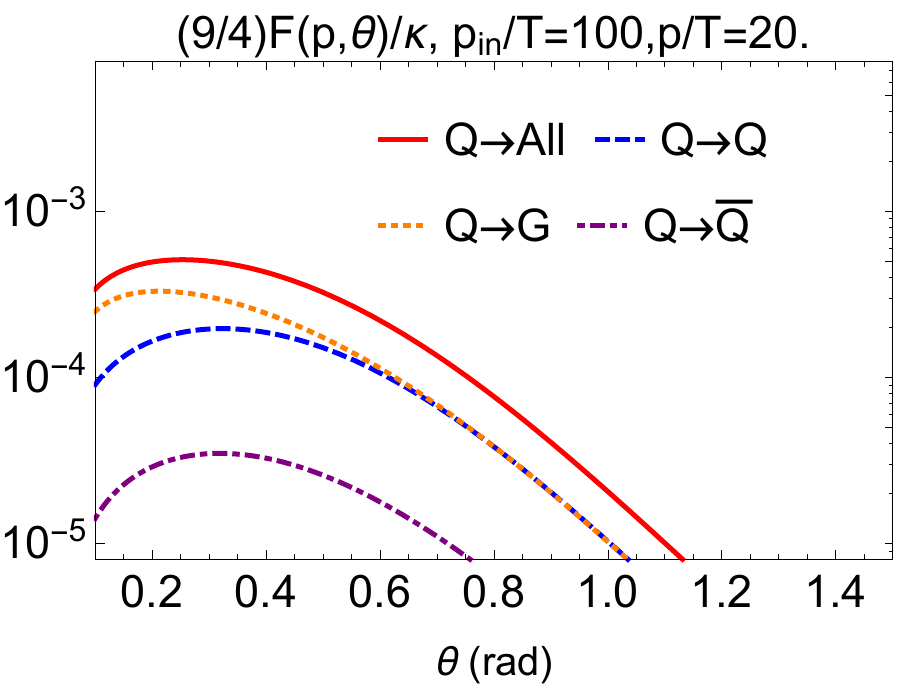}
\caption{
\label{fig:rate-fixPf}
The probability distributions $F^{C\to \all}\(p,\theta\)$ divided by $\kappa=g^{4} T \CT$ plotted as functions of $\theta$ for an incident gluon with $\pinit/T=100$ ($C=G$, upper row)  and for an incident quark with $\pinit/T=100$ ($C=Q$, lower row). From left to right, the columns correspond to choosing $p/T=80$, 40 and 20. 
  }
\end{figure}

To complement Fig.~\ref{fig:rate-fixtheta}, which illustrates 
the dependence of $F^{G\ra\all}(p,\theta)$ on $p$ with fixed $\theta$,
in the top row of Fig.~\ref{fig:rate-fixPf} we  show the dependence of $F^{G\ra\all}(p,\theta)$ on $\theta$ at three fixed values of $p/T$. 
In another words, in Fig.~\ref{fig:rate-fixPf}
we are looking into the angular distribution of an outgoing parton with 
a fixed $p/T$, 
considering three different values of $p/T$, namely 80, 40 and 20.  We have chosen an incoming gluon with $\pinit/T=100$ in all three panels. 
In the second row of Fig.~\ref{fig:rate-fixPf}, we show results for an incoming quark with
the same $\pinit/T$. As before, we see that after, after multiplying by the ratio of Casimirs 9/4, $F^{Q\ra\all}(p,\theta)$ is reasonably similar to $F^{G\ra\all}(p,\theta)$.   
From our results with $p/T=80$,  we see that when we look at outgoing partons whose
energies are not much lower than those of the incident parton, smaller values of
the scattering angle $\theta$ are favored and the scattered parton is dominantly the
same type as the incident parton. In contrast, in our results at smaller $p/T
$ we see a much broader $\theta$ distribution and, in particular at larger values of $\theta$,
we see comparable contributions from quarks or antiquarks and gluons in the final state, confirming that the detected parton was a parton from the medium that was struck by
the incident parton.

%
\begin{figure} 
\centering
\includegraphics[height=0.35\textwidth]{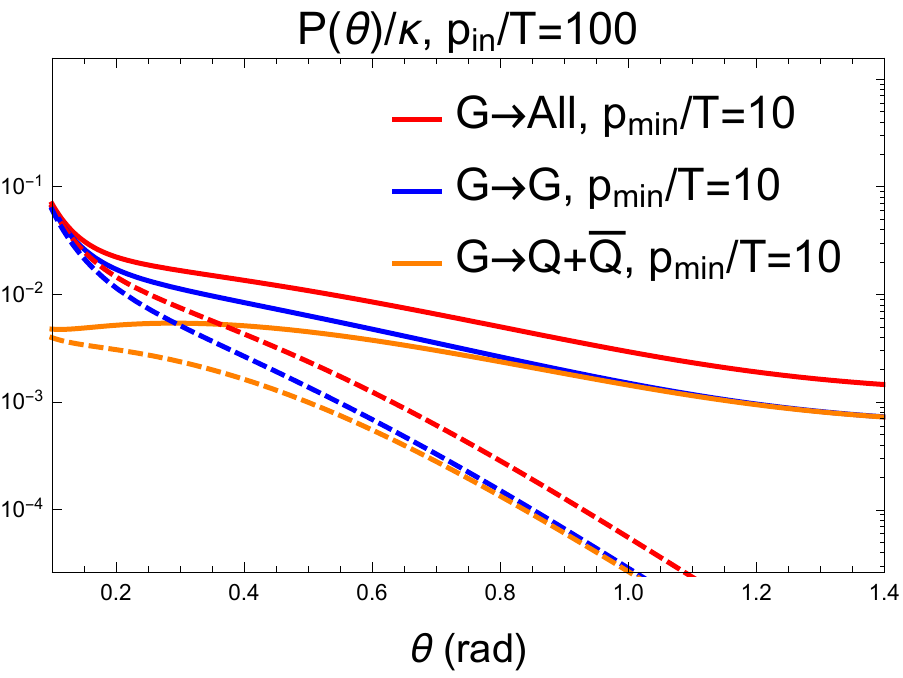}
\includegraphics[height=0.35\textwidth]{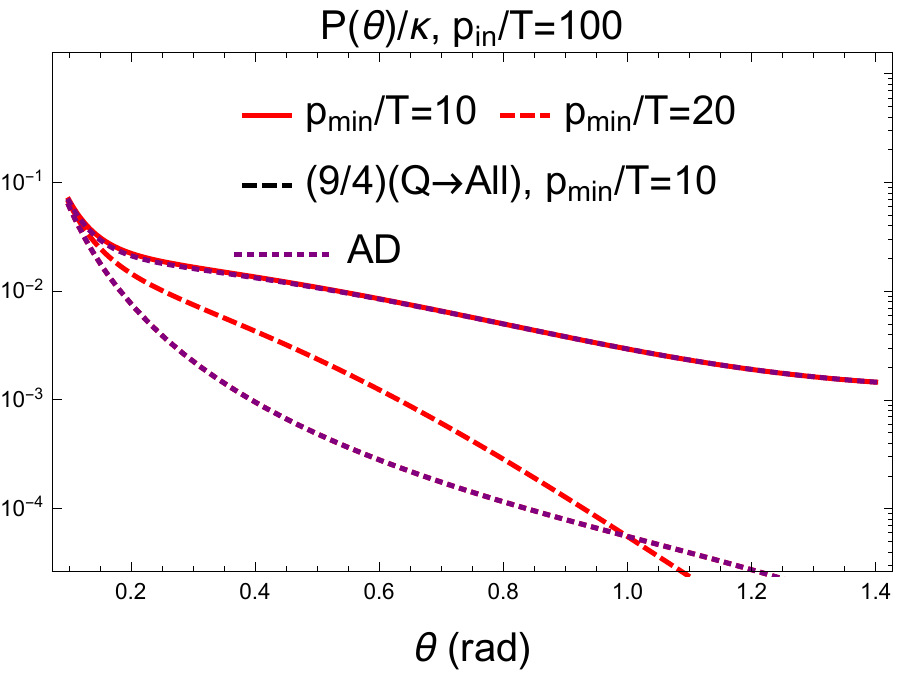}
\includegraphics[height=0.35\textwidth]{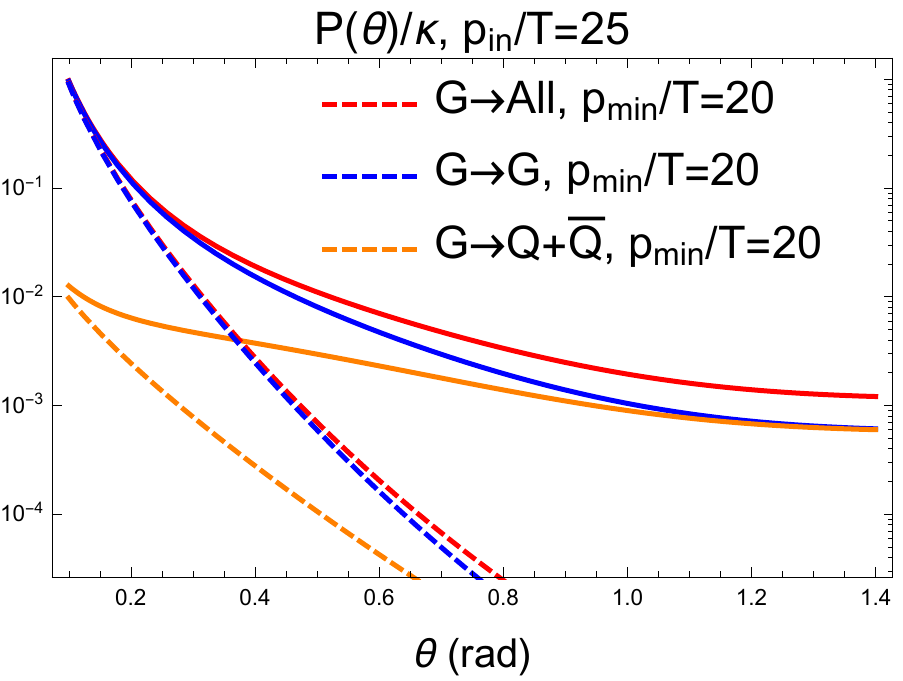}
\includegraphics[height=0.35\textwidth]{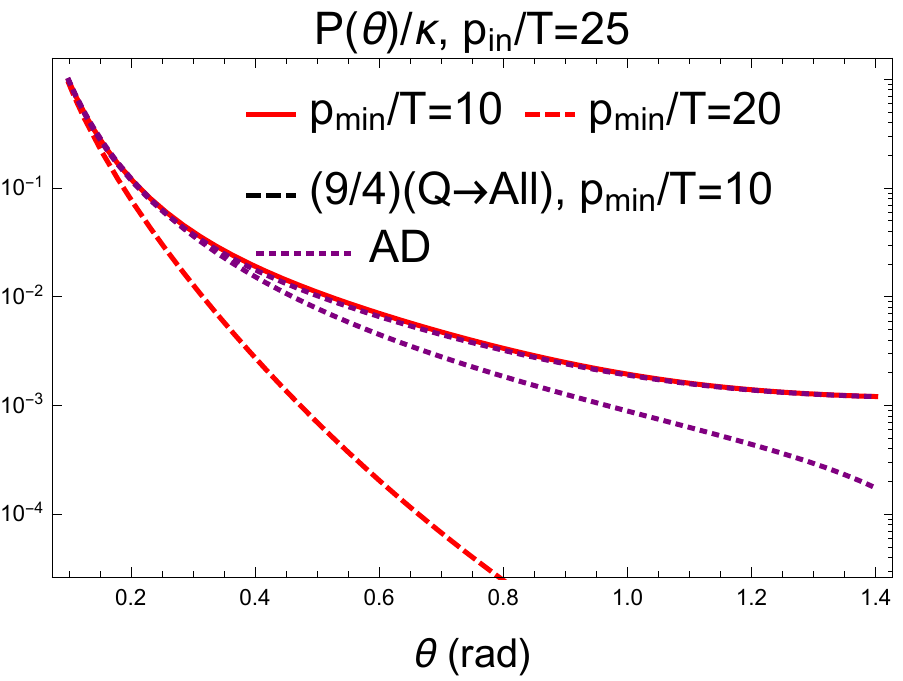}
\caption{
\label{fig:dPdThetaPiT}
The probability distribution $P\(\theta\)$ divided by $\kappa$ for an incident gluon with $\pinit/T=100$ (upper row) and $\pinit/T=25$ (lower row).
In the two panels in the left column, the solid curves correspond to choosing $p_{\min}$,
the lower limit on the integration over $p$ in Eq.~\eqref{P-Theta}, to take the
value $p_{\min}/T=10$ while the dashed curves correspond to choosing $p_{\min}/T=20$.
In addition to plotting the probability distribution for finding any outgoing parton at a given
$\theta$ as the red curves, we also present its breakdown into the cases of an outgoing gluon (blue curves) and an outgoing quark or antiquark (orange curves). 
In the right column, 
we plot $P(\theta)$ for an incident gluon (red) as well as for an incident quark times $9/4$ (black dashed curves) as well as the $\theta\ll 1$ result $P^{{\rm AD}}(\theta)$ from Eq.~\eqref{PAD-theta} first obtained by Arnold and Dogan~\cite{Arnold:2008zu}. 
 }
\end{figure}

We now present our results for the probability distribution 
$P\(\theta\)$, which we obtain by integrating $F^{C\ra \all}\(p,\theta\)$ over $p$, following Eq.~\eqref{P-Theta}.
In the top-left panel of Fig.~\ref{fig:dPdThetaPiT},
we plot $P(\theta)$ for an incident gluon with $\pinit/T=100$. 
Since the integration \eqref{P-Theta} depends on a somewhat arbitrary choice of $p_{\min}/T$, 
we will consider two different choices, $p_{\min}/T=10$ and $p_{\min}/T=20$, 
and check the sensitivity of $P(\theta)$ to this variation in this choice. 
We observe that for sufficiently small $\theta$, 
$P(\theta)$ is insensitive to the choice of $p_{\min}/T$. 
This is to be expected,  given our discussion of $F^{C\to\all}(p,\theta)$: recall that it is peaked at $p\sim \pinit \gg p_{\min}$ for small $\theta$, meaning that where we place $p_{\min}$ does not matter much in this case. 
However, when we choose a
 larger value of $\theta$
the magnitude of $P(\theta)$ becomes much smaller if we increase $p_{\min}/T$ from $10$ to $20$. 
This is also expected since at large $\theta$ we have seen that  $F(\theta, p)$ is a rapidly 
decreasing function of $p$. 
In the bottom-left panel of the figure, we see similar behavior in the case in which the
incident gluon has $\pinit/T=25$. When $\theta$ is not small, $P(\theta)$ is highly
suppressed when we choose $p_{\min}/T=20$. This is  no surprise since for this choice $p_{\min}$ is
close to $\pinit$, meaning that  the phase space included in the integration \eqref{P-Theta} is quite restricted. 
In both the panels in the left column of Fig.~\ref{fig:dPdThetaPiT}, 
we have in addition plotted $P(\theta)$ for an outgoing gluon, $G\to G$,
and for an outgoing quark or antiquark, $G\to Q$.
At small angles Rutherford-like scattering dominates and since the
incident parton is a gluon we see that the probability to find an outgoing gluon is much
greater than that for an outgoing quark or antiquark.
At larger angles Rutherford-like scattering is no longer dominant, the parton
that is detected most likely comes from the medium, and we see that the probability to find 
an outgoing quark or antiquark becomes comparable to the probability to find an outgoing gluon.

In the right panels of Fig.~\ref{fig:dPdThetaPiT}, 
we compare $P(\theta)$ for an incident gluon with that for an incident quark with the same choice of $\pinit/T$ and $p_{\min}/T$ multiplied by $C_{A}/C_{F}$. 
We observe that, after taking into account the appropriate Casimir scaling factor, $P(\theta)$ is almost identical for both cases.

As we discussed in Section~\ref{sec:qualitative}, 
the transverse momentum distribution due to a single binary scattering $\PT_{{\rm single}}(q_{\perp})$ has been obtained previously in the small $\theta$ limit \eqref{limit}~\cite{Arnold:2008zu,DEramo:2012uzl}.
 If in addition $q_{\perp}\gg T$, $\PT_{{\rm single}}(q_{\perp})$
 reduces to the expression first derived by Arnold and Dogan (AD) in Ref.~\cite{Arnold:2008zu} which we shall denote  $\PT^{{\rm AD}}_{{\rm single}}(q_{\perp})$ and which we provide
 explicitly in Eq.~\eqref{PAD}. (See also Ref.~\cite{DEramo:2012uzl}).) 
In the small $\theta$ limit, we can convert $\PT^{\rm AD}_{\rm single}(q_\perp)$ to a probability distribution for the angle $\theta$ that we shall denote by $P^{{\rm AD}}(\theta)$ 
using the Jacobian~\eqref{FullJacobian}. We obtain
 \begin{eqnarray}
\label{PAD-theta}
P^{{\rm AD}}(\theta)
&=&
\[\, {\CJ}^{-1}_{\perp}\,\PT^{{\rm AD}}_{\single}\(\qperp=\pinit\sin\theta\)\]
\nonumber \\
&=&\kappa\, C_{A}\,\zeta(3)\, \(\frac{4N_{c}+3 N_{f}}{4\pi^{3}}\)\,\(\frac{T}{\pinit}\)^{2}\, \cos\theta\, \(\frac{1}{\sin\theta}\)^{3}\, 
\end{eqnarray}
where $\zeta(3)\approx 1.202$ is the Riemann zeta function.
Here, the incident parton is a gluon;
for the case of an incident quark, one has to replace $C_{A}$ with $C_{F}$ in Eq.~\eqref{PAD-theta}. 
In the two panels in the right column of Fig.~\ref{fig:dPdThetaPiT}, 
we have compared $P(\theta)$ with $P^{{\rm AD}}(\theta)$ extrapolated to finite $\theta$. 
We observe that, as expected, $P^{{\rm AD}}(\theta)$ agrees very well with $P(\theta)$ at small $\theta$. 
However, 
the large-angle tail of $P(\theta)$ is much fatter than that of  $P^{{\rm AD}}(\theta)$ when $\pinit/T=100$ for all $p_{\min}/T$ under consideration, as well as when $\pinit/T=25$ for $p_{\min}/T=10$. 
This implies that when $\pinit \gg p_{\min}$, 
it is important to include all $2\to 2$ scattering processes as we have done, not only the Rutherford-like scattering process that  dominates at small $\theta$.

The results that we have illustrated in this Section are the principal results of our
calculation.  We have presented them here upon dividing $F(p,\theta)$ and $P(\theta)$ by $\kappa\equiv g_s^4 T \Delta t$.  This is the appropriate form in which to provide them
to anyone incorporating them in a future jet Monte Carlo calculation, since the values of the coupling $g_s$
and the time-step $\Delta t$ will be provided by that calculation and in such a calculation the local 
value of $T$ will come from the description of the expanding cooling droplet of QGP which the Monte Carlo jet is traversing.
As described in the Introduction, we also wish to provide some qualitative guidance for the
planning of future experiments and for how to use future precise, high statistics, suitably
differential measurements of jet substructure modification in heavy ion collisions to
find the scatterers within the QGP liquid.  To this end, in Section~\ref{sec:NhardEstimates} we
shall illustrate our results for $P(\theta)$ and its integral $N_{\rm hard}(\theta_{\min})$ using
phenomenologically motivated values for various input parameters including $\kappa$.
First, though, in the next Section we shall discuss the regime of validity of our calculation.

\subsection{Regime of validity of the calculation}
\label{sec:validity}

In this Section we pause to discuss the domain of applicability of the calculations presented
in the previous Section.  We have assumed that single scattering dominates, neglecting
multiple scattering.  This assumption is valid when 
$N_{\hard}(\theta_{\min})$ is much smaller than one, a criterion that depends on
the value chosen for $\kappa$.  We therefore leave the assessment of this criterion
to Section~\ref{sec:NhardEstimates}, in particular to
Fig.~\ref{fig:Phenomenology}.
We shall focus here on a different limitation of our calculation.
Since we are neglecting all medium-effects in the QCD matrix elements for $2\lra 2$ collisions, 
our results are trustable only in the kinematic regime in which the energy and momentum transferred between the incident parton and the parton from the medium off which it scatters are both much larger than the Debye mass $m_{D}$. That is, our results are
trustable only in the regime where
\begin{eqnarray}
\label{cond1}
-\tnew  \gg  m^{2}_{D}\quad {\rm and} \quad -\unew \gg m^2_D\ .
\end{eqnarray}
Here, we will denote the square of the four momentum difference between the incident parton and the detected outgoing parton and  that between the incident parton and the undetected parton by $\tnew$ and $\unew$ respectively, as in Section~\ref{sec:integration}. 
By using Eq.~\eqref{qt-in-theta}, in which $\tnew$ is  expressed in terms of 
$\pinit$, $p$ and $\theta$,
we can determine the region in the  $(\theta,p/T)$ plane where
the condition $-\tnew \gg m^{2}_{D}$ is satisfied for any given $\pinit$ and  $m_{D}$. 
Furthermore, 
$\unew$ can be written as
\begin{eqnarray}
\label{unew-full}
-\unew =  2 \pinit \, p_{X} \(1-\cos\theta_{X}\)\, ,
\end{eqnarray}
where $p_{X}$ and $\theta_{X}$ are determined from transverse momentum conservation and energy conservation, respectively, and are given by
\begin{equation}
\label{X-conservation}
k_{\perp}=p\, \sin\theta-p_{X}\, \sin\theta_{X}\, ,
\qquad
\pinit+k=p+p_{X}\, , 
\end{equation}
where $k_{\perp}$ denotes the transverse momentum of the thermal scatterer.
While in general $\unew$ also depends on  the magnitude of the
momentum of the parton from the thermal medium $k=|\vk|$, 
we can express $\unew$ in terms of $\pinit$,  $p$, and $\theta$ 
for any value of $\theta$ that is not too small because the 
characteristic values of $k_{\perp}$ and $k$ are of the order of $T$. 
First, 
since $p\gg T$, 
the transverse momentum of the outgoing parton, $p\sin\theta$, will be much larger than $T$ when $\theta$ is not too small. 
To balance such a large transverse momentum, we need to have $p_{X}\,\sin\theta_{X}\approx p\,\sin\theta$. 
Second, we have observed from our study of $F^{C\to \all}(p,\theta)$ in Section~\ref{sec:FandP} that when the momentum transfer is large, the energy transfer in a binary collision is also likely to be  large, i.e. $\o \gg T$. 
We therefore have from energy conservation~\eqref{X-conservation} that $p_{X}\approx \pinit-p=\o$.  
Combining the above two approximations and substituting into Eq.~\eqref{unew-full}, we obtain
\begin{eqnarray}
\label{unew}
\unew \approx -2 \pinit \, \[\(\pinit-p\)-\sqrt{\(\pinit-p\)^{2}-\(p\sin\theta\)^{2}}\]\, ,
\end{eqnarray}
from which we can determine the region in the $(\theta,p/T)$ plane where the condition $-\unew \gg m^{2}_{D}$ is satisfied.


\begin{figure} 
\centering
\includegraphics[height=0.32\textwidth]{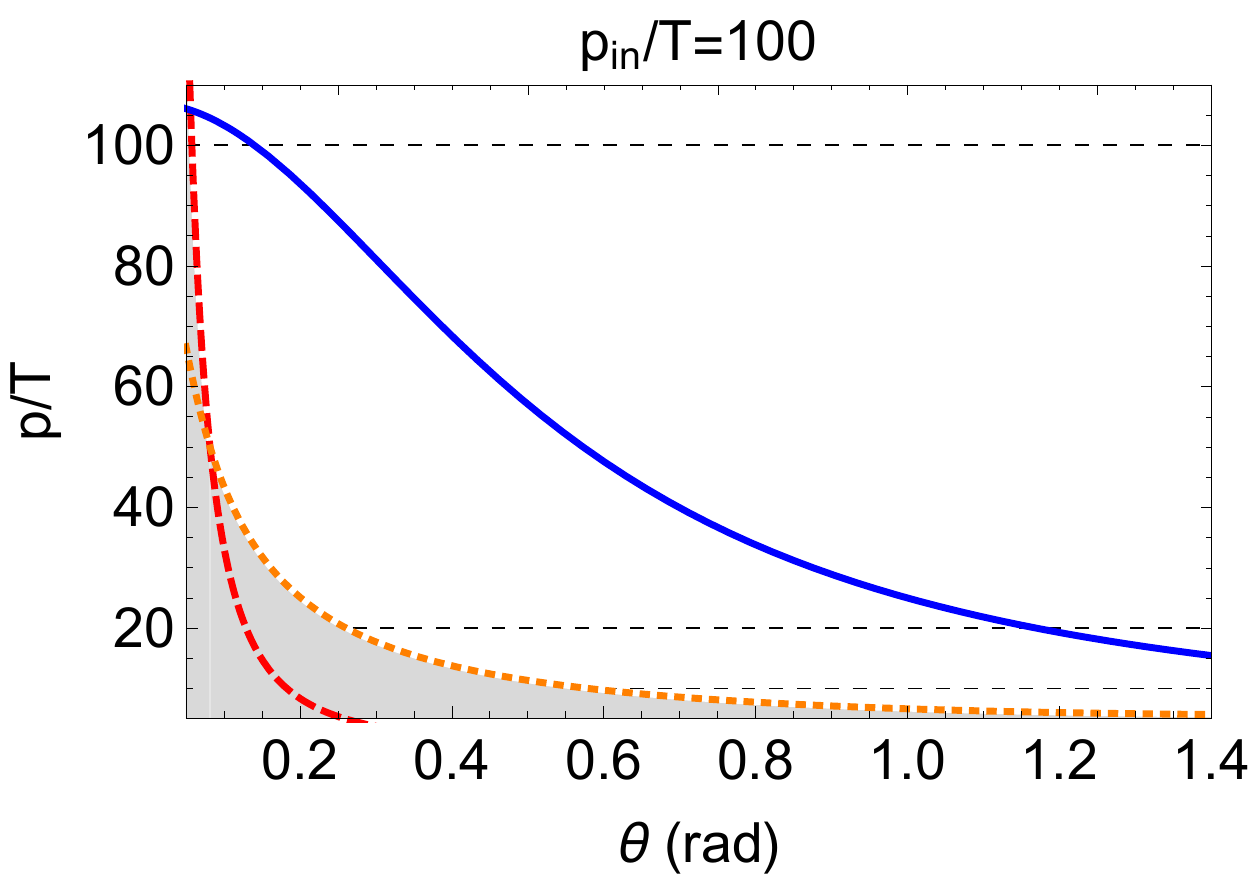}
\includegraphics[height=0.32\textwidth]{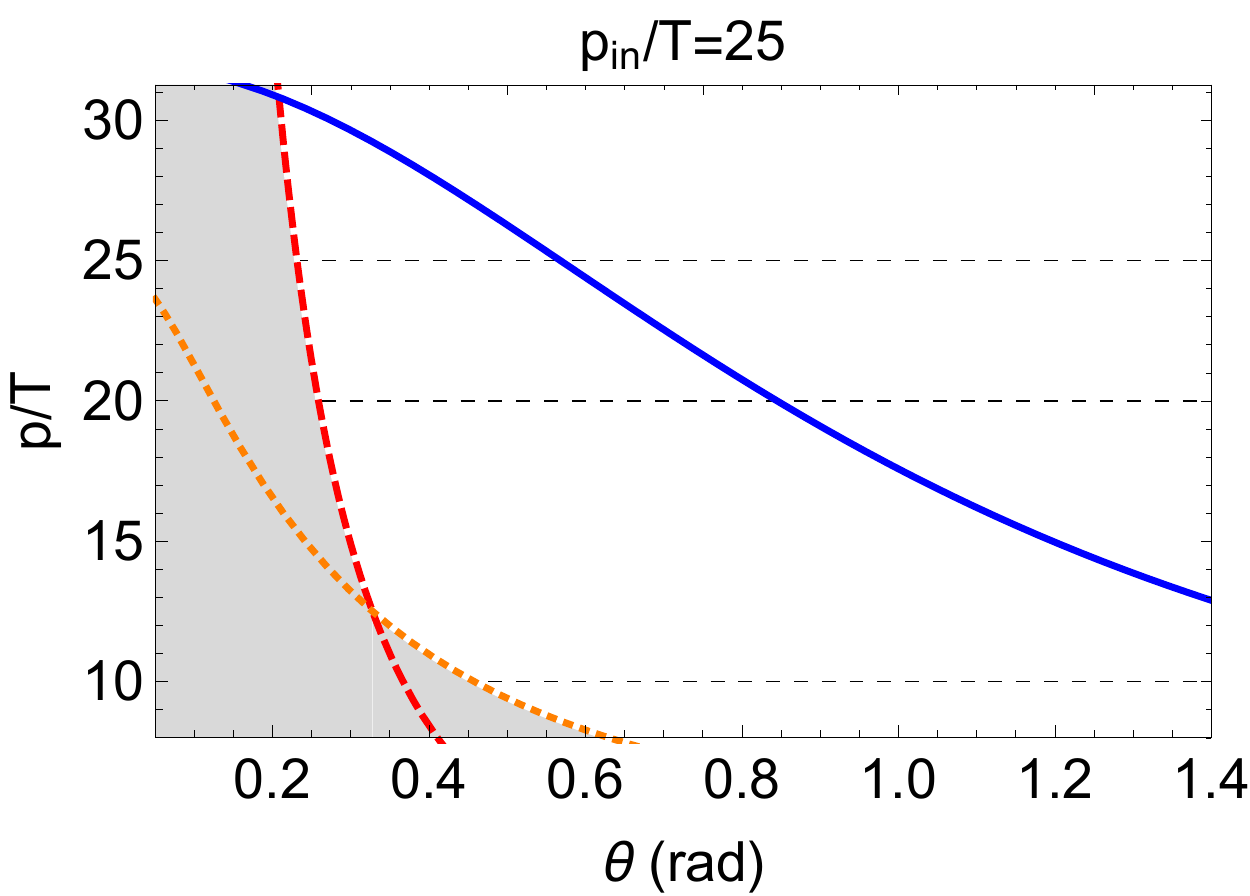}
\caption{
\label{fig:Thetamin}
The red and orange curves illustrate the boundary of the region in the  $(\theta,p/T)$ plane,
analogous to what is often called the Lund plane,
defined by the conditions \eqref{cond1} where medium effects can be neglected in the matrix elements for $2\lra 2$ scattering processes,
as we do in our calculations.
The red dashed curve and the orange dotted correspond to  $-\tnew=10 m^{2}_{D}$ 
and $-\unew =10 m^{2}_{D}$, respectively. 
Our calculations are valid in the region above both these curves, and should not
be relied upon quantitatively in the shaded region.
In plotting the curves in the left panel (right panel) we have chosen an incident parton
with $\pinit/T=100$ ($\pinit/T=25$).
The horizontal black dashed lines in both panels show the location of $p=\pinit$, $p=20 T$ and $p=10 T$, 
the latter two corresponding to the two different choices of $p_{\min}$ that we employed
in our evaluation of $P(\theta)$ in Fig.~\ref{fig:dPdThetaPiT}. 
The solid blue curves in both panels are determined by the 
condition $k_{\min}=7 T$ where $k_{\min}$ is the minimum possible value of the
energy of a parton in the medium that, when struck by a parton with incident energy $\pinit$, can yield an outgoing parton at a given point in the $(\theta,p/T)$ plane. $k_{\min}$ is given by the
expression \eqref{k-min}, and we have used 
$\pinit/T=100$ (left panel) or $\pinit/T=25$ (right panel) in our numerical evaluation of $k_{\min}$. 
All our results become smaller and smaller farther and farther above the blue curves.
Hence, our calculations are valid and our results are not small  in the region below the blue curves and above the red and orange curves.
  }
\end{figure}

In Fig.~\ref{fig:Thetamin}, 
we illustrate the regimes in the $(\theta,p/T)$ plane where the conditions \eqref{cond1} are satisfied for incident partons with  $\pinit/T=100$ and $\pinit/T=25$. 
We use the standard expression for Debye mass squared:
\begin{equation}
\label{mD}
m^{2}_{D}
= \frac{g_s^{2}}{3}\, \(N_{c}+\frac{N_{f}}{2}\)\, T^{2}\, ,
\end{equation}
choosing $N_c=N_f=3$ and, as described in the next Section, choosing $g_s=1.5$. 
The red dashed and orange dotted curves are determined by solving $-\tnew =10 m^{2}_{D}$ 
and $ -\unew =10 m^{2}_{D}$, respectively. 
We observe that the conditions \eqref{cond1} are
satisfied for sufficiently large $\theta$,  although how large $\theta$ needs to be depends on the values of $\pinit/T$ and $p/T$. 

The blue curves in Fig.~\ref{fig:Thetamin} do not represent limits on the validity of our calculation. However, above the blue curves the results that we obtain must be small in magnitude, for the following reason.
For scattering processes to yield outgoing partons with values of $(\theta,p/T)$
above the blue curves, the only partons from the medium that can contribute are those
with energies $k$ greater than $7 T$, whose $n_a(k)$ in (\ref{eq:BEandFD}) are smaller than $10^{-3}$.  For this reason,
the probability for scattering events that yield outgoing partons above the blue curves 
must be small.  Hence, the regime in the $(\theta,p/T)$ plane where medium effects
can be neglected in the matrix elements for $2\lra 2$ scattering as we do and where our calculations 
yield a significant scattering probability is the region above the red and orange curves and below
the blue curves.

\subsection{Estimating $P(\theta)$ and $N_{{\rm hard}}(\theta_{\min})$ for phenomenologically motivated inputs}
\label{sec:NhardEstimates}

In Fig.~\ref{fig:dPdThetaPiT} in Section~\ref{sec:FandP},
we have evaluated $P(\theta)/\kappa$.
By dividing the probability distribution $P(\theta)$ by $\kappa$ we obtained and plotted
 $\kappa$-independent results.  And, as we noted in Section~\ref{sec:FandP}, this is
 the form of our results that we should provide for use in a 
 future jet Monte Carlo analysis, which is the path to phenomenologically relevant predictions
 for experimental observables. 
 It may also be interesting to study the importance of processes in which a photon is radiated~\cite{Fries:2002kt} as well as $2\rightarrow 3$ scattering processes in future phenomenological studies.
 This is for the future.
 In the present paper, we would like to get at least a qualitative sense
 of $P(\theta)$ for incident partons with several values of $\pinit$.
This means that we need to 
input phenomenologically motivated values of $g_{s}$, $\CT$, and $T$ --- and hence $\kappa$.  

Since we are interested in those binary collisions with characteristic momentum transfer 
which is of the order $10$~GeV, following Ref.~\cite{PDG}
we will use $g_{s}=1.5$ as our benchmark value in the following analysis.
Of course in reality $g_s$ runs, meaning that in a future calculation that goes beyond
tree-level one should allow $g_s$ to depend on the momentum transfer in a particular
collision.  Working at tree level as we do, it is consistent just to pick a value of $g_s$, 
and we shall choose $g_s=1.5$.
We shall pick $T=0.4$~GeV as the temperature of our brick of QGP and $\CT=3$~fm as the time that a parton spends in our brick of QGP. 
With these choices of parameters, $\kappa \approx 30$. (The actual value is $30.84$, but this would be misplaced precision. We shall use $\kappa=30$ in plotting results in this Section.) 
While we should only expect our calculation to be quantitatively reliable for $g_{s}\ll 1$, 
we hope our results with $g_{s}=1.5$ will be of qualitative value in estimating the magnitude of $P(\theta)$ as well as its $\theta$-dependence. 
(We also note that $g_{s}=1.5$ corresponds to $\a_{{\rm QCD}}\approx 0.18$, in many contexts a weak coupling.)
Of course, any reader who has their own preferred values of $g_s$, $T$ and $\Delta t$ that they
like to use to make phenomenologically motivated estimates should feel free to do so. Our result for $P(\theta)$ is simply proportional to $\kappa = g_s^4 T\Delta t$.

\begin{figure}
%
%
\includegraphics[height=0.35\textwidth]{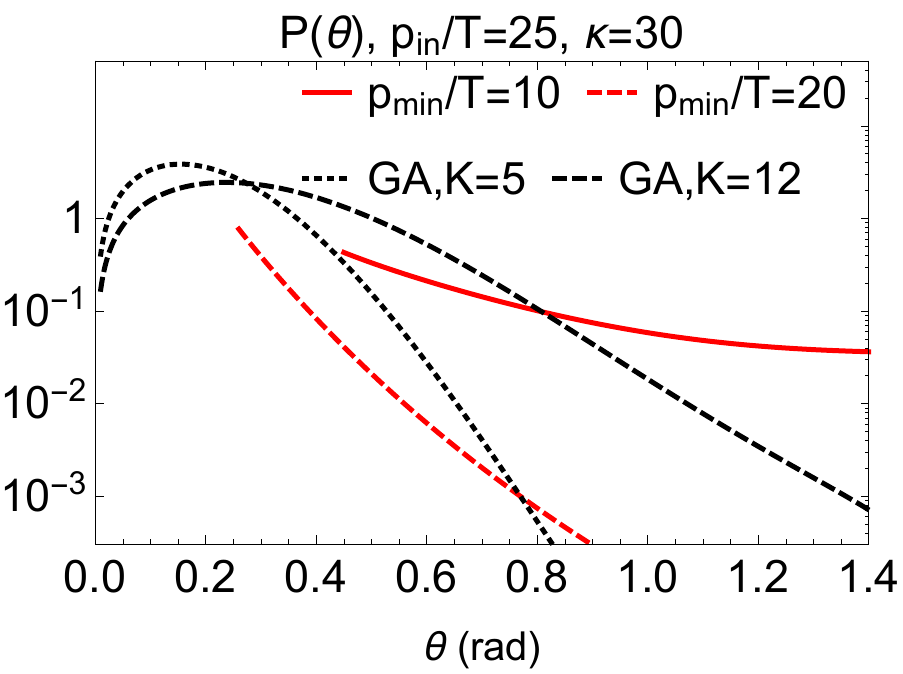}
\includegraphics[height=0.35\textwidth]{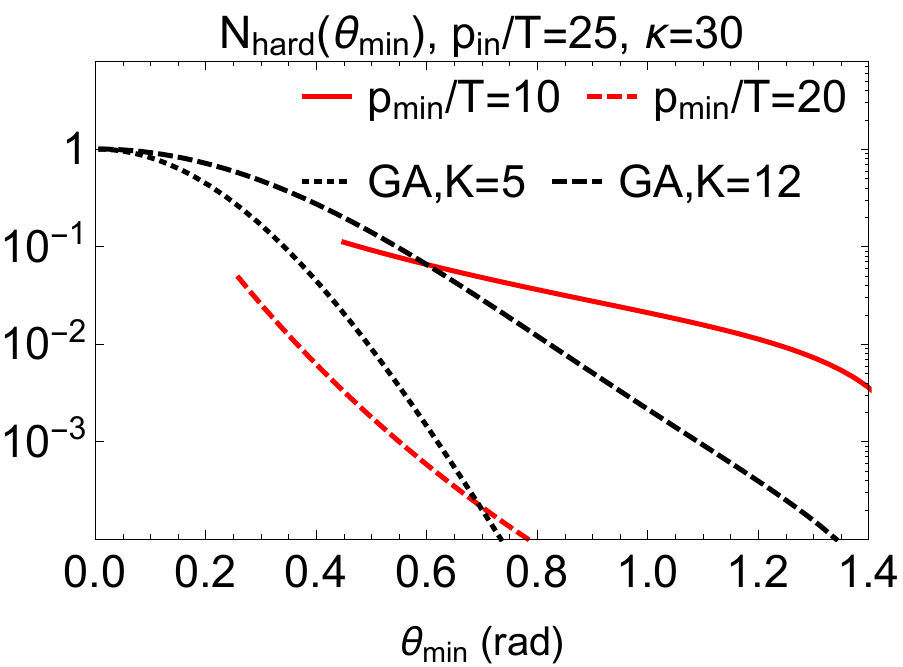}
%
%
\centering
\includegraphics[height=0.35\textwidth]{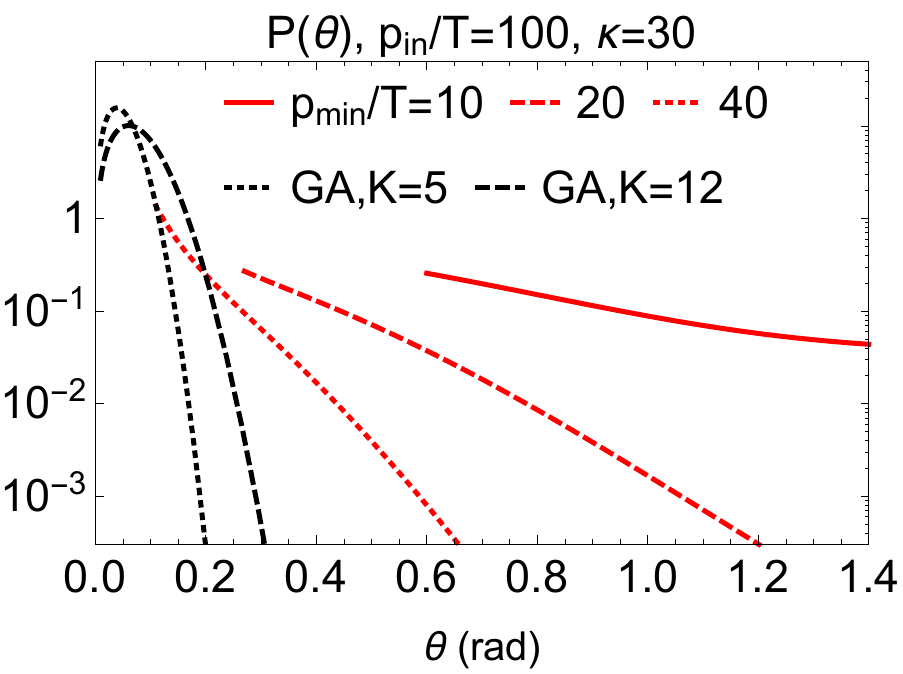}
\includegraphics[height=0.35\textwidth]{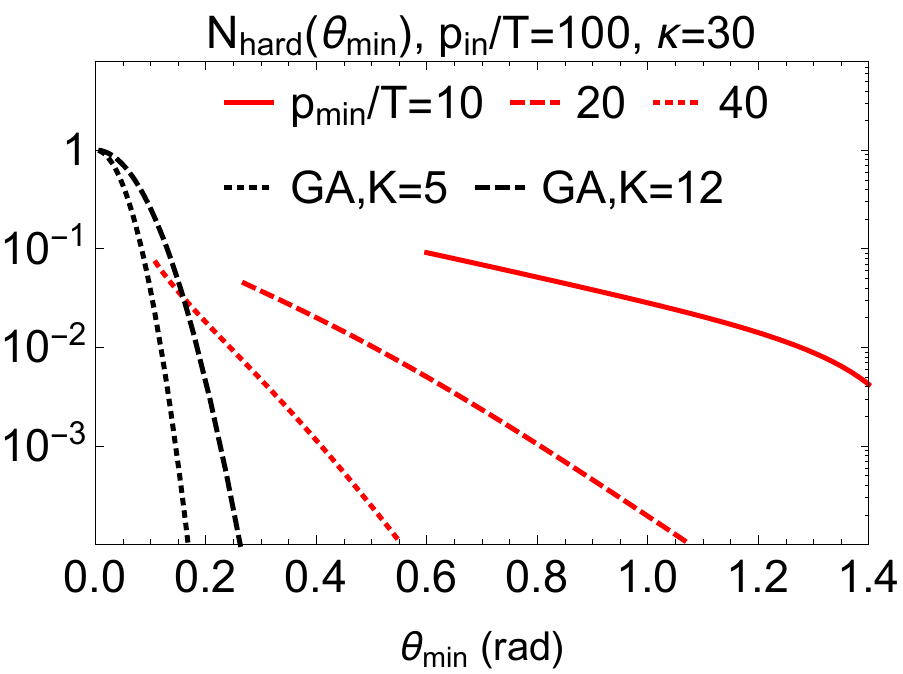}
%
%
%
\includegraphics[height=0.35\textwidth]{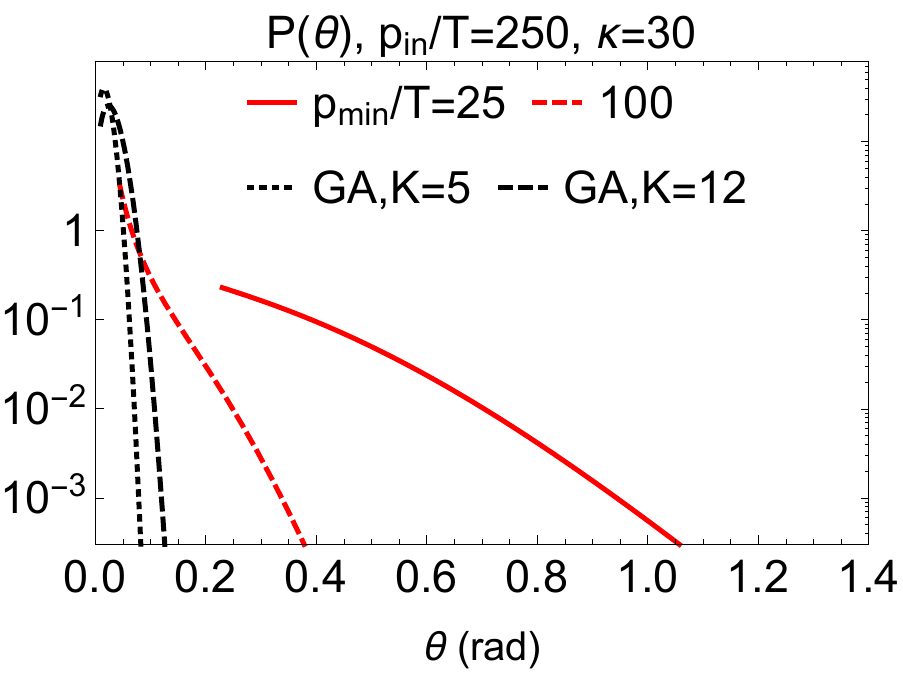}
\includegraphics[height=0.35\textwidth]{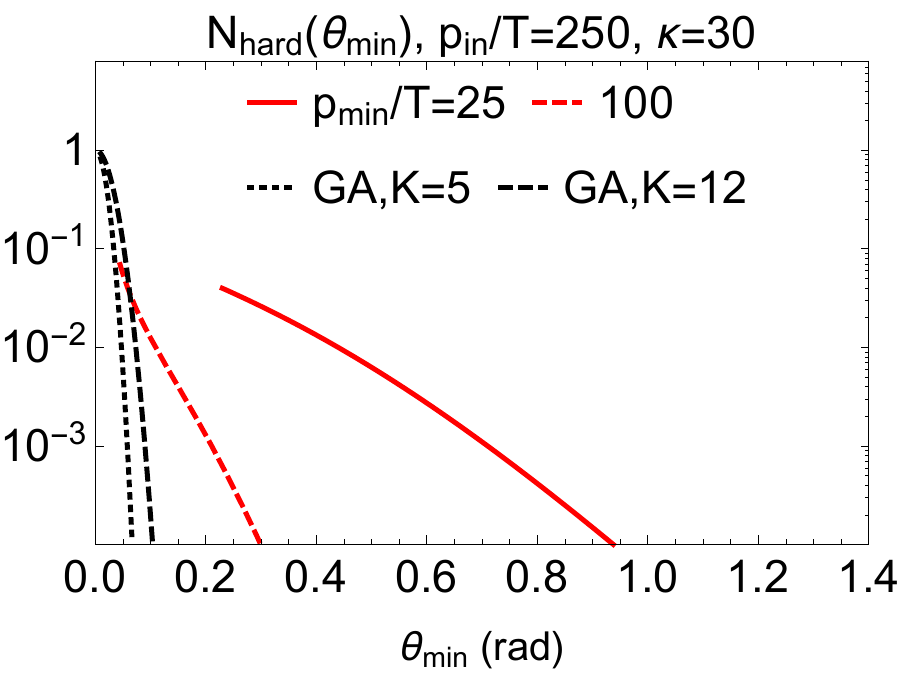}
\caption{
\label{fig:Phenomenology}
 $P(\theta)$ (left column) and $N_{\hard}(\theta_{\min})$ (right column)
for an incident gluon with $\pinit/T=25$ (upper row) and $\pinit/T=100$ (middle row).
In the top four panels, the solid red curves (dashed red curves) show our results when we include
all partons with $p>p_{\min}=10 \,T$ ($20 \,T$). 
In the middle panels, the dotted red curves
show our results when we only include partons with $p>p_{\min}=40\, T$.
In the lower panels we consider an incident gluon with $\pinit/T=250$ that yields a scattered
parton with $p>p_{\min}=25 \,T$ or $100\,T$.
We have set $\kappa=30$, corresponding to $g_s\approx 1.5$, $T=0.4~\GeV$ and $\CT=3$~fm, 
as discussed in the text. 
For comparison, 
we plot $P^{{\rm GA}}(\theta)$ from Eq.~\eqref{P-GA} for $K=5$ and 12 (black dotted and black dashed curves, respectively).   
}
\end{figure}

We will concentrate on the case where the incident parton is a gluon. 
We plot $P(\theta)$ in the left column of Fig.~\ref{fig:Phenomenology}
for $\pinit/T= 25$ (upper left) and 100 (middle left), in each case for $p_{\min}/T=10$ and 20.
These curves correspond to results shown in Fig.~\ref{fig:dPdThetaPiT}, multiplied 
by $\kappa=30$. Taking $T=0.4$~GeV, they correspond to incident gluons
with $\pinit=10$~GeV and 40 GeV and scattered partons with $p>4$~GeV and 8 GeV.
In the lower left panel, we plot $P(\theta)$ for $\pinit/T=250$, corresponding to $\pinit=100$~GeV,
for scattered partons with $p>10$~GeV and $p>40$~GeV.
As we have demonstrated in Fig.~\ref{fig:dPdThetaPiT}, 
$P(\theta)$ for an incident quark can be well described by multiplying $P(\theta)$ for an incident gluon by the ratio of Casimirs $C_F/C_A=4/9$.

In the right column of Fig.~\ref{fig:Phenomenology},
we integrate $P(\theta)$ over $\theta$ and obtain $N_{\hard}(\theta_{\min})$, 
defined in Eq.~\eqref{N-min}. 
(Since $P(\theta)$ drops very quickly for large values of $\theta$, when we evaluate  $N_{\hard}(\theta_{\min})$ numerically we stop the integration in Eq.~\eqref{N-min} at $\theta=1.5$.)
Among the quantities that we can calculate, $N_{\hard}(\theta_{\min})$  is perhaps the most useful
for the purpose of obtaining a qualitative sense of how large the effects of
point-like scatterers in the QGP will be.  For example, reading from
the dashed red curve in the middle-right panel of Fig.~\ref{fig:Phenomenology},
we see that if an incident gluon with $\pinit =100 T=40$~GeV traverses 3 fm of QGP with a temperature 
of 0.4 GeV, the probability that a parton with an energy $p>p_{\min}=20 T=8$~GeV is detected at some angle $\theta>0.8$ is around $1/1000$, while this probability rises to around $1/100$ for
detection at an angle $\theta>0.5$, and 
the probability that a parton with $p>p_{\min}=10 T=4$~GeV is 
detected at an angle $\theta>0.8$ 
is  around 1/20.
This gives a sense of the probability of kicking partons to these angles and as such
is helpful in making qualitative assessment of how small (meaning how improbable) the effects
that will need to be looked for via detecting suitable modifications to jet substructure
observables may be.
We would be happy to provide curves depicting our results for 
$N_{\hard}(\theta_{\min})$ or $P(\theta)$ for different choices of $\pinit$, $p_{\min}$, $T$,
$\Delta t$ and $g_s$.

In the middle row of Fig.~\ref{fig:Phenomenology}, where we consider incident
partons with $\pinit=100\, T$, we have also included results where we
only count scattered partons with $p>p_{\min}=40\, T$ (the red dotted curves).
This allows us to look at the dependence of our results on $\pinit$ in two ways.
If we compare 
the red solid curves above ($\pinit=25 \,T$ and $p_{\min}=10\, T$)  
to
the red dotted curves in the middle panels ($\pinit=100\, T$ and $p_{\min}=40\, T$) 
we see that
increasing $\pinit$ while increasing $p_{\min}$ proportionally rapidly reduces the probability
for large angle scattering.  This corresponds to increasing the momentum transfer in the binary collision, and is qualitatively as one would expect based upon intuition from Rutherford scattering.
On the other hand, if we compare the solid red curves in the top and middle panels, or the dashed red curves in the top and middle panels, we see that increasing $\pinit$ while keeping $p_{\min}$ fixed results in a much
smaller change in the probability for large angle scattering.  This corresponds to the observation
that the probability for kicking a parton with $p\gtrsim p_{\min}$ for some fixed $p_{\min}$ out of the medium
at some fixed large angle $\theta$ increases slowly with increasing  $\pinit$. This further highlights
the importance in our results of processes other than Rutherford scattering where what is detected is a parton that was kicked out of the medium.

In Fig.~\ref{fig:Phenomenology}, 
we have only plotted our results (the red solid, red dashed and red dotted curves) 
for $P(\theta)$ and $N_{\hard}(\theta_{\min})$
at large enough values of $\theta$ and $\theta_{\min}$ that the 
the condition \eqref{cond1} is satisfied. 
As we discussed in Section~\ref{sec:validity}, our calculation breaks down 
at smaller values of $\theta$.
For example, for $\pinit/T=100$ and $p_{\min}/T=20$ 
we observe from Fig.~\ref{fig:Thetamin} that the orange curve (determined by $(-\unew)=10 m^{2}_{D}$) intersects with $p/T=20$ at $\theta=0.27$, 
meaning the condition \eqref{cond1} will be satisfied for $\theta\geq 0.27$.  
We have therefore plotted $P(\theta)$ and  $N_{\hard}(\theta_{\min})$ for $\theta \geq 0.27$ and $\theta_{\min}\geq 0.27$ respectively. 

Our results can also only be trusted where $N_{\hard}(\theta_{\min})\ll 1$, since if $N_{\hard}(\theta_{\min})$ approaches 1 this tells us that we cannot neglect multiple scattering.
Including only single scattering, as we have done, is only valid where
$N_{\hard}(\theta_{\min})\ll 1$.
We see in the right column of Fig.~\ref{fig:Phenomenology}  that, for the values of parameters used,
$N_{\hard}(\theta_{\min}) < 0.1$ wherever we have shown our results, e.g.~wherever we
have plotted the red solid or dashed curves.  This means that, 
for $\kappa=30$, everywhere that the condition  \eqref{cond1} is satisfied
we also have $N_{\hard}(\theta_{\min}) < 0.1$.  If we had chosen a larger value of $\kappa$ this
would not have been the case, and we would have needed to enforce a separate constraint.

At values of $\theta$ and $\theta_{\min}$ that are smaller than those for
which we have plotted our results for $P(\theta)$ and $N_{\hard}(\theta_{\min})$, multiple scattering
will become important, making the calculation much more difficult.
At small enough angles, where many scatterings contribute, the
result will simplify as the probability distribution
for the transverse momentum transfer $\PT(\qperp)$
becomes a Gaussian at small enough $\qperp$~\cite{DEramo:2012uzl}.
As we noted in the Introduction, this is also the result that must
be obtained in the regime in which the momentum transfer
is small enough that the hard parton sees the QGP only
as a  liquid, without resolving the partons within it.
The transverse momentum picked up by an energetic
parton traversing a strongly coupled liquid is 
Gaussian distributed.  Hence, whether we think of this
from the perspective of a hard parton 
traversing a strongly coupled liquid
or from the perspective of multiple scattering in a weakly coupled
plasma, at small $\qperp$ we expect $\PT(\qperp)$ to take the form
\begin{eqnarray}
\label{PT-Gaussian}
\PT^{{\rm GA}}(\qperp)
=
\frac{4\pi}{\hat{q}\, L}\, e^{-\frac{\qperp^{2}}{\hat{q} L}}\ ,
\end{eqnarray}
where we have written the width of the Gaussian as $\hat q L$, 
denoting the mean transverse momentum squared picked up
per distance travelled by $\hat q$ as is conventional.
The physics of multiple soft scattering in a weakly coupled plasma
or the physics of how an energetic probe ``sees'' a liquid then
determine the value of the parameter $\hat q$.
Following Ref.~\cite{Casalderrey-Solana:2016jvj}, 
it is convenient to introduce a dimensionless parameter $K$ to parametrize 
the magnitude of $\hat{q}$ via
\begin{eqnarray}
\hat{q} = K \, T^{3}\,  .
\end{eqnarray}
We can then use Eq.~\eqref{FullJacobian}) from Appendix C
to convert $\PT^{\rm GA}(\qperp)$ in Eq.~\ref{PT-Gaussian} to a probability
distribution $P^{\rm GA}(\theta)$ for the angle $\theta$, obtaining
\begin{eqnarray}
\label{P-GA}
P^{{\rm GA}}(\theta)
&=&\[{\CJ}^{-1}_{\perp}\,\PT^{{\rm GA}}(\qperp=\pinit\, \sin\theta) \]
\no \\
&=& \(\frac{2\, \sin\theta\cos\theta}{K(T/\pinit)^{2}T\CT}\)\,
\exp\(-\frac{\(\sin\theta\)^{2}}{K(T/\pinit)^{2}T\CT}\) \, ,
\end{eqnarray}
where we have used the approximation $\qperp \approx \pinit \sin \theta$, 
valid for small $\theta$ where $p\approx \pinit$.

Hence, the behavior that we expect for $P(\theta)$ is that it should
take the form (\ref{P-GA}) at small $\theta$, for some value of $K$, and
should then have a tail at larger angles $\theta$ that is due to single scattering
of partons in the QGP, a tail that we have calculated and that is illustrated by
the red curves in Fig.~\ref{fig:Phenomenology}.
To get a sense of how this might look, in Fig.~\ref{fig:Phenomenology} in addition
to plotting the results of our calculations, in red, we have plotted $P^{\rm GA}(\theta)$
from (\ref{P-GA}) for two benchmark values of $K$, namely $K=5$ and $K=12$.
($K=5$ is the value obtained by the JET collaboration~\cite{Burke:2013yra}
upon comparing calculations of observables sensitive to parton energy loss in a weakly coupled framework in which $K$ controls energy loss as well as transverse momentum broadening.
$K=12$ is half of the value found for an energetic parton
traversing the strongly coupled plasma of ${\cal N}=4$ SYM theory~\cite{Liu:2006ug,DEramo:2010wup,DEramo:2012uzl}; since this theory has 
more degrees of freedom than QCD, its strongly coupled plasma would have a larger value of $K$ than the strongly coupled QGP.)

Plotting $P^{\rm GA}(\theta)$ in addition to our own results in Fig.~\ref{fig:Phenomenology} is useful for two
reasons.  First, it helps us to imagine how these quantities may behave in a more complete calculation, following one of the black curves at small angles and then behaving along 
the lines of our results in red at large angles where single Moli\`ere scattering off partons in the QGP
dominates.  Second, by comparing the red curves to the black curves we can
get a sense of at how large values of $\theta$ 
single hard scattering off partons in the QGP  is likely to dominate over multiple soft scattering
or the physics of the strongly coupled liquid. 
From the middle panels of Fig.~\ref{fig:Phenomenology}
we see that the situation is rather clean for incident partons with $\pinit=100\, T=40$~GeV: as long
as we look at partons that scatter into a direction that deviates from the direction of the incident
parton by $\theta>0.3$, we will be seeing Moli\`ere scattering.  And, the probability for
scattering at these angles can be quite substantial.  
If it proves possible to look at the scattering of even higher energy jet partons, for example as in the bottom panels of Fig.~\ref{fig:Phenomenology} where we take $\pinit=250 \,T =100$~GeV, 
Moli\`ere scattering and multiple soft scattering or the physics of the
strongly coupled liquid separate even further. And, the probabilities for
seeing large angle scattering remain quite significant as long as one
looks for scattered partons with $p>p_{\min}$ for a small enough $p_{\min}$, for example
$p_{\min}=25 \,T=10$~GeV as in the solid red curves in the bottom panels of Fig.~\ref{fig:Phenomenology}.
The situation is less clear when we look at incident partons with $\pinit =25 T=10$~GeV, in
the top panels of Fig.~\ref{fig:Phenomenology}.  We see there that
in order to see a red curve above the black curves at a probability above $10^{-3}$ we need
to look at the solid red curves, meaning we need to look at scattered partons
with energies down to $p_{\min}=10 T = 4$~GeV and we need to look at rather large
angles.  It will be hard to separate final state hadrons coming from scatterings
with these parameters from final state hadrons coming from the wake that the jet leaves
behind in the plasma.

To the extent that one can draw conclusions from a calculation of
scattering off a brick of plasma with $T=0.4$~GeV and $\Delta t=3$~fm, our results suggest
that experimentalists should look for observables sensitive to phenomena along the following lines:
40 GeV partons within a jet scatter off a parton in the plasma, yielding partons
with energies greater than 8 GeV at angles $\theta > 0.5$ with probability $1/100$ and
at angles $\theta>0.8$ with probability $1/1000$.  We would be happy to work with
anyone planning future experiments to provide them with results along these lines
for other values of the various parameters.  But, the real path to predictions for observables
is to take our results, formulated as in Section~\ref{sec:FandP}, and
to incorporate  them into a jet Monte Carlo analysis that also includes a 
realistic description of the expanding cooling droplet of plasma produced
in a heavy ion collision.

\section{Summary and outlook
\label{sec:outlook}
}

We have analyzed the thought experiment depicted in Fig.~\ref{fig:kinematic} in which an incident parton (quark, antiquark or gluon) with energy $\pinit$ 
traverses a brick of QGP with some thickness $L$ and some constant temperature $T$ and 
computed the probability distribution $F(p,\theta)$ for finding a parton (quark, antiquark or gluon) subsequently with an energy $p$ that has been scattered by an angle $\theta$ relative to the
direction of the incident parton.
By integrating over $p$ we obtain $P(\theta)$, the probability for finding a parton with $p>p_{\min}$  scattered by $\theta$, and
then by integrating over $\theta$ we obtain $N_{\rm hard}(\theta_{\min})$, the number of hard partons scattered by
an angle $\theta>\theta_{\min}$.
We only consider binary collision processes in which the incident parton strikes a single parton from the medium, once.
Because we neglect multiple scattering, our results are relevant only in the kinematic regime in which $N_{\rm hard}$ turns
out to be small, which means 
at large momentum transfer, and in particular at large values of $\theta$.  Because we are focusing on binary collisions with a large momentum transfer, for our medium we choose 
a gas of massless quarks, antiquarks and gluons with Fermi-Dirac or Bose-Einstein momentum distributions.  
Although we have ensured that we work only in a regime in which the momentum transfer in the binary collisions
that we analyze is large enough 
that it is reasonable to neglect the Debye masses of the partons in the plasma, 
choosing their momentum distributions as if they were a noninteracting gas is relevant only as a simple benchmark.
Ultimately, we look forward to the day when experimental measurements that are sensitive to
the Moli\`ere scattering that we have analyzed can be used, first of all, to provide tangible evidence that the liquid QGP that we see today really is made of point-like quarks and gluons when probed at high momentum transfer and, second of all, via deviations from 
predictions based upon our calculations, to learn about the actual momentum distributions of these quarks and gluons.
This would realize the vision of using the scattering of jet partons to learn about the microscopic structure of liquid QGP and would be analogous to learning about the parton distribution
functions for QGP.

Realizing this vision will require incorporating the results of our calculations within jet Monte Carlo analyses in which realistic jets
are embedded within realistic hydrodynamic models for the
expanding cooling droplets of QGP produced in heavy ion collisions.  
Our results as we have obtained them here are based upon a thought experiment and cannot be compared directly to 
experimental data.  It would be interesting to use comparisons between our results and results from
Monte Carlo analyses in which binary collisons are already included (set up with jets probing a static brick like ours)
to identify observable consequences of large-angle scattering.  With a view toward 
Monte Carlo calculations which 
do not currently include binary collisions, we have presented our results in Section~\ref{sec:FandP} in a form in which
they could be incorporated into such analyses.

We note that we have worked only to leading order in perturbative QCD. This can certainly be improved upon
in future work.  However, it is our sense that incorporating these results in more realistic (Monte Carlo) modeling
of jets probing more realistic (hydrodynamic) droplets of QGP
is a more immediate priority than pushing our
``brick calculation'' beyond leading order.

Although the road ahead toward quantitative comparison to experimental measurements is a long one, our present results
can already be used to reach several interesting qualitative conclusions.  Perhaps the most interesting aspect of
our results from a theoretical
perspective is the importance of channels that are not Rutherford-like.  It is only at small angles $\theta$ (where high momentum transfer requires large $\pinit$, as in previous calculations done in the $\pinit\to\infty$ limit) where the dominant binary collision process is 
the Rutherford-like process where the parton that is detected is the incident parton, scattered by an angle $\theta$.  
We have checked that our results reproduce the results of previous calculations in this regime.
At the larger values of $\theta$ that are of interest, though, processes in which the detected parton is either a parton from the
medium that received a kick or a parton that was produced in the collision (cf $gg\leftrightarrow q \bar q$)
are much more important.  Consequently, also, we realize that at the values of $\theta$ that are of interest it is 
important to look for scattered partons that are still hard but that have substantially smaller energy than
the incident parton.

Even though quantitative predictions for experimental measurements await further steps down the road ahead as we
have discussed, the second place where our results are of qualitative interest is in the context of gauging what sorts of observables
experimentalists should aim to measure.  To get a sense of this, in Section~\ref{sec:NhardEstimates}
we have considered a brick of plasma that is 3 fm thick and that has a temperature $T=0.4$~GeV, and have set $g_s=1.5$,
corresponding to $\alpha_{\rm QCD}\approx 0.18$.  (This exercise can easily be redone with other
values of these parameters.) With these values, we find that it would be quite a challenge
to look for the Moli\`ere scattering of jet partons that have $\pinit=10$~GeV before they scatter.  Doing so would require
looking for observables that are sensitive to scattered partons with energies down to 4 GeV, and even if that were possible
it would be hard to differentiate between partons scattering off particulate structures within the liquid QGP and partons picking 
up a Gaussian distribution of transverse 
momentum just from soft interactions with the liquid QGP.
The picture is much more promising if instead we look for the Moli\`ere scattering of jet partons that have $\pinit=40$~GeV (or more).
before they scatter.  Moli\`ere scattering is the dominant contribution if we look for scattering with $\theta>0.3$.
And, although these processes are rare (they have to be rare in the regime in which they are the dominant 
contribution), the relevant probabilities are not tiny, given the high statistics data sets for jets in
heavy ion collisions anticipated in the 2020s.
For an incident parton with $\pinit=40$~GeV, the probability of seeing a scattered parton with $p>8$~GeV deflected by $\theta>0.5$ ($\theta>0.8$) is around $1/100$ $(1/1000)$.
Getting a sense of the kinds of values of $\pinit$, $p$ and $\theta$ where one should look, and a sense of the scale of
the probability for the Moli\`ere scattering that one is looking for, should be of value both to experimentalists planning
future measurements and to theorists exploring which jet substructure observables may be the most promising to measure.

\acknowledgments
This work was supported in part by the Office of Nuclear Physics of the U.S. Department of Energy under Contract Number~DE-SC0011090~(KR, YY) and by Istituto Nazionale di Fisica Nucleare (INFN) through the ``Theoretical Astroparticle Physics'' (TAsP) project~(FD).
KR gratefully acknowledges the hospitality of the CERN Theory Group.
We  thank Jorge Casalderrey-Solana, Leticia Cunqueiro, Peter Jacobs, Aleksi Kurkela, Volker Koch, Yen-Jie Lee, Guilherme Milhano, Dani Pablos, Gunther Roland, Wilke van der Schee, Xin-Nian Wang, Urs Wiedemann, Bowen Xiao, Feng Yuan and Korinna Zapp for helpful conversations.

\begin{appendix}

\section{Full Boltzmann Equation}
\label{app:Boltz}

In this Appendix, we present a full derivation of the Boltzmann equation describing the evolution of the phase space density. After presenting the general formalism, we show how we recover \Eq{eq:finaldistr} in the limit of a single binary collision.  The expression (\ref{eq:finaldistr}) 
is then the starting point for the derivation of all of our results.

Beginning with greater generality than in \Eq{eq:finaldistr}, we define the phase-space distribution as follows
\be
\begin{split}
\ftotal_{a}({\bs p}, \lambda_a, \chi_a)  \; \equiv \;  & \text{Phase-space probability of finding a parton of species $a$ ($u, d, s, \bar{u}, \bar{d}, \bar{s}$ or $g$)} \\ &
\text{with momentum ${\bs p}$, helicity $\lambda_a$ and color state $\chi_a$.} 
\end{split}
\label{eq:ftildedef}
\ee
This function depends on the time $t$, but we leave this dependence out of our notation for
the present. The Boltzmann equation describing the time-evolution of this phase-space distribution takes the schematic form
\be
\frac{\partial \ftotal_{a}({\bs p}, \lambda_a, \chi_a)}{\partial t} = \mathcal{C}_{a}[\ftotal_{a}({\bs p}, \lambda_a, \chi_a)] \ .
\label{eq:BE}
\ee
On the left-hand side, we have the time derivative of the phase-space distribution. On the right-hand side, we have the reason why such a function evolves with time: (binary) collisions. The collision operator $\mathcal{C}_{a}$ is a functional that depends on the phase-space distribution of the parton $a$ under consideration.

The collision operator has two distinct contributions that we denote via
\be
\mathcal{C}_{a}[\ftotal_{a}({\bs p}, \lambda_a, \chi_a)] = 
\mathcal{C}^{(+)}_{a}[\ftotal_a({\bs p}, \lambda_a, \chi_a)] - 
\mathcal{C}^{(-)}_{a}[\ftotal_a({\bs p}, \lambda_a, \chi_a)] \ ,
\ee
because there are two different ways to alter the distribution:
\begin{itemize}
\item a binary collision produces the parton $a$ with momentum $\bs p$ in the final state, which is accounted for by $\mathcal{C}^{(+)}_{a}[\ftotal_{a}({\bs p}, \lambda_a, \chi_a)]$ that appears with a plus sign;
\item a parton $a$ with momentum $\bs p$ in the initial state is involved in a binary collision, which is accounted for by $\mathcal{C}^{(-)}_{a}[\ftotal_{a}({\bs p}, \lambda_a, \chi_a)]$ that appears with a minus sign.
\end{itemize}
We are interested only in the phase space distribution for the momentum, meaning that later in our derivation we will average over the helicity and color states.

\subsection{Collision Operator for a Specific Binary Process}

The expression in \Eq{eq:BE} is very general. Once we have a specific theory for the interactions mediating the binary collisions (in our  calculation, QCD), we can derive an 
explicit expression for the collision operator.  In this Appendix we shall not specialize that far, considering here a specific binary process
\be
a_{(\bs p)} \; b_{(\bs k)} \;\;\; \leftrightarrow \;\;\; c_{(\bs p^\prime)} \; d_{(\bs k^\prime)} \ .
\ee
In our derivation, we account for this process going both from left to right and from right to left. In the former case, it contributes to $\mathcal{C}^{(-)}_{a}$ (it can destroy a parton $a$ with the given momentum $\bs p$), whereas in the latter case it can contribute  $\mathcal{C}^{(+)}_{a}$. 
The explicit expressions for both contributions are given by:
\be
\begin{split}
\left. \mathcal{C}^{(-)}_{a}[\ftotal_{a}({\bs p}, \lambda_a, \chi_a)] \right|_{a b \leftrightarrow c d}= 
\frac{1}{1 + \delta_{c d}}  \sum_{\lambda_{b c d} \chi_{bcd}}  & \,
\int_{\phaseAll}  \,  \left| M_{a b \rightarrow c d} \right|^2 \, \ftotal_{a}({\bs p}, \lambda_a, \chi_a) \ftotal_{b}({\bs k}, \lambda_b, \chi_b) \\ &
\left[ 1 \pm \ftotal_{c}({\bs p}^\prime, \lambda_c, \chi_c) \right] \left[ 1  \pm \ftotal_{d}({\bs k}^\prime, \lambda_d, \chi_d) \right]  \ .
\end{split}
\ee
\be
\begin{split}
\left. \mathcal{C}^{(+)}_{a}[\ftotal_{a}({\bs p}, \lambda_a, \chi_a)] \right|_{a b \leftrightarrow c d} = 
\frac{1}{1 + \delta_{c d}}  \sum_{\lambda_{b c d} \chi_{bcd}}  & \,
\int_{\phaseAll}  \, \left| M_{a b \rightarrow c d}\right|^2 \, 
\ftotal_{c}({\bs p}^\prime, \lambda_c, \chi_c) \ftotal_{d}({\bs k}^\prime, \lambda_d, \chi_d) \\ &
\left[ 1 \pm \ftotal_{a}({\bs p}, \lambda_a, \chi_a) \right] \left[ 1  \pm \ftotal_{b}({\bs k}, \lambda_b, \chi_b)  \right]  \ .
\end{split}
\ee
Here, the sign of the $\pm$ in a term like $\left[ 1 \pm \ftotal_c \right]$ is positive for bosons and negative for fermions, and these factors describe the Bose enhancement or Pauli blocking for the particles produced in the final state.
Note that we are using the short-handed notation
\begin{eqnarray}
\label{intphasespaceapp}
\int_{\phaseAll} &\equiv& \frac{1}{2 p}\, \int \frac{d^{3}\vk}{2k \, \phase}\,\int \frac{d^{3}\vp'}{2p' \, \phase}\,\int \frac{d^{3}\vk'}{2k' \, \phase}\, 
\no\\ &\times& (2\pi)^{4}\, \d^{(3)}\(\vp+\vk-\vp'-\vk'\)\, \d\(p+k-p'-k'\) \ .
\end{eqnarray}
The squared matrix elements $\left| M_{a b \rightarrow c d} \right|^2$ are for a given polarization and color configuration, and we explicitly sum over such configurations for the states $b$, $c$, $d$. The prefactor with the $\delta_{cd}$ accounts for the case where $c$ and $d$ are identical particles, where we must not double count. Upon assuming CP invariance, valid in particular for strong interactions, we have the identity
\be
\left| M_{a b \rightarrow c d} \right|^2 = \left|M_{c d \rightarrow a b} \right|^2 \equiv \left|M_{a b \leftrightarrow c d} \right|^2  \ .
\ee
Thus we can combine the two contributions together, and write the collision operator as
\be
\begin{split}
\left. \mathcal{C}_{a}[\ftotal_{a}({\bs p}, \lambda_a, \chi_a)] \right|_{a b \leftrightarrow c d} = & \, 
\frac{1}{1 + \delta_{cd}} \sum_{\lambda_{b c d} \chi_{bcd}}  \int_{\phaseAll} \left| M_{a b \leftrightarrow c d} \right|^2 \\ &
\left\{\ftotal_{c}({\bs p}^\prime, \lambda_c, \chi_c) \ftotal_{d}({\bs k}^\prime, \lambda_d, \chi_d) \left[ 1 \pm \ftotal_{a}({\bs p}, \lambda_a, \chi_a) \right] \left[ 1  \pm \ftotal_{b}({\bs k}, \lambda_b, \chi_b)  \right]+ \right. \\ & \left. - \ftotal_{a}({\bs p}, \lambda_a, \chi_a) \ftotal_{b}({\bs k}, \lambda_b, \chi_b) \left[ 1 \pm \ftotal_{c}({\bs p}^\prime, \lambda_c, \chi_c) \right] \left[ 1  \pm \ftotal_{d}({\bs k}^\prime, \lambda_d, \chi_d) \right] \right\}\ .
\end{split}
\ee
The total collision operator appearing in the Boltzmann equation for species $a$ is then the sum of all the individual ones accounting for each binary collision process in which $a$ is involved:
\be
\mathcal{C}_{a}[\ftotal_{a}({\bs p}, \lambda_a, \chi_a)] = \sum_{n} \left. \mathcal{C}_{a}[\ftotal_{a}({\bs p}, \lambda_a, \chi_a)] \right|_{n} \ ,
\ee
where $n$ is the index labeling the different processes (e.g. $n = a b \leftrightarrow c d$).

\subsection{Average over helicity and color states}

We are not interested in keeping track of helicities and colors, since they cannot be resolved by the detector. We will average over them by introducing a new distribution
\be
\ftotalbis_{a}({\bs p}) \equiv \frac{1}{\nu_a} \sum_{\lambda_a \chi_a} \ftotal_{a}({\bs p}, \lambda_a, \chi_a)  \ .
\label{eq:faver}
\ee
The degeneracy factor $\nu_a$ is the sum of all helicity and color configurations.
Upon applying this definition to the Boltzmann equation in \Eq{eq:BE} we find
\be
\frac{\partial \ftotalbis_{a}({\bs p})}{\partial t} =
\frac{1}{\nu_a} \sum_{\lambda, \chi} \frac{\partial \ftotal_{a}({\bs p}, \lambda_a, \chi_a)}{\partial t}  =
\frac{1}{\nu_a} \sum_n \sum_{\lambda_a \chi_a}  \left. \mathcal{C}_{a}[\ftotal_{a}({\bs p}, \lambda_a, \chi_a)] \right|_{n}  \ .
\label{eq:BE2}
\ee
Focusing on a specific binary process $n = a b \leftrightarrow c d$, we can then write the explicit expression
\be
\begin{split}
\left. \frac{\partial \ftotalbis_{a}({\bs p})}{\partial t} \right|_{a b \leftrightarrow c d} =  & \, 
\frac{1}{\nu_a} \frac{1}{1 + \delta_{cd}}  \sum_{\lambda_a \chi_a} \sum_{\lambda_{b c d} \chi_{bcd}}  \int_{\phaseAll} \left| M_{a b \leftrightarrow c d} \right|^2 \\ & \left\{\ftotal_{c}({\bs p}^\prime, \lambda_c, \chi_c) \ftotal_{d}({\bs k}^\prime, \lambda_d, \chi_d) \left[ 1 \pm \ftotal_{a}({\bs p}, \lambda_a, \chi_a) \right] \left[ 1  \pm \ftotal_{b}({\bs k}, \lambda_b, \chi_b)  \right]+ \right. \\ & \left. - \ftotal_{a}({\bs p}, \lambda_a, \chi_a) \ftotal_{b}({\bs k}, \lambda_b, \chi_b) \left[ 1 \pm \ftotal_{c}({\bs p}^\prime, \lambda_c, \chi_c) \right] \left[ 1  \pm \ftotal_{d}({\bs k}^\prime, \lambda_d, \chi_d) \right] \right\}\ .
\end{split}
\ee
Finally, we replace all the distributions occurring on the right-hand side with the those averaged over polarizations and colors as defined in \Eq{eq:faver}. In doing so, we are assuming that the medium has no net polarization and no net color charge. We also average over the helicity and color state of the incoming parton probing the medium. We end up with the expression
\be
\left. \frac{\partial \ftotalbis_{a}({\bs p})}{\partial t} \right|_{a b \leftrightarrow c d} = \left. \widetilde{\mathcal{C}}_{a}[\ftotalbis_{a}({\bs p})] \right|_{a b \leftrightarrow c d} \ ,
\ee
where we have defined the collision operator accounting for the process $a b \leftrightarrow c d$ by
\be
\begin{split}
\left. \widetilde{\mathcal{C}}_{a}[\ftotalbis_{a}({\bs p})] \right|_{a b \leftrightarrow c d} \equiv  & \, 
\frac{1}{\nu_a} \frac{1}{1 + \delta_{cd}}  \int_{\phaseAll} \left| \mathcal{M}_{a b \leftrightarrow c d} \right|^2 \\ & \left\{ \ftotalbis_{c}({\bs p}^\prime) 
\ftotalbis_{d}({\bs k}^\prime) \left[ 1 \pm \ftotalbis_{a}({\bs p}) \right] \left[ 1  \pm \ftotalbis_{b}({\bs k})  \right] + \right. \\ & \left. - \ftotalbis_{a}({\bs p}) \ftotalbis_{b}({\bs k}) \left[ 1 \pm \ftotalbis_{c}({\bs p}^\prime) \right] \left[ 1  \pm \ftotalbis_{d}({\bs k}^\prime) \right] \right\}\ .
\label{eq:collisionoperator}
\end{split}
\ee
Here, we have introduced the matrix elements in the form that we use them in Section~\ref{sec:formalism}, namely
\be
\left| \mathcal{M}_{a b \leftrightarrow c d} \right|^2 \equiv \sum_{\lambda_{a b c d} \chi_{a bcd}} \left|M_{a b \leftrightarrow c d} \right|^2 \ ,
\ee
summed over initial {\it and final} polarizations. For the QCD processes of interest to us, these matrix elements are given in Table~\ref{tab:QCDprocesses} of Section~\ref{sec:matrixelements}. The full evolution of the averaged phase space distribution reads
\be
 \frac{\partial \ftotalbis_{a}({\bs p})}{\partial t} = \sum_n \left. \widetilde{\mathcal{C}}_{a}[\ftotalbis_{a}({\bs p})] \right|_{n} \ ,
\ee
with the sum accounting for all possible processes affecting the phase space distribution of the parton $a$.

\subsection{Single Scattering Approximation}

The results found so far allow for the possibility of multiple binary collisions.
Next, we make the further assumption that the incoming probe scatters off a constituent of the medium just once before escaping on the opposite side. In order to so do, we find it convenient to employ the decomposition
\be
\ftotalbis_{a}({\bs p}) \equiv n_{a}({\bs p}) + f_{a}({\bs p}) \ ,
\label{eq:decomposition}
\ee
where the ``soft'' thermal part $n_a$ is constant in time, and the residual piece can be interpreted as the ``hard'' part of the distribution, describing energetic partons. The collision operator for a specific binary process $a b \leftrightarrow c d$, whose explicit expression is given in \Eq{eq:collisionoperator}, can then be simplified as follows. First, we observe that once we employ the decomposition in \Eq{eq:decomposition} the contribution with only thermal distributions vanishes because of the detailed balance principle. 
Next, we observe that we are only interested in collisions in which an energetic parton
collides with a soft parton from the medium. (If we included many collisions, somewhere downstream from the first collision an energetic parton might collide with another energetic parton. This is impossible in the first collision, which for us is the only collision.) We furthermore observe
that in the ``hard region'' of phase space (i.e. $\bs p \gg T$) where we shall focus,
we have $n_{a}({\bs p}) \ll 1$ and $f_a\ll 1$ also.
Looking at the second and third lines in \Eq{eq:collisionoperator}, describing the process $c d \leftrightarrow a b$, 
we find that via these considerations they simplify:
\be
\ftotalbis_{c}({\bs p}^\prime) \ftotalbis_{d}({\bs k}^\prime) \left[ 1 \pm \ftotalbis_{a}({\bs p}) \right] \left[ 1  \pm \ftotalbis_{b}({\bs k})  \right] \; \rightarrow \;
\left[n_{c}({\bs p}^\prime) f_{d}({\bs k}^\prime) + f_{c}({\bs p}^\prime) n_{d}({\bs k}^\prime)\right] \left[ 1 \pm n_{b}({\bs k}) \right] \ ,
\ee
and
\be
\ftotalbis_{a}({\bs p}) \ftotalbis_{b}({\bs k}) \left[ 1 \pm  \ftotalbis_{c}({\bs p}^\prime) \right] \left[ 1  \pm \ftotalbis_{d}({\bs k}^\prime)  \right] \; \rightarrow \;
f_{a}({\bs p}) n_{b}({\bs k})  \left[ 1 \pm n_{c}({\bs p^\prime}) \pm n_{d}({\bs k^\prime}) \right]   \ .
\ee
Upon making this single scattering assumption, and upon noting that the medium thermal distribution functions for our brick of noninteracting QGP are known and independent of the time, the 
Boltzmann equation takes the form
\be
\frac{\partial f_{a}({\bs p})}{\partial t} = \sum_n  C_{a}[f_{a}({\bs p})] \Bigr|_{n} \ .
\label{eq:finalBE}
\ee
The sum still runs over all the different binary processes involving species $a$, and the collision operator takes the final form
\be
\begin{split}
 C_{a}[f_{a}({\bs p})]  \Bigr|_{a b \leftrightarrow c d} =  & \, 
\frac{1}{\nu_a} \frac{1}{1 + \delta_{cd}}  \int_{\phaseAll} \left| \mathcal{M}_{a b \leftrightarrow c d} \right|^2 \\ & \left\{ \left[n_{c}({\bs p}^\prime) f_{d}({\bs k}^\prime) + f_{c}({\bs p}^\prime) n_{d}({\bs k}^\prime)\right] \left[ 1 \pm n_{b}({\bs k}) \right] + \right. \\ & \left. - f_{a}({\bs p}) n_{b}({\bs k})  \left[ 1 \pm n_{c}({\bs p^\prime}) \pm n_{d}({\bs k^\prime}) \right]  \right\}\ .
\label{eq:collisionoperator2}
\end{split}
\ee

We can now solve the Boltzmann equation (\ref{eq:finalBE}), within the single scattering approximation. Upon considering the system for a short time interval $\Delta t$ (much shorter than the typical scattering time), the solution to the Boltzmann equation in \Eq{eq:finalBE} takes the form
\be
f_{a}({\bs p} , t_I + \Delta t) = f_{a}({\bs p} , t_I) + \Delta t \sum_n C_{a}[f_{a}({\bs p}, t_I)] \Bigr|_{n} \  ,
\ee
where we have now added explicit mention of the time dependence to our notation. 
Focusing on just a single binary process $a b \leftrightarrow c d$, the solution reads
\be
\label{eq:A24}
\begin{split}
& \, f_{a}({\bs p} , t_I + \Delta t) = \\ & f_{a}({\bs p}, t_I) \left\{ 1 - \frac{\Delta t}{\nu_a} \frac{1}{1 + \delta_{cd}}  \int_{\phaseAll} \left| \mathcal{M}_{a b \leftrightarrow c d} \right|^2 n_{b}({\bs k})  \left[ 1 \pm n_{c}({\bs p^\prime}) \pm n_{d}({\bs k^\prime}) \right]   \right\} + \\ & 
\frac{\Delta t}{\nu_a} \frac{1}{1 + \delta_{cd}}  \int_{\phaseAll} \left| \mathcal{M}_{a b \leftrightarrow c d} \right|^2 \left[n_{c}({\bs p}^\prime) f_{d}({\bs k}^\prime, t_I) + f_{c}({\bs p}^\prime, t_I) n_{d}({\bs k}^\prime)\right] \left[ 1 \pm n_{b}({\bs k}) \right] \ . 
\end{split}
\ee
The probability for the parton $a$ to have momentum $\bs p$ at the time $t_I + \Delta t$, namely the left-hand side of the above equation, is the sum of two contributions, the two terms on the right-hand side. First, we could already have a parton $a$ with momentum $\bs p$ at the initial time $t_I$ and then have no further momentum transfer. Or we could achieve a momentum $\bs p$ at the time $t_I + \Delta t$ by a binary scattering. 
In this paper, we only care about the latter, since we 
are studying binary collisions with large momentum transfer resulting in the presence of a parton
with a large angle deflection with respect to the incoming direction. 
That is, we shall always choose $\bs p$ to point in a direction that differs
from that of the incident parton by some large angle $\theta$, meaning that there is no parton $a$
with momentum $\bs p$ at $t_I$.   
Thus, for our purposes we need only consider the contribution in the last line
of Eq.~(\ref{eq:A24}), which then becomes our \Eq{eq:finaldistr} in the main text after summing
appropriately over different processes.
This is the key result of this Appendix, and the starting point for our analysis in Section~\ref{sec:formalism}.

\section{Phase space integration
\label{sec:delta-int}
}

\subsection{The derivation of Eqs.~\eqref{ave-alpha1} and \eqref{ave-alpha2}}
\label{sec:int-derivation}

We present a detailed derivation of Eqs.~\eqref{ave-alpha1} and \eqref{ave-alpha2} in this Appendix. 
We begin with the desired phase space integration domain \eqref{intphasespace}:
\begin{eqnarray}
\label{delta-int1}
I_{\rm phase}\equiv \int_{\phaseAll} &=&
\frac{1}{2 p}\,\int\, \frac{d^{3}\vp'}{\(2\pi\)^{3} 2p'}\, \int\, \frac{d^{3}\vk'}{\(2\pi\)^{3} 2k'}\, \int\, \frac{d^{3}\vk}{\(2\pi\)^{3}2k}\, 
\no \\
&\times&
(2\pi)^{4}\, \d^{(3)}\(\vp+\vk-\vp'-\vk'\)\,\d\(p+k-p'-k'\)
\no \\
 &=&
 \frac{1}{\(2\pi\)^{3}}\, 
\int^{\infty}_{0} d q_{1}\, q^{2}_{1}\, \int^{1}_{-1} d\cos\theta_{pq_{1}}\,\int^{\infty}_{0} dk'\, k'^{2} \int^{1}_{-1}\, d\cos\theta_{k' q_{1}} \, \int^{2\pi}_{0}\, \frac{d\phi_{1}}{2\pi}\, 
\no \\
&\times&\(\frac{1}{2 p}\)\, \(\frac{1}{2 k}\)\, \(\frac{1}{2 p'}\)\, \(\frac{1}{2 k'}\)\, \d\(k'+p'-k-p\)\, ,  
\end{eqnarray}
where we used the spatial delta function to perform the integration over $d^{3}\vk$ and then shifted variables $\vp'$ to $\vq_{1}\equiv \vp'-\vp$, 
where $\phi_{1}$ is the angle between the $(\vq_{1}, \vp)$ plane and the $(\vq_{1}, \vk')$ plane, 
and where  $\cos\theta_{k' q_{1}}$ and $\cos\theta_{p q_{1}}$ denote the angles between $\vk'$ and $\vq_{1}$ and between $\vp$ and $\vq_{1}$, respectively. 
The integration over the azimuthal angle of $\vq_{1}$ has been performed trivially. 
To further integrate over the remaining delta function in \eqref{delta-int1}, 
we follow the integration technology of Ref.~\cite{Baym:1990uj} (see also Refs.~\cite{Moore:2001fga,Arnold:2003zc,Moore:2004tg}) and consider the identity 
\begin{eqnarray}
\label{delta-Baym}
\d\(k'+p'-k-p\)=\int^{\infty}_{-\infty} d\o_{1} \, \delta \(\o_{1}-\(p-p'\)\)\, \delta \(\o_{1}-\(k'-k\)\)\, . 
\end{eqnarray}
The two delta functions in \eqref{delta-Baym} can be recast as
\begin{eqnarray}
\label{delta-cosp}
\delta \(\o_{1}-\(p-p'\)\)&=& \frac{p'}{p q_{1}}\, \delta \(\cos\theta_{p'q_{1}}- \frac{2\o_{1} p+\o^{2}_{1}-q^{2}_{1}}{2 p q_{1}}\)\, , 
\\
\label{delta-cosk}
\delta \(\o_{1}-\(k'-k\)\)&=& \frac{k}{k' q_{1}}\, \delta \(\cos\theta_{ k'q_{1}}-  \frac{2\o k'+\o^{2}_{1}-q^{2}_{1}}{2 k' q_{1}}\)\, ,
\end{eqnarray}
where we have used kinematic relations
\begin{eqnarray}
\label{pk-kinematic}
p'&=& \sqrt{p^{2}+q^{2}_{1} + 2 p\, q_{1}\cos\theta_{p q_{1}}}\, , 
\qquad
 k = \sqrt{\(k'\)^{2}+q^{2}_{1} + 2 k'\, q_{1}\cos\theta_{k' q_{1}}}\, , 
\end{eqnarray}
which follow from the definition of $\cos\theta_{p q_{1}}$ and $\cos\theta_{k'q_{1}}$. 
Substituting Eq.~\eqref{delta-cosp} and Eq.~\eqref{delta-cosk} into Eq.~\eqref{delta-Baym} and then substituting Eq.~\eqref{delta-Baym} into Eq.~\eqref{delta-int1}, 
we have:
\begin{eqnarray}
\label{delta-int2}
I_{\rm phase}&=& \frac{1}{16\(2\pi\)^{3} p^{2}}\, \int^{\infty}_{-\infty}d\o_{1}\int^{\infty}_{0}dq_{1}\int^{1}_{-1}d\cos\theta_{p q_{1}}\int^{\infty}_{0}dk'\int^{1}_{-1}d\cos\theta_{k'q_{1}}\, 
\no \\
&\times&
\delta \(\cos\theta_{p'q_{1}}- \frac{2\o_{1} p+\o^{2}_{1}-q^{2}_{1}}{2 p q_{1}}\)\, 
\delta\(\cos\theta_{ k'q_{1}}-  \frac{2\o_{1} k'+\o^{2}_{1}-q^{2}_{1}}{2 k' q_{1}}\)\, . 
\end{eqnarray}
The integration over $\cos\theta_{p q_{1}}$ and $\cos\theta_{k' q_{1}}$ in Eq.~\eqref{delta-int2} can be performed trivially when the following kinematic constraints are satisfied:
\begin{eqnarray}
\label{constrainp}
-1&\leq& \frac{2\o_{1} p+\o^{2}_{1}-q^{2}_{1}}{2 p q_{1}} \leq 1\, , 
\\
\label{constraink}
-1&\leq& \frac{2\o_{1} k'+\o^{2}-_{1}q^{2}_{1}}{2 k' q_{1}} \leq 1\, .
\end{eqnarray}
The constraints \eqref{constrainp} and \eqref{constraink} imply that $|\o_{1}|\leq q_{1}\leq 2p+\o_{1}$ 
and $k'\geq \frac{q_{1}-\o_{1}}{2}$.
We consequently have
 \begin{eqnarray}
 \label{delta-int2}
 I_{\rm phase}
 &=&
 \frac{1}{16\,\(2\pi\)^{3}p^{2}}\, \int^{\infty}_{-\infty} d\o_{1}\,\int^{2p+\o_{1}}_{|\o_{1}|}dq_{1} \int^{\infty}_{\(q_{1}-\o_{1}\)/2}\, dk'\, \int^{2\pi}_{0}\frac{d\phi_{1}}{2\pi}\, 
 \no \\
 &=&
 \frac{1}{16\,\(2\pi\)^{3}p}\, \int^{1}_{-1}d\cos\(\Delta\theta_{1}\)\,
 \int^{q_{1}}_{-q_{1}} d\o_{1}\, \(\frac{p'}{q_{1}}\) \int^{\infty}_{\(q_{1}-\o_{1}\)/2}\, dk'\, \int^{2\pi}_{0}\frac{d\phi_{1}}{2\pi}\, , 
 \end{eqnarray}
where $\Delta\theta_{1}=\theta_{1}-\theta$ and $\theta_{1}$ denotes the angle between
the directions of $\vp'$ and ${\bs p}_{\rm in}$. 
Here, we used the relation 
\begin{eqnarray}
d\, q_{1} = \frac{p p'}{q_{1}}\, d\cos\(\Delta\theta_{1}\)\, ,
\end{eqnarray}
which follows from the fact that
\begin{eqnarray}
q^{2}_{1} = p^{2}+\(p'\)^{2}-2 p p'\cos\(\Delta\theta_{1}\)\, .  
\end{eqnarray}
 
We now substitute Eq.~\eqref{delta-int2} into Eq.~\eqref{ave-def-a} to obtain:
\begin{eqnarray}
\label{ave-alpha-anow}
\<\(n\)\>_{D, B}&=& 
 \frac{p\sin\theta }{16\,\(2\pi\)^{5} \,T}\,\int^{\infty}_{-\infty} d\o_{1}\, \int^{-1}_{1}d\cos\(\Delta\theta_{1}\) \,\(\frac{p'}{q_{1}}\) \int^{\infty}_{\(q_{1}-\o_{1}\)/2} dk'\, \int^{2\pi}_{0}\frac{d\phi_{1}}{2\pi}\,
\no\\
&\times&\,
 \Big| \Mnew^{(\a)}\(t,u\)\Big|^{2}\, f_{I}(\vp') n_{D}(k')\, \[1\pm n_{B}(k'+\o_{1})\] \, ,
\end{eqnarray}
To proceed, we express $f_{I}(\vp')$ in Eq.~\eqref{eq:BC} as a function of $\Delta\theta_{1}$ and $\o_{1}$: 
\begin{eqnarray}
\label{fI-1}
f_{I}(\vp') =\,  \frac{1}{V} 
\(\frac{\(2\pi\)^{2}}{\pinit^{2}}\)\, \delta\(\o- \o_{1}\)\,\delta\[\cos\(\theta-\Delta\theta_{1}\)-1\]\, .
\end{eqnarray}
Therefore, the integration over $\o_{1}$ and $\Delta\theta_{1}$ in Eq.~\eqref{ave-alpha-anow} can be performed directly after substituting Eq.~\eqref{fI-1} into Eq.~\eqref{ave-alpha-anow}. 
As a result, 
we replace $\o_{1}$ with $\o$, $q_{1}$ with $q$, $\Delta \theta_{1}$ with $\theta$ and identify $t,u$ with $\tnew,\unew$ as defined in Eq.~\eqref{t1}. 
After relabeling the dummy integration variables $k'$ with $\kT$ and $\phi_{1}$ with $\phi$,
we eventually arrive at Eq.~\eqref{ave-alpha1}.

The derivation of Eq.~\eqref{ave-alpha2} follows similar steps.

\subsection{Integration over $\phi$
\label{sec:phi-int}
}
In this subsection, we demonstrate how to integrate over $\phi$ in Eq.~\eqref{ave-alpha1} and Eq.~\eqref{ave-alpha2} analytically. 

We begin with the observation that $|\CM^{(n)}(\tnew,\unew)|^{2}$ and $|\CM^{(n)}(\unew,\tnew)|^{2}$ can always be decomposed as: 
\begin{eqnarray}
\label{CM-decomposition}
\frac{1}{g_s^4}\Big|\Mnew^{(n)}(\tnew,\unew)\Big|^{2} &=& \sum_{i}\, c^{(n)}_{i}\, m_{i}(\tnew,\unew)\, ,
\qquad
\frac{1}{g_s^4}\Big|\Mnew^{(n)}(\unew,\tnew)\Big|^{2} = \sum_{i}\, \widetilde{c}^{(n)}_{i}\, m_{i}(\tnew,\unew)\, ,
\end{eqnarray}
where we have introduced:
\begin{eqnarray}
\label{m-def}
m_{1} &=&  \(\frac{\snew}{\tnew}\)^{2}
\, , \qquad
m_{2} = \(\frac{\snew}{\tnew}\)
\, , \qquad
m_{3}=1
\, , \qquad
m_{4}=\(\frac{\tnew}{\snew}\)
\, , \qquad
m_{5}=\(\frac{\tnew}{\snew}\)^{2}\, , 
\no
\\ 
m_{6}&=&
\(\frac{\tnew}{\snew+\tnew}\)=-\frac{\tnew}{\unew}
\, , \qquad
m_{7}=\(\frac{\tnew}{\snew+\tnew}\)^{2}
=\(\frac{\tnew}{\unew}\)^{2}\, ,
\end{eqnarray}
and $i$ is summed from 1 to 7.
As a reminder, $\unew=-\snew-\tnew$. 
Here the coefficients $c_{i}^{(n)}$ and $\widetilde{c}_{i}^{(n)}$, with $i=1,2,\ldots, 7$, only depend on $N_{c}$ (i.e. the representation of color gauge group). 
Consequently, 
we have from Eq.~\eqref{CM-decomposition} that:
\begin{eqnarray}
\label{dphi-results}
\frac{1}{g_s^4} \int^{2\pi}_{0} \frac{d\phi}{2\pi}\,\Big|\Mnew^{\(n\)}(\tnew,\unew)\Big|^{2} &=& \sum_{i}\, c^{(n)}_{i}\, M_{i}\(\pinit,p,q, \kT\)\, , 
\no \\
\frac{1}{g_s^4} \int^{2\pi}_{0} \frac{d\phi}{2\pi}\,\Big|
\Mnew^{\(n\)}(\unew,\tnew)\Big|^{2}  &=& \sum_{i}\, \widetilde{c}^{(n)}_{i}\, M_{i}\(\pinit,p,q, \kT\)\, ,
\end{eqnarray}
where 
\begin{eqnarray}
\label{dphi-M}
M_{i}(\pinit,p,q, \kT)&\equiv&\int^{2\pi}_{0} \frac{d\phi}{2\pi}\, m_{i}(\tnew,\unew)\, 
\end{eqnarray}
will be obtained analytically below, in Eq.~\eqref{s-int}.
Substituting Eq.~\eqref{dphi-results} into Eq.~\eqref{ave-alpha1} and Eq.~\eqref{ave-alpha2}, 
we then have:
\begin{eqnarray}
\label{ave-alpha-final}
\< \(n\)\>_{D,B}
&=&
\frac{1}{16\, \(2\pi\)^{3}}\(\frac{p \sin\theta}{\pinit\,q\, T}\)
\sum_{i}\,c^{(n)}_{i}\,\,\int^{\infty}_{\rm k_{\rm min}}\, d\kT\, n_{D}(\kT)\, \[1\pm n_{B}(k_{X}) \]\,
 M_{i}(\pinit, p,q, \kT)\, , 
 \no \\
\label{ave-alpha1-final}
\< \(\widetilde{n}\)\>_{D, B}
&=&\frac{1}{16\, \(2\pi\)^{3}}\(\frac{p \sin\theta}{ \pinit\,q\, T}\)
\sum_{i}\,\widetilde{c}^{(n)}_{i}\,\int^{\infty}_{\rm k_{\rm min}}\, d\kT\, n_{D}(k_{i})\, \[1\pm n_{B}(k_{X}) \]\, M_{i}(\pinit, p,q,\kT)  \, ,\no \\
\end{eqnarray}
where $k_{X}=\kT+\o$.

We now determine the explicit expression for $M_{i}(\pinit,p,q,\kT)$ in Eq.~\eqref{dphi-M}.
To save notation, 
we rewrite Eq.~\eqref{s1} as
\begin{eqnarray}
\label{su-AB}
\snew=\(-\frac{\tnew}{2 q^{2}}\) \[\, A-B\cos\phi\, \]\, , 
\qquad
\unew=-\snew-\tnew=
\(\frac{\tnew}{2 q^{2}}\) \[\, A_{u}-B\cos\phi\, \]\, , 
\end{eqnarray}
where
\begin{eqnarray}
\label{A-B-define}
A\(\pinit,p,q, \kT\) &\equiv& \[ \(\pinit+p\)\,\(\kT+k_{X}\)+q^{2}\]\, , 
\\
A_{u}\(\pinit,p,q, \kT\) &\equiv& A\(\pinit,p,q, \kT\)- 2 q^{2}
=\[ \(\pinit+p\)\,\(\kT+k_{X}\)-q^{2}\]\, ,  
\\
B\(\pinit,p,q, \kT\) &\equiv& \sqrt{\(4\pinit\,p+\tnew\)\,\(4 \kT k_{X}+\tnew\)}\, .
\end{eqnarray}
We then have:
\begin{eqnarray}
\label{s-int}
M_{1}&=&
\int^{2\pi}_{0} \frac{d\phi}{2\pi}\, \(\frac{\snew}{\tnew}\)^{2}
=
\(\frac{1}{4 q^{4}}\)\int^{2\pi}_{0} \frac{d\phi}{2\pi}\, \(A-B\cos\phi\)^{2}
= 
\frac{1}{8 q^{4}}\, \(2 A^{2} + B^{2}\)\, ,
\no \\
M_{2}
&=& \int^{2\pi}_{0} \frac{d\phi}{2\pi}\, \(\frac{\snew}{\tnew}\)
=
\(-\frac{1}{2 q^{2}}\)\, \int^{2\pi}_{0} \frac{d\phi}{2\pi}\, \(A-B\cos\phi\)
= \(-\frac{1}{2 q^{2}}\)\, A\,  ,
\no \\
M_{3}&=&
\int^{2\pi}_{0} \frac{d\phi}{2\pi}\, 1 = 1\, ,
\no \\
M_{4}&=&
\int^{2\pi}_{0} \frac{d\phi}{2\pi}\, \(\frac{\tnew}{\snew}\)
=
\(-2q^{2}\)\int^{2\pi}_{0} \frac{d\phi}{2\pi}\, \frac{1}{A- B\cos\phi}
= \frac{-2 q^{2}}{\sqrt{A^{2}-B^{2}}}\, , 
\no \\
M_{5}&=&
\int^{2\pi}_{0} \frac{d\phi}{2\pi}\, \(\frac{\tnew}{\snew}\)^{2}
=
\( 4q^{4}\)\int^{2\pi}_{0} \frac{d\phi}{2\pi}\, \frac{1}{\(A- B\cos\phi\)^{2}}
= \frac{ 4 q^{4}}{\(A^{2}-B^{2}\)^{3/2}}\, , 
\no \\
M_{6}&=&
\int^{2\pi}_{0} \frac{d\phi}{2\pi}\, \(-\frac{\tnew}{\unew}\)
=
\(-2q^{2}\)\int^{2\pi}_{0} \frac{d\phi}{2\pi}\, \frac{1}{A_{u}- B\cos\phi}
= \frac{-2 q^{2}}{\sqrt{A^{2}_{u}-B^{2}}}\, , 
\no \\
M_{7}&=&
\int^{2\pi}_{0} \frac{d\phi}{2\pi}\, \(\frac{\tnew}{\unew}\)^{2}
=
\( 4q^{4}\)\int^{2\pi}_{0} \frac{d\phi}{2\pi}\, \frac{1}{\(A_{u}- B\cos\phi\)^{2}}
= \frac{ 4 q^{4}}{\(A^{2}_{u}-B^{2}\)^{3/2}}\, . 
\end{eqnarray}
With the explicit expressions for $M_{i}$  in Eq.~\eqref{s-int} in hand, 
there is only one integration (over $k_{T}$) remaining in each of the two expressions 
in Eq.~\eqref{ave-alpha-final} that must be performed numerically,
as we advertised earlier.

\section{Comparison with previous results
\label{sec:compare}
}

\subsection{The relation between $\PT(\qperp)$ and $P(\theta)$ 
\label{sec:relation}
}

To elucidate the connection with previous studies~\cite{Aurenche:2002pd,Arnold:2008zu,DEramo:2012uzl} in which  the two-dimensional probability distribution
for the transverse momentum of the outgoing parton,  $\PT(\qperp)$, 
has been computed, we need to relate this 
quantity, normalized as
\begin{eqnarray}
\label{PT-normalization}
\int \frac{d^{2}\qperp}{(2\pi)}\, \PT(\qperp)
=\int \frac{d \qperp}{2\pi}\, \qperp\, \PT(\qperp) =1\, ,
\end{eqnarray}
to the probability distribution 
$P(\theta)$ for the angle $\theta$ that we compute.
Since the previous studies all work in a limit in which $\pinit$ is large
and $\theta$ is small, energy loss is negligible in these studies,
i.e. $p\approx \pinit$, and hence 
\begin{eqnarray}
\label{qperp-pi-relation}
\qperp=\pinit\, \sin\theta\, .
\end{eqnarray}
We shall also need to take into account  the Jacobian $\CJ_{\perp}$ defined through the relation:
\begin{eqnarray}
\label{FullJacobian}
\int^{\pi}_{0} d\theta\, \int^{\infty}_{0} dp
= \int \frac{d^{2}\qperp}{\(2\pi\)^{2}}\int dp\,  \CJ_{\perp}(p, \qperp)\, , 
\quad
\CJ_{\perp}
&=& \(\frac{2\pi}{\qperp}\) \frac{1}{\sqrt{p^{2}-\qperp^{2}}}
= \frac{2\pi}{p^{2}\sin\theta\cos\theta}\, .
\end{eqnarray}
We shall use this expression, with $p$ replaced by $\pinit$, in Eqs.~(\ref{PAD-theta}) and (\ref{P-GA}).

It simplifies the explicit comparisons that we shall make in Section~\ref{sec:comparison}
if there we work in the small-$\theta$ limit in which
$q_\perp \approx \pinit \theta$ 
and $\CJ_{\perp}$  reduces to
\begin{eqnarray}
\label{J-perp}
\CJ_{\perp}\approx \tilde{\CJ}_{\perp}
= \frac{2\pi}{p^{2}_{i}\,\theta}\, .
\end{eqnarray}
In Section~\ref{sec:comparison} our goal will be to check whether the following relation holds:
\begin{eqnarray}
\label{PTheta-relation}
\lim_{\theta\to 0}\[ \tilde{\CJ}_{\perp}\,P(\theta)\] =
\PT_{\single}(\qperp=\pinit\, \theta)\, ,
\end{eqnarray}
where $P(\theta)$ is the result of our calculation and $\PT_{\single}(\qperp)$
is one of the results from Refs.~\cite{Aurenche:2002pd,Arnold:2008zu,DEramo:2012uzl}
for $\PT(\qperp)$ due to a single binary collision.

\subsection{Previous results, compared to ours
\label{sec:comparison}
}

The expression for $\PT(\qperp)$ due to a single binary collision, $\PT_{\single}(\qperp)$, has been obtained 
in the limit $m_{D}\ll\qperp \ll T$,  
by Aurenche, Gelis and Zaraket (AGZ)~\cite{Aurenche:2002pd}, who showed that (in our notation)
\begin{eqnarray}
\label{PAGZ}
\PT^{{\rm AGZ}}_{\single}\(\qperp\) &=& 
\kappa\,C_{A}\,\(\frac{m^{2}_{D}}{g^{2}_{s}}\)\, \frac{1}{q^{4}_{\perp}}\, 
\end{eqnarray}
in this regime, and in the limit $\qperp \gg T$
by Arnold and Dogan (AD)~\cite{Arnold:2008zu}, who showed that
\begin{eqnarray}
\label{PAD}
\PT^{{\rm AD}}_{\single}\(\qperp\) &=& 
\kappa\, C_{A}
\(4N_{c}+3N_{f}\)\, 
 \(\frac{\zeta(3)\, T^{2}}{2\pi^{2 }}\)\,
  \frac{1}{q^{4}_{\perp}}
\end{eqnarray}
in this regime.
Each of these expressions is a limiting case of the more general expression
for
$\PT_{\single}(\qperp)$ computed by
D{'}Eramo, Lekaveckas, Liu and Rajagopal (DLLR)~\cite{DEramo:2012uzl}. In the limit $q_{\perp}\gg m_{D}$ their result can be written as (see Eq.~(5.2) and Eq.~(5.15) of Ref.~\cite{DEramo:2012uzl}):
\begin{eqnarray}
\label{P-DLLR}
\PT^{{\rm DLLR}}_{\single}(q_{\perp})
= \frac{2\kappa\, C_{A}}{g_s^2 T} \, \int \frac{d \o }{2\pi}\, \left[ 1+n_{\rm B.E.}(\o)\right] \, \frac{\(\Im\Pi^{L}-\Im\Pi^{T}\)
}
{ q^{2}_{\perp}\, q^{2}}\, ,
\end{eqnarray}
where $\Im \,\Pi^{T,L}$ are the imaginary parts of the gluon longitudinal and transverse self energy in QGP. 
To obtain Eq.~\eqref{P-DLLR}, 
we have used the relation $q_{z}\approx \o$ which is valid in the limit \eqref{limit}. (See Eq.~\eqref{q-limit} below.)
After evaluating the integration in Eq.~\eqref{P-DLLR} explicitly by substituting the appropriate expressions for the self-energies $\Pi^{T,L}$, 
Eqs.~\eqref{PAGZ} and \eqref{PAD} are each reproduced upon
taking the appropriate limits, as demonstrated in Ref.~\cite{DEramo:2012uzl}. 
The above expressions \eqref{PAGZ}, \eqref{PAD}, \eqref{P-DLLR} were all
obtained for an incident gluon. 
Those for an incident quark/antiquark differ only in that $C_{A}$ must then be replaced by
$C_{F}$ in each case. 

We wish to use (\ref{PTheta-relation}) to compare our results $P(\theta)$ in the limit $\theta\to 0$ with the results
from previous studies above.
In order to obtain $P(\theta)$ in the small $\theta$ limit from our calculation, 
we first need to consider $F^{C\to \all}(p,\theta)$ in this limit. 
As we have already observed in our results as presented in Sec.~\ref{sec:FandP}, and 
as we shall check explicitly later, when $\theta$ is small
the integration over $p$ in Eq.~\eqref{P-Theta} is dominated by $p\approx\pinit$, namely 
$\o/\pinit \ll 1$. 
We therefore wish to analyze $F^{C\to \all}(p,\theta)$ in the limit \eqref{limit}
and subsequently evaluate $P(\theta)$ using Eq.~\eqref{P-Theta}:
\begin{eqnarray}
\label{P-Theta-small}
P(\theta)
=\int^{\pinit-p_{\min}}_{-\infty}\, d\o\, F^{C\to \all}(p=\pinit-\o,\theta)
=\int^{\infty}_{-\infty}\, d\o\, F^{C\to \all}(p=\pinit-\o,\theta)\, ,
\end{eqnarray}
where we have first changed the integration variable from $p$ to $\o$ and then changed the upper limit of the integration range from $\pinit-p_{\min}$ to $\infty$ since $\pinit\gg \o, p_{\min}$.
For later use, we also note that in the limit \eqref{limit} the expression \eqref{qt-in-theta}
for $\tnew$ simplifies:
\begin{eqnarray}
\label{tnew-limit}
\tnew=-2p\,\pinit\(1-\cos\theta\)
\approx - \pinit^{2}\,\theta^{2}
\approx -q^{2}_{\perp}\, .
\end{eqnarray}
Consequently, 
we have from the first equation in \eqref{t1} that 
\begin{equation}
\label{q-limit}
q^{2}=\o^{2}+\tnew \approx \o^{2}+q^{2}_{\perp}\, , 
\end{equation}
and consequently $\o \approx q_{z}$ as we mentioned earlier.

In the limit \eqref{limit}, 
$\tnew$ vanishes as 
$\tnew \approx - \pinit^2 \theta^{2}$ (see Eq.~\eqref{tnew-limit}) while $\snew$ remains finite.
This implies that $m_{1}=\(\snew/\tnew\)^{2}$ will diverge as $1/\theta^{4}$ and therefore will be dominant over the other terms in Eq.~\eqref{CM-decomposition}:
\begin{eqnarray}
\Mnew^{(n)}(\tnew,\unew) \approx  c^{(n)}_{1}\, \(\snew/\tnew\)^{2}\, ,
\qquad
\Mnew^{(n)}(\unew,\tnew) \approx  \widetilde{c}^{(n)}_{1}\, \(\snew/\tnew\)^{2}\, .
\end{eqnarray}
Consequently, Eq.~\eqref{ave-alpha1} and Eq.~\eqref{ave-alpha2} become:
\begin{eqnarray}
\label{alpha-small}
\<\(n\)\>_{D,B}
&\approx& c^{(n)}_{1} 
\<\<\(\frac{\snew}{\tnew}\)^{2} \>\>_{D,B}\, ,
\qquad
\<\(\widetilde{n}\)\>_{D,B}
\approx
 \widetilde{c}^{(n)}_{1}
 \<\<\(\frac{\snew}{\tnew}\)^{2} \>\>_{D,B}\, ,
\end{eqnarray}
where we have introduced the notation
\begin{eqnarray}
\label{st-ave-def}
\<\<\(\frac{\snew}{\tnew}\)^{2}\>\>_{D,B}
&\equiv& \frac{1}{16\(2\pi\)^{2} q\, T}\, \(\frac{p^{2}_{i}\theta}{2\pi}\)
\, \int^{\infty}_{k_{\min}}\, d\kT\, n_{D}(\kT)\(1\pm n_{B}(k_{X})\)\,\frac{1}{p^{2}_{i}}\int \frac{d\phi}{2\pi}\, 
\(\frac{\snew}{\tnew}\)^{2}\, ,
\nonumber\\
&=&\frac{1}{32\, q^{5} T}\frac{1}{\(2\pi\)^{2}} \, \(\frac{p^{2}_{i}\theta}{2\pi}\)
\nonumber\\
&\times&
\, \int^{\infty}_{k_{\min}}\, d\kT\, n_{D}(\kT)\[1\pm n_{B}(\kT+\o)\]\,
H(\kT, q, \o)\, ,
\end{eqnarray}
with 
\begin{eqnarray}
\label{m1-small}
H(\kT,q,\o)\equiv \frac{2 q^4}{p^{2}_{i}}\int \frac{d\phi}{2\pi}\, 
\(\frac{\snew}{\tnew}\)^{2} \ ,
\end{eqnarray}
and where $\tilde s$ and $\tilde t$ are to be expressed using Eqs.~\eqref{t1} and \eqref{s1}.
Using the first equation in Eq.~\eqref{s-int} and the behavior of $A,B$ defined in Eq.~\eqref{A-B-define} upon taking the limit \eqref{limit}, namely
\begin{eqnarray}
\label{AB-approx}
\frac{A^{2}}{p^{2}_{i}}\approx
4(2\kT+\o)^{2}
\, , 
\qquad
\frac{B^{2}}{p^{2}_{i}}\approx\, 
 4\[ 4 \kT \(\kT+\o\) + \o^{2}-q^{2}\]\, ,
\end{eqnarray} 
we have
\begin{eqnarray}
\label{H-def}
H(\kT, q, \o)\approx \(12 \kT^{2} +12 \kT\, \o+3 \o^{2}- q^{2}\)\, . 
\end{eqnarray}

Next, the nonzero values of $c^{(n)}_{1}$ (and $\widetilde{c}^{(n)}_{1}$) can be computed straightforwardly through their definitions \eqref{CM-decomposition}:
\begin{eqnarray}
\label{t-coff}
c^{\(1,2\)}_{1} &=&\widetilde{c}^{\(1,2\)}_{1} 
= c^{(3,4,5,6)}_{1}
=\frac{16 C^{2}_{F} d^{2}_{F}}{d_{A}}
=8 C_{F} d_{F}
\nonumber\\
c^{(9,10)}_{1}
&=& 16 C_{A}\, C_{F}\, d_{F}
=8 C_{A}\, d_{A}\,  ,
\qquad
c^{(11)}_{1}=\widetilde{c}^{(11)}_{1}
= 16 C^{2}_{A}\, d_{A} 
=16 N_{c}\, C_{A}\, d_{A}\, , 
\end{eqnarray}
where we have used the relation $\(C_{F}\, d_{F}/d_{A}\)=1/2$ and $C_{A}=N_{c}$.
One important consequence of Eq.~\eqref{t-coff}, 
in particular the fact that $c^{(8)}_{1} = \widetilde{c}^{(3,6,7,8,9)}_{1}=0$,
can be found by substituting these generic results 
together with Eq.~\eqref{alpha-small} which is valid in the small-$\theta$ limit into
Eqs.~\eqref{FQQ}, \eqref{FQG},   \eqref{FGQ}, \eqref{FGG}, \eqref{FQQbar} and 
discovering that  
$F^{G\to G}\(p, \theta\)\gg F^{G\to Q}\(p, \theta\),F^{G\to \bar{Q}}\(p, \theta\)$ and $F^{Q\to Q}\(p, \theta\)\gg F^{Q\to G}\(p, \theta\), F^{Q\to \bar{Q}}\(p, \theta\)$.
 That is, 
 \begin{eqnarray}
 F^{G\to\all}\(p, \theta\) &\approx&  F^{G\to G}\(p, \theta\)\ , \nonumber\\
  F^{Q\to\all}\(p, \theta\) &\approx&  F^{Q\to Q}\(p, \theta\)\ , \nonumber\\
  F^{{\bar Q}\to\all}\(p, \theta\) &\approx&  F^{{\bar Q}\to {\bar Q}}\(p, \theta\)\, . 
 \end{eqnarray}
This simply reflects the fact that in the small-$\theta$ limit, Rutherford-like scattering (in which the parton that is detected is the incident parton after scattering) 
is much more important than other channels. 
We will focus on $F^{G\to G}(p,\theta)$ and  $F^{Q\to Q}(p,\theta)$ from now on and write explicit expressions for them by substituting Eqs.~\eqref{alpha-small} and \eqref{t-coff}  into Eq.~\eqref{FGG}, obtaining
\bes
\begin{eqnarray}
\label{FGG-small}
F^{G\to G}\(p, \theta\)
&=&
\frac{\kappa
}{\nu_{g}}\, 
\[ 2 N_{f}\, c^{(9)}_{1}\, \<\<\(\frac{\snew}{\tnew}\)^{2}\>\>_{Q,Q}  + 
c^{(11)}_{1}\,\<\<\(\frac{\snew}{\tnew}\)^{2}\>\>_{G,G} \]
\no\\
&=&\frac{\kappa
}{2\, d_{A}}\, 
\[ 2 N_{f} \(8 C_{A}\, d_{A}\) \<\<\(\frac{\snew}{\tnew}\)^{2}\>\>_{Q,Q}  + 
\(16 N_{c}\, C_{A}\, d_{A}\)\<\<\(\frac{\snew}{\tnew}\)^{2}\>\>_{G,G} \]
\no \\
&=&8 C_{A}\,\kappa
\[ N_{f}\,\<\<\(\frac{\snew}{\tnew}\)^{2}\>\>_{Q,Q} + N_{c}\, \<\<\(\frac{\snew}{\tnew}\)^{2}\>\>_{G,G} \]\, , 
\end{eqnarray}
and into Eq.~\eqref{FQQ}, obtaining
\begin{eqnarray}
\label{FQQ-small}
F^{Q\to Q}(p, \theta)
&=& \frac{\kappa
}{\nu_{q}}
\{ 
\[c^{(1)}_{1}+c^{(3)}_{1}+2\(N_{f}-1\)\,c^{(4)}_{1}\]
\<\<\(\frac{\snew}{\tnew}\)^{2}\>\>_{Q,Q}
+c^{(9)}_{1}\, \<\<\(\frac{\snew}{\tnew}\)^{2}\>\>_{G,G}
\}
\no \\
&=& \frac{\kappa}{2\,d_{F}}
\[ 
2\, N_{f}\,\(8 C_{F} d_{F}\)
\<\<\(\frac{\snew}{\tnew}\)^{2}\>\>_{Q,Q}
+\(16 C_{A}\, C_{F}\, d_{F}\)\, \<\<\(\frac{\snew}{\tnew}\)^{2}\>\>_{G,G}
\]
\no \\
&=&8 C_{F}\,\kappa
\[ N_{f}\,\<\<\(\frac{\snew}{\tnew}\)^{2}\>\>_{Q,Q} + N_{c}\, \<\<\(\frac{\snew}{\tnew}\)^{2}\>\>_{G,G} \]\, , 
\end{eqnarray}
\ees
where we have used $\nu_{q}=2 d_{F}, \nu_{g}=2 d_{A},C_{A}=N_{c}$.
Comparing Eq.~\eqref{FQQ-small} with Eq.~\eqref{FGG-small}, 
we obtain the relation 
\begin{eqnarray}
\label{casimir}
F^{Q\to Q}(p,\theta)
= \frac{C_{F}}{C_{A}}\, 
F^{G\to G}(p,\theta)\, . 
\end{eqnarray}
We can now compute the left-hand-side of (\ref{PTheta-relation}) for an incident gluon by substituting Eq.~\eqref{FGG-small} into Eq.~\eqref{P-Theta-small}. We find that
\begin{eqnarray}
\label{PG-small-final}
\lim_{\theta\to 0} \,\tilde{\CJ}_{\perp}\, P(\theta)&=&
8\, \kappa\, C_{A}\, \tilde{\CJ}_{\perp}\,\int^{\infty}_{-\infty}\, d\o\, 
\[ N_{c}\, \<\<\(\frac{\snew}{\tnew}\)^{2}\>\>_{G,G} +N_{f}\,\<\<\(\frac{\snew}{\tnew}\)^{2}\>\>_{Q,Q} \]\, 
\no\\
 &=&
\frac{\kappa\, C_{A}}{8\pi T}\, 
\int^{\infty}_{-\infty}\frac{d\o}{2\pi}\, \frac{1}{q^{5}}\,\int^{\infty}_{k_{\min}}\kT\, H(\kT, q,\o)\, 
\no \\
&\times&
\Bigl[ N_{c}\, n_{\rm B.E}(\kT)\, \[1+n_{\rm B.E}(\kT+\o)\]+N_{f}\, n_{\rm F.D}(\kT)\, \[1-n_{\rm F.D}(\kT+\o)\]\Bigr]
\, ,\nonumber\\
\end{eqnarray}
where we have used Eq.~\eqref{st-ave-def}.
For an incident quark, the resulting $P(\theta)$ can be obtained by replacing $C_{A}$ with $C_{F}$ thanks to the relation \eqref{casimir}. 
Eq.~\eqref{PG-small-final} is a central result of this Appendix, as it will allow us to compare  our
results to those obtained previously in the limits in which such comparisons can be made.

In order to compare our result to the AGZ result \eqref{PAGZ}~\cite{Aurenche:2002pd}
we must evaluate our expression \eqref{PG-small-final} in the limit $\o, \qperp \ll T$.
We see from Eq.~\eqref{q-limit} that in this limit $q\ll T$.
Since the characteristic $\kT$ is of the order of $T$,
we can set $\o=0$ in $n_{B.E}(\kT+\o)$ and $n_{F.D.}(\kT+\o)$ in Eq.~\eqref{PG-small-final}. 
Furthermore, $k_{\min}= 0$ in Eq.~\eqref{PG-small-final}. 
From Eq.~\eqref{H-def}, we see that in this limit
we can also replace $H$ in Eq.~\eqref{PG-small-final} with
\begin{eqnarray}
\label{H-limit}
H(\kT, \o, q) \approx12 \kT^{2}\, .
\end{eqnarray}
The integration in Eq.~\eqref{PG-small-final} can then be evaluated analytically by using 
\begin{eqnarray}
\int^{\infty}_{0}
 d\kT\, n_{\rm B.E}(\kT)
 \[1+n_{\rm B.E}(\kT)\] \, \kT^{2}&=&\frac{\pi^{2}}{3}T^{3}\, , 
\no\\
\int^{\infty}_{0} d\kT\, n_{F.D}(\kT)\[1-n_{F.D}(\kT)\] \, \kT^{2}
&=&\frac{\pi^{2}}{6}T^{3}\, , 
\end{eqnarray}
and
\begin{eqnarray}
\label{o-int}
\int^{\infty}_{-\infty}\, d\o\, \frac{1}{q^{5}}
=\int^{\infty}_{-\infty}\, d\o\, \frac{1}{\(q^{2}_{\perp}+\o^{2}\)^{5/2}}
= \frac{4}{3\, q^{4}_{\perp}}\, .
\end{eqnarray}
As a result, we have:
\begin{eqnarray}
\label{AGZ-Theta}
\lim_{\theta\to 0} \tilde{\CJ}_{\perp} \,P(\theta)
= \frac{1}{3}\kappa\,C_{A}\, 
\(N_{c}+N_{f}/2\)\, \frac{T^{2}}{q^{4}_{\perp}}
= \kappa\, C_A \frac{\(m^{2}_{D}/g_s^{2}\)}{q^{4}_{\perp}} 
\end{eqnarray}
where the Debye mass $m_D$  is given by Eq.~\eqref{mD}.
 We observe that, as advertised earlier, Eq.~\eqref{AGZ-Theta} is equivalent to the 
 AGZ result \eqref{PAGZ} through the relation \eqref{PTheta-relation}.   
 It is worth noting that the dominant contribution to the integration in Eq.~\eqref{o-int} comes from $\o \sim q_{\perp}\ll \pinit$, 
 which justifies taking the limit $\o/\pinit \ll 1$ in $F^{C\to\all}\(p,\theta\)$.

We now turn to comparing our result \eqref{PG-small-final}  to the DLLR result \eqref{P-DLLR}~\cite{DEramo:2012uzl}. 
To simplify the discussion, 
we will only include the contribution coming from thermal scatterers which are gluons. 
This amounts to setting $N_{f}=0$ in Eq.~\eqref{PG-small-final}, obtaining
\begin{eqnarray}
\label{PG-small-final-G}
\lim_{\theta\to 0} \CJ_{\perp} \,P(\theta) &=&
\frac{\kappa\, C_{A}
}
{8\pi T}\, 
\int^{\infty}_{-\infty}\frac{d\o}{2\pi}\,\frac{1}{q^{5}} \int^{\infty}_{k_{\min}}\kT\, H(\kT, q,\o)\, 
\, N_{c}\, n_{\rm B.E.}(\kT)\, \[1+n_{\rm B.E.}(\kT+\o)\]\, .\nonumber\\
\end{eqnarray}
Correspondingly, we will only include the contribution to the gluon self-energy $\Pi^{L,T}$ in
Eq.~\eqref{P-DLLR} that comes from gluon loops,
and show that the resulting $\PT^{{\rm DLLR}}_{\single}(q_{\perp})$ is equivalent to Eq.~\eqref{PG-small-final-G} through the relation \eqref{PTheta-relation}. 
The comparison upon including the contribution coming from fermionic thermal scatterers (quark and antiquark) is quite similar. 

To proceed, we write the explicit expressions for $\Im\, \Pi_{L}$ and $\Im\, \Pi_T$ coming from the gluon loop as given in 
Ref.~\cite{DEramo:2012uzl}:
\begin{eqnarray}
\label{PiL}
\frac{\Im\, \Pi_{L}}{g^{2}_{s}}
= \(\frac{N_{c}}{8\pi}\) \(\frac{q^{2}_{\perp}}{q^{3}}\)
\,\Biggl\{\int^{\infty}_{(q-\o)/2}d\kT\, n_{B.E.}(\kT)\, \[\(2\kT+\o\)^{2}-2 q^{2}\] - \(\o\to -\o\)\Biggr\}\, ,\nonumber\\
\end{eqnarray}
and 
\begin{eqnarray}
\label{PiT}
\frac{\Im\,\Pi_{T}}{g^{2}_{s}} =  -\(\frac{N_{c}}{16\pi}\) \(\frac{q^{2}_{\perp}}{q^{3}}\) 
\,\Biggl\{\int^{\infty}_{(q-\o)/2}d\kT\, n_{\rm B.E.}(\kT)\, \[ \(2\kT+\o\)^{2}-q^{2}\] - \(\o\to -\o\)\Biggr\}.\nonumber\\
\end{eqnarray}
The contribution from fermion loops can be obtained by replacing $n_{\rm B.E.}$ with $n_{\rm F.D.}$ and replacing $N_{c}$ with $N_{f}$. 
Adding Eq.~\eqref{PiL} and Eq.~\eqref{PiT}, we have:
\begin{eqnarray}
\label{self-energy}
[ 1 &+& n_{\rm B.E.}(\o) ] \frac{\(\Im\Pi_{L}-\Im\Pi_{T}\)}{g^{2}_{s}}
=
 \no \\
&=&\(\frac{N_{c}}{16\pi}\) \(\frac{q^{2}_{\perp}}{q^{3}}\)\, \[1+n_{\rm B.E.}(\o)\]
\,\Biggl\{\int^{\infty}_{(\o-\kT)/2}\, d\kT\, n_{\rm B.E.}(k)\, \[ H(\o, q, \kT)- \(\o\to -\o\)\]\Biggr\}
\no \\
&=&
\(\frac{N_{c}}{16\pi}\)\, \(\frac{q^{2}_{\perp}}{q^{3}}\)\, 
\,\int^{\infty}_{k_{\min}}\, d\kT\, n_{\rm B.E}(\kT)\, \[1+n_{\rm B.E}\, \(\kT+\o\)\]\,  H(\o, q, \kT)\, ,
\end{eqnarray}
where $H(\o, q, \kT)$ is given by Eq.~\eqref{H-def} and where we have used the identity
\begin{eqnarray}
\[1+n_{\rm B.E}(\o)\]\,\[n_{\rm B.E}(k)-n_{\rm B.E.}(k+\o)\]
= n_{\rm B.E}(k)\, \[1+n_{\rm B.E}\, (k+\o)\]\, .
\end{eqnarray}
Finally, 
we substitute Eq.~\eqref{self-energy} into the DLLR result \eqref{P-DLLR}. 
It is now transparent that our expression \eqref{PG-small-final-G} is equivalent to Eq.~\eqref{P-DLLR} through the the relation~\eqref{PTheta-relation}.

Noting that it has been demonstrated in Ref.~\cite{DEramo:2012uzl} that the AD result (\ref{PAD}) 
is obtained from the DLLR result (\ref{P-DLLR}) in the $q_\perp \gg T$ limit, this concludes
our verification that our result, in particular in the form \eqref{PG-small-final}, 
reduces to the previously known AGZ, AD and DLLR results in the appropriate limits.

 \end{appendix}

\bibliographystyle{JHEP}
\bibliography{MoliereQGP}
\end{document}